\documentclass[pdflatex, sn-mathphys, referee]{sn-jnl}
\usepackage[english]{babel}
\usepackage[utf8]{inputenc}
\usepackage{pdfpages} 

\usepackage{amsthm}
\usepackage{mathtools}
\usepackage{physics}
\usepackage{xcolor}
\usepackage{graphicx}
\usepackage{placeins}
\usepackage[T1]{fontenc}
\usepackage{lipsum}
\usepackage{csquotes}

\begin{document}
\title{Universal quasi-particle kinetics control the cell death decision}

\author[1,2]{Felix Meige$^\$ $}
\author[3,4]{Lina Hellwig$^\$ $}
\author[5]{Harald Stachelscheid}
\author[3,4,6]{Philipp Mergenthaler}
\equalcont{These authors contributed equally as senior authors}
\author*[1,2]{Steffen Rulands}
\equalcont{These authors contributed equally as senior authors\\
\hspace{-2cm}$^\$$These authors contributed equally as first authors}

\email{rulands@lmu.de}

\affil[1]{Max Planck Institute for the Physics of Complex Systems, Noethnitzer Str. 38, 01187 Dresden, Germany}
\affil[2]{Ludwigs-Maximilians-Universität München, Arnold Sommerfeld Center for Theoretical Physics, Theresienstr. 37, 80333 München, Germany}
\affil[3] {Charité - Universitätsmedizin Berlin, Center for Stroke Research Berlin, Charitéplatz 1, 10117 Berlin, Germany}
\affil[4] {Charité - Universitätsmedizin Berlin, Department of Neurology with Experimental Neurology, Charitéplatz 1, 10117 Berlin, Germany}
\affil[5] {Berlin Institute of Health at Charité - Universitätsmedizin Berlin, Core Unit pluripotent Stem Cells and Organoids, Charitéplatz 1, 10117 Berlin, Germany}
\affil[6] {Radcliffe Department of Medicine, University of Oxford, Oxford, UK}

\date{\today} 

\abstract{
Understanding how fluctuations propagate across spatial scales is central to our understanding of inanimate matter from turbulence to critical phenomena. In contrast to physical systems, biological systems are organized into a hierarchy of processes on a discrete set of spatial scales: they are compartmentalized. Here, we show that dynamic compartmentalization of stochastic systems leads to emergent, quasi-particle-like kinetics which are used by cells to perform key biological functions. Specifically, we derive a general theory that predicts the emergence of a single degree of freedom irrespective of system specifics. We obtain equations of motion and response characterising its unique kinetic properties. We experimentally demonstrate the biological relevance of quasi-particle kinetics in the decision of cells to commit suicide (apoptosis). Using fluorescent microscopy, we show that the response of cells to apoptotic stimuli exhibits quasi-particle like kinetics which establish a low-pass filter for cellular stress signals. By highlighting that cells manipulate how noise and signals propagate across spatial scales, our work reveals a new mechanism of cell fate decision-making.
}


\maketitle

Understanding how fluctuations propagate across spatial scales is the foundation of theories of inanimate matter. For example, in turbulence velocity fluctuations induced on the macroscopic scale propagate to the microscopic scale and dissipate there \cite{kolmogorov_dissipation_1997,sano_universal_2016,yamada_anatomy_2008}. In critical phenomena, fluctuations propagate self-similarly from the microscopic to the macroscopic scale, leading to divergences in thermodynamic response functions \cite{goldenfeld_lectures_2019,bi_density-independent_2015,bialek_statistical_2012}. Understanding these phenomena required theoretical methodologies in the form of renormalization group theory \cite{yeomans_statistical_1992,cavagna_natural_2023,caballero_bulk_2018} or coupled modes theory \cite{christopoulos_coupled-mode-theory_2016,janssen_mode-coupling_2018,zanotto_perfect_2014}, which allow for calculating the emergent effect of all intermediary scales on physical properties of interest. 

\begin{figure}
    \centering
\includegraphics[width=1.0\linewidth]{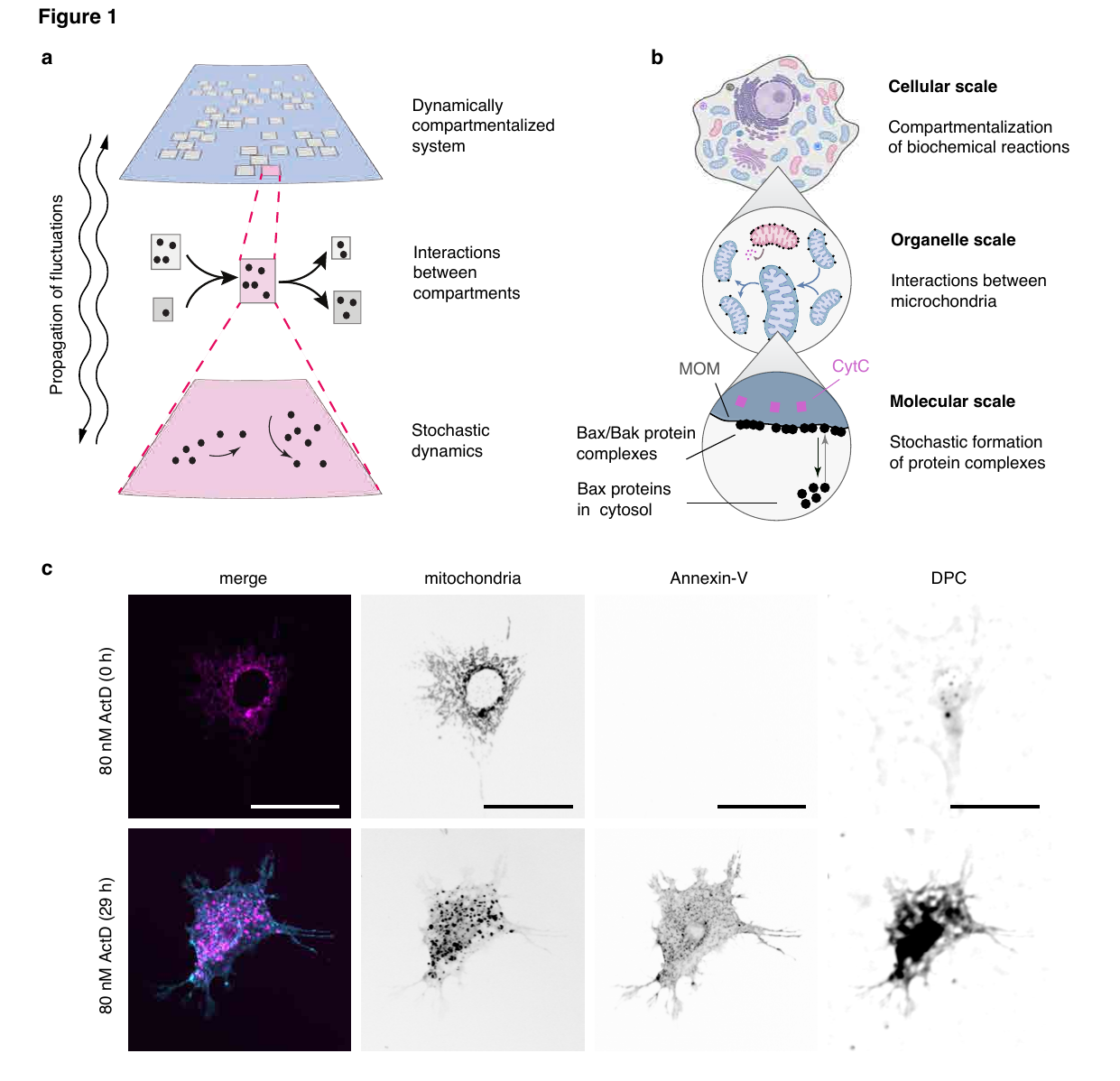}
    \caption{\textbf{a} Schematic representing a dynamically compartmentalized system. \textbf{b} Schematic depicting the regulation of apoptosis with mitochondria-bound Bax. \textbf{c} Fluorescence microscopy images of a mouse embryonic fibroblast (MEF) cell undergoing apoptosis after treatment with 80\,nM actinomycin D (ActD). Shown are timepoints immediately after addition of ActD (0h, no apoptotic activity) and after 29h with highly fragmented mitochondria and Annexin V staining as a marker of apoptosis. Mitochondria are visualized by TMRE, apoptotic activity by Annexin V staining, DPC (digital phase contrast, representing cell borders), merge (TMRE and Annexin V channels), scale bars: 50 µm.
    This cell is taken from the full data set in Fig. 5c.}
    \label{fig:model}
\end{figure}

In contrast to inanimate matter, biological systems are organized into a discrete hierarchy of non-equilibrium processes on vastly different spatial scales \cite{patalano_self-organization_2022,meigel_controlling_2024,duso_stochastic_2020,anderson_stochastic_2023}. Molecular processes are often embedded into subcellular structures termed organelles and cells interact to form complex organs that constitute organisms. Biological systems have explicit mechanisms that propagate fluctuations and signals across scales. As an example, the decision of cells to commit suicide (i.e. apoptosis) is mediated  by complexes of  apoptosis-regulating proteins such as Bax \cite{newton_cell_2024}  (Fig.~\ref{fig:model}b). Bax is recruited from the cytosol to the outer membrane of subcellular compartmens termed mitochondria (mitochondrial outer membrane, MOM) at a rate that depends on the cell’s stress level \cite{kale_bcl-2_2018,lovell_membrane_2008}. In the MOM, these proteins undergo stochastic dynamics that lead to the formation of pores. These pores release the protein cytochrome C (CytC) from mitochondria, thereby irreversibly executing apoptosis \cite{czabotar_mechanisms_2023,kale_bcl-2_2018}. The mitochondria themselves are highly dynamic organelles that undergo rapid fusion and fission \cite{tabara_molecular_2024}, which leads to concentration changes of protein complexes on their membranes. This gives rise to a two-way propagation of fluctuations and signals between the molecular and the organelle scale (Fig.~\ref{fig:model}b).  Figure~\ref{fig:model}c shows fluorescent microscopy images of the morphological changes of mitochondria and early apoptotic features of a cell subject to an apoptotic stimulus.  Beyond the fundamental biological mechanisms, understanding the regulation of apoptosis is fundamental to understanding development of organisms \cite{newton_cell_2024}, and the development of disease \cite{vitale_apoptotic_2023}.

Here, we show that compartmentalized systems exhibit an emergent degree of freedom whose kinetic properties are used by cells to regulate cell decisions.
We study a paradigmatic class of compartmentalized stochastic systems, in which stochastic processes are embedded into interacting compartments. We derive the emergent degrees of freedom that arise from the propagation of fluctuations across the hierarchy of scales and experimentally demonstrate that cells make use of this to regulate apoptosis. Specifically, we develop a general theory of the propagation of noise and signals in a general class of dynamically compartmentized systems. We identify an emergent quasi-particle-like degree of freedom describing the kinetics of molecular pathways on the cellular scale. By deriving effective equations of motion, we show that this quasi-particle differs qualitatively in the response dynamics predicted by the biochemistry of the signaling pathway alone. We demonstrate the functional relevance of the quasi-particle in the context of apoptosis, where we show the emergence of a kinetic low-pass filter that allows cells to distinguish slow, biologically relevant changes that require the cell to commit suicide from fast, irrelevant fluctuations. We validate our findings experimentally by quantifying cell death in cell cultures subject to varying duration and strengths of apoptotic stimuli.

\section*{Results}

\subsection*{Emergent quasi-particles in compartmentalized systems}

To derive an effective theory of compartmentalized systems, we define a paradigmatic, yet general, theoretical framework for compartmentalized stochastic processes. We consider a system divided into $N$ compartments that are defined by their respective sizes $s_i(t)$. Each compartment hosts a stochastic process represented by the intrinsic variable $c_i(t)$ within each compartment. In analogy to biochemical systems, we refer to $c_i(t)$ as a concentration. $c_i(t)$ can be any stochastic process with multiplicative Gaussian white noise and arbitrary noise amplitude. In this case, the time evolution of $c_i(t)$ follows a Langevin equation of the form $\dot{c_i}=F(c_i)+D(c_i)\xi$ with a deterministic ''force'' $F$, noise amplitude $D(c_i)$ and noise $\xi$. Most biochemical systems fall into this category of stochastic processes. Compartments themselves undergo compartment dynamics like compartment growth, compartment fission, or compartment fusion. For homeostatic conditions, among all compartment processes, compartment fusion, and fission are the dominant contributions to fluctuations in the concentrations (Supplemental Theory~2.3). The state of the system is then defined by the combination of concentrations and compartment sizes for each compartment, $\{c_i(t), s_i(t)\}$.

Compartment dynamics affect the size of each compartment, but they also lead to fluxes in the concentrations as particles are exchanged between compartments. This leads to a nontrivial feedback between processes on the compartment scale and processes on the microscopic scale - fluctuations and signals propagate across scales.

In physics, one is usually interested in macroscopic observables, like pressure or conductivity. We therefore usually integrate out microscopic degrees of freedom. In biology, the molecular scale is important and target point of many experimental measurements and medical therapies. We therefore aim to derive an effective theory on the smallest scale. To this end, we integrate out probability fluxes between the microscopic and the compartment scale to derive an effective description of the time evolution of the probability that a randomly chosen compartment has a concentration $c$ at a given time $t$,
\begin{equation}
    \partial_t f(c,t) = \partial_c \left[F(c) + \partial_c \Phi(c)\right] f(c,t) + \partial^2_c D(c) f(c,t)\label{eq:vlasov}
\end{equation}
Here, the first term on the right-hand side describes the deterministic drift of the probability density function. This drift happens with a velocity that is proportional to the deterministic force $F(c)$ and additional feedback from the compartment scales. This feedback is given by an additional effective interaction potential $\Phi=\Lambda \int \text{d} c' f(c',t) ( c'-\bar{c})^2$ which describes an attractive force to the mean in the concentration space. The second, diffusive term on the right-hand side describes the dispersive spreading of $f$ due to noise.
\begin{figure}
    \centering
    \includegraphics[width=1\linewidth]{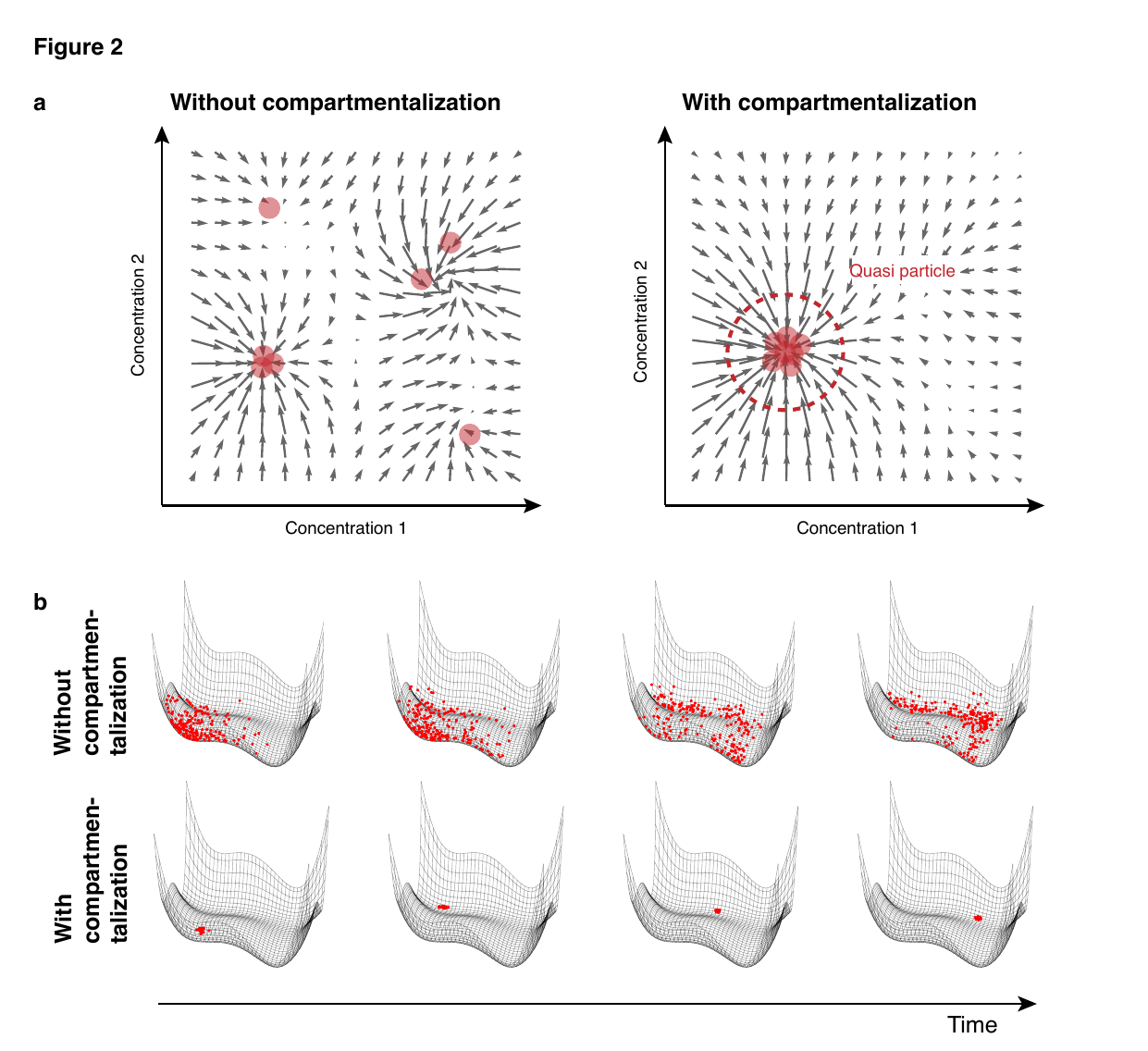}
    \caption{\textbf{a} Schematic depiction of the emergence of a quasi-particle in the example of a two-component system. Arrows show the direction of a force field and red dots the concentrations of individual compartments. Without dynamic compartmentalization concentrations take values corresponding to local attractors of the force field. Dynamic compartmentalization leads to an additional force and the localization of concentrations around a single point in concentration space. \textbf{b} Schematic time evolution of compartment concentrations for a two-component system with potential dynamics in a simulation. Each red dot denotes the concentrations in a compartment.}
    \label{fig:quasi-particle}
\end{figure}
Equation~\eqref{eq:vlasov} therefore describes the balance between the dispersion of the probability density function due to the diffusion term and its contractions due to the effective interactions between compartments. As a result, one intuitively expects the probability density function to obtain a fixed shape with constant variance. In plasma physics, Eq.~\eqref{eq:vlasov} is known as the Vlasov-McKean equation, which admits solutions where the probability density function localizes (see also \cite{meigel_controlling_2024} and  the Supplemental Theory for a derivation). Analogous to localization phenomena observed in condensed matter physics \cite{anderson_absence_1958,anderson_basic_2018,romatschke_collective_2003,gupta_world_2017}, we refer to this localized mode as a quasi-particle. (see the Supplemental Theory~2.5. for a detailed discussion of the use of this term).

Therefore, the time evolution of the stochastic multi-scale system is effectively described by a single degree of freedom, which is the position of the quasi-particle in concentration space. We therefore next derive equations of motion for the quasi-particle.  The kinetics of the quasi-particle is effectively described in terms of its “position” $x$ as the geometric median of the distribution of compartment concentrations, $f(c,t)$, and its “internal deformation” as the non-parametric skew, $s= \bar{c} - x$. In the limit of weak forces $F(c)/\Lambda\ll1$ and weak noise $D(c) \rightarrow0$ to the highest order in the derivatives of $f(c,t)$, the equations of motion for the quasi-particle read $\dot{x}=F(x) + \Lambda s$, and $\dot{s}=-\Lambda s-\gamma F''(x)$ (Supplemental Theory~2.6.). $\gamma$ is a parameter that is proportional to $D(x)\Lambda^{-1}$ and for times much longer than the time scales much longer than $\Lambda^{-1}$ it quantifies the width of the distribution $f(c,t)$. Therefore, the deterministic motion in concentration space is complemented by an effective drift proportional to the non-parametric skew $s$. This drift takes a value that exponentially approaches a steady state given by the second derivative of the force $F(x)$. For time scales much longer than the time scale of the compartment dynamics the equation of motion of the quasi-particle simplifies to
\begin{equation}
    \dot{x} = F(x) - \gamma F''(x) \, .\label{eq:motion-limit}
\end{equation}
With this equation of motion, we have reduced the dynamics of a high-dimensional multi-scale system to the time evolution of a single degree of freedom, $x$. 
The simplicity of the equation of motion allows graphical integration and thus offers an intuitive approach for predicting the emergent consequences of dynamic compartmentalization. Moreover, despite its simplicity, the equation of motion quantitatively agrees with results from numerical simulations of the full stochastic dynamics (Supplemental Theory Fig.~S11).

\begin{figure}
    \centering
    \includegraphics[width=1\linewidth]{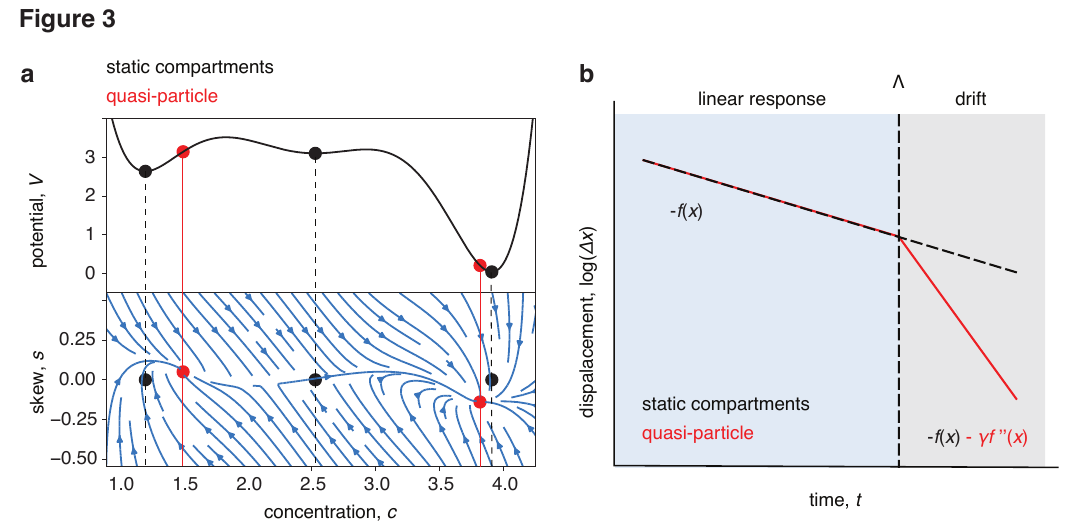}
    \caption{\textbf{a} Phase portrait (bottom) for the quasi-particle kinetics for a potential of the form given in the top. Block dots are fixed points of a non-compartmentalized, well-mixed system and red points denote fixed points of the compartmentalized system. Blue arrows depict solutions of the equations of motion for $x$ and $s$ according to Eq.~(\ref{eq:motion-limit}). \textbf{b} Schematic depiction of the temporal regimes of the nonlinear response of the quasi-particle (red line). The dashed line is the corresponding response of a non-compartmentalized system. See also the Supplemental Theory figure Fig.~S5.}
    \label{fig:QP-kinetics}
\end{figure}

One can gain intuition about the kinetics of the quasi-particle and the unusual appearance of the second derivative of the force by formal analogy to a simple mechanical system of three equal masses coupled by overdamped springs with spring constant $\Lambda/3$ (Supplemental Theory Figure Fig.~{S4}). In this system, the position of the central mass is equivalent to $x$ and its deflection from the center of mass reflects $s$. The force acting on the central mass is a combination of forces from the left and right masses resembling the discrete second derivative of the force field equivalent to the third derivative of the potential. Notably, with Eq.~\eqref{eq:motion-limit}, the time evolution of the stochastic multi-scale system has been reduced to an ordinary differential equation that can be solved graphically or by using standard tools from dynamical systems theory.

The dynamics of the quasi-particle defined by Eq.~\eqref{eq:motion-limit} exhibit a qualitatively different behavior compared to non-compartmentalized systems. First, the stationary states of the quasi-particle position $\dot{x}=0$ do not necessarily coincide with the fixed points of $F(x)$ (Fig.~\ref{fig:QP-kinetics}a). In general, since the quasi-particle distorts the force $F(x)$ on length scales smaller than its variance and deepens it on much longer length scales, the number of metastable states is reduced compared to the number of fixed points of $F(x)$ (Supplemental Theory~2.8.). In equilibrium biochemical systems this implies that steady states do not coincide with the minima of the Gibbs free energy defined by the chemical reactions.

Secondly, dynamically compartmentalized systems respond differently to perturbations. Specifically, the linearized response kinetics of the quasi-particle to perturbations in the force $F(x)$ exhibit two temporal regimes. On short time scales, the response kinetics is dominated by a fast relaxation of $s$ to its steady state value. On time scales much longer than $1/\Lambda$, the dynamics are governed by nonlinear dependencies in force (Fig.~\ref{fig:QP-kinetics}c). This behavior can be understood intuitively in the spring system analogy, where perturbations of the position of the central mass on time scales much smaller than the time scale associated with the spring constant (retardation time, $1/\Lambda$) lead to a deflection independent of the positions of the other masses akin to a point mass in a potential (Supplemental Theory Figure Fig.~S4). Perturbations on time scales much slower than the retardation time are propagated between the masses and lead to a joint response depending on non-local properties of the potential. 

Taken together, dynamically compartmentalized systems show qualitatively altered kinetics compared to non-compartmentalized systems. This holds for systems in which compartment dynamics happen on timescales faster than macroscopic changes in concentration. This raises the question of whether this behavior is used by cells to perform biological functions.

\subsection*{Quasi-particle kinetics in apoptosis}
To test whether the emergent quasi-particle kinetics is used by cells to perform a biological function we now test these predictions in the context of a paradigmatic cell fate decision, that of apoptosis. Apoptosis is executed by the permeabilization of the MOM leading to release of the protein CytC from the mitochondrial intermembrane space, which triggers the disintegration of the cell \cite{czabotar_mechanisms_2023,kale_bcl-2_2018}. The release of CytC is facilitated by protein complexes of the apoptosis executing  proteins Bax, which is recruited to the MOM, and Bak, which is resident in the MOM. Both proteins undergo stochastic kinetics that can lead to the formation of pores in the MOM \cite{kale_bcl-2_2018,lovell_membrane_2008}. Notably, mitochondria are highly dynamic organelles that undergo continuous cycles of fusion and fission, which lead to the redistribution of Bax complexes among mitochondria. Therefore, apoptosis is an example of a dynamically compartmentalized system.
Mitochondria undergo fusion and fission on a time scale of seconds to minutes \cite{bhola_spatial_2009,berman_bcl-xl_2009}. This is comparable to the time scale of Bax concentration changes on the MOM to lead to MOM permeabilization (MOMP) \cite{bhola_spatial_2009,rehm_dynamics_2009}. Because of this, the kinetic parameters of apoptosis regulation fall into a regime where the quasi-particle kinetics, Eq.~\eqref{eq:motion-limit}, is applicable.

\begin{figure}
    \centering
    \includegraphics[width=1\linewidth]{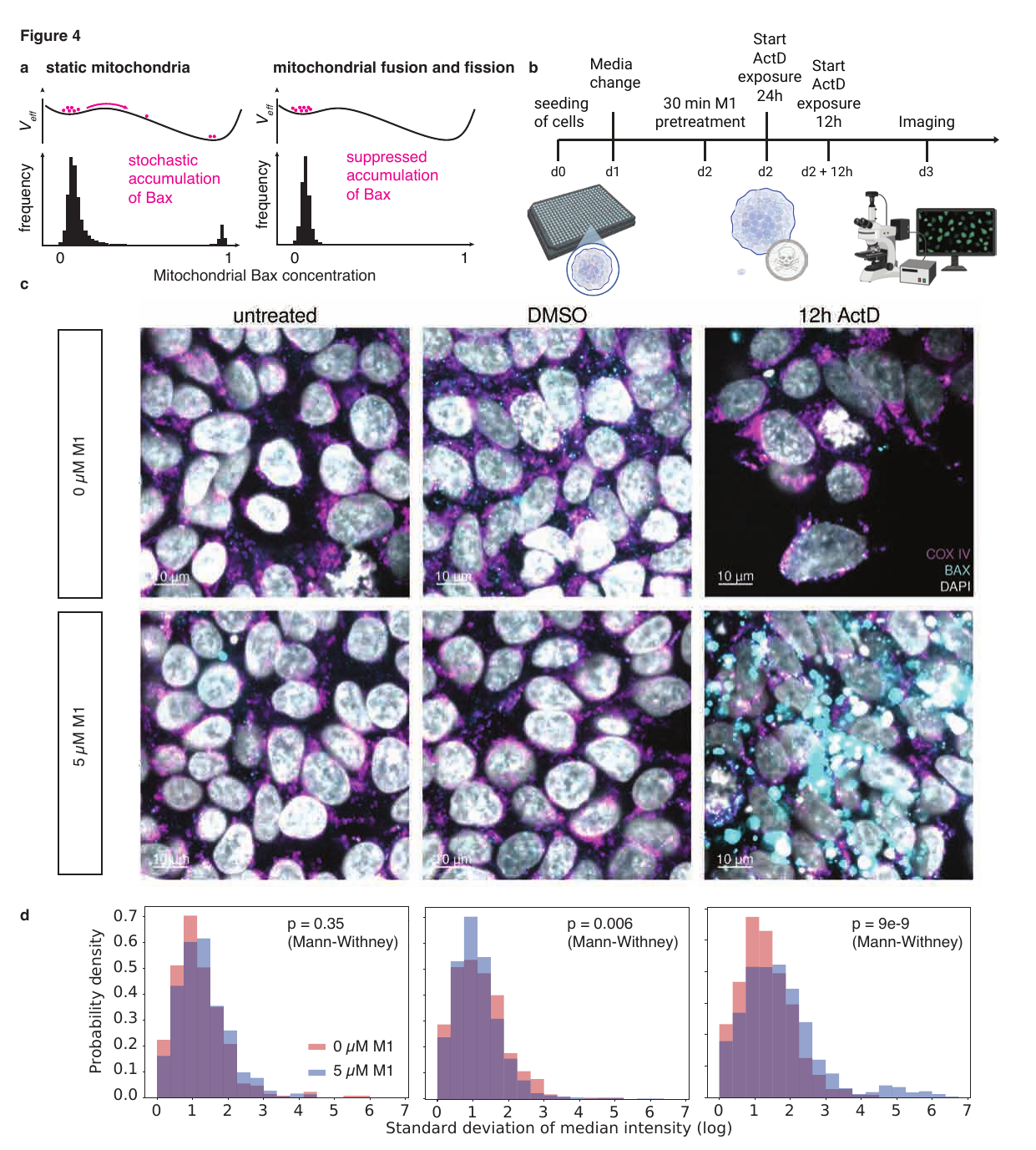}
    \caption{\textbf{a} Prediction for the distribution of mitochondrial Bax concentrations for static and interacting mitochondria. The top row represents an effective potential governing the Bax kinetics.  \textbf{b} Experimental setup \textbf{c} Microscopy images showing cells stained with a markers for mitochondria (CoxIV, magenta) and Bax (cyan). Cell nuclei a marked by DAPI (grey scale) \textbf{d} Probability distributions of Bax intensities for signals that colocalize with CoxIV. The panels represent the same treatments as the corresponding columns in c.}
    \label{fig:experiment-localization}
\end{figure}

By using the known biochemistry of Bax complex formation~\cite{kale_bcl-2_2018,lovell_membrane_2008,billen_bcl-xl_2008} we can derive the force term $F(x)$ of Eq.~\eqref{eq:motion-limit} and the kinetic parameters of mitochondrial dynamics~\cite{cagalinec_principles_2013,eisner_mitochondrial_2014,bhola_spatial_2009} fix the value of $\gamma$. This force term takes the form of a bistable system \cite{cui_two_2008,chen_modeling_2007,suen_mitochondrial_2008,spencer_measuring_2011} with one stable state at a low Bax concentration and a further stable state at a high Bax concentration, separated by unstable steady state positioned at an intermediary concentration (Fig.~\ref{fig:experiment-localization}a, Supplemental Theory~3.2.).  

To investigate whether apoptotic decision making involves quasi-particle dynamics, we performed two sets of experiments in which we compared cell cultures with dynamic mitochondria and cell cultures in which mitochondrial fission was inhibited. In the first experiment, we aimed to test the static properties of the quasi-particle to find out whether a quasi particle forms by the localization of Bax concentrations in concentration space, as illustrated in Fig.~{\ref{fig:quasi-particle}~b}. In the context of the translocation dynamics of Bax to the mitochondrial membrane, we predict reduced variability of Bax concentration across different mitochondria inside the same cell, when cells are subjected to apoptotic stimuli.  

To test this, we quantified Bax localization on mitochondria in fluorescent microscopy experiments. Human induced pluripotent stem cell (hiSPC) cultures were treated with the Bax-acitvating apoptotic stimulus (actinomycin D, Act D) for 12 or 24 hours, with the mitochondrial intermembrane space and Bax fluorescently labeled (see Experimental Methods). Although this method cannot fully label the outer mitochondrial membrane, CoxIV-labeled pixels were used to robustly identify a subset of mitochondrial regions. In turn, the CoxIV co-localized median Bax intensity provides a robust measure of membrane-localized Bax, and the CoxIV area-weighted standard deviation over co-localized median Bax concentrations within the same cell serves as a robust approximation of Bax concentration variability. To assess the effect of multi-scale feedback, we proceed analogously with a second cell culture that was treated before the experiment with the functional mitochondrial fission inhibitor hydrazone M1 \cite{wang_small_2012}, which effectively diminishes quasi-particle dynamics.

We observed both for ActD untreated and 12h treated cells that the variability of membrane-bound Bax was significantly increased (two-sided Mann-Whitney U-test p < 0.05) when cells were pretreated with M1 Fig.~\ref{fig:experiment-localization}c,~d. Specifically, we found that the M1-treated cells comprise a sub-population with strongly increased Bax variability inside a single cell which we did not observe in cells not treated with M1. These cells correspond to cells with Bax puncta only on a subset of CoxIV label mitochondrial regions. The absence of this sub-population in cells not treated with M1 indicates that mitochondrial dynamics is necessary to localize subcellular mitochondrial Bax concentrations in the  concentration phase space, a key characteristic of the quasi-particle predicted by Eq.~\eqref{eq:vlasov}.

\begin{figure}
    \centering
    \includegraphics[width=1\linewidth]{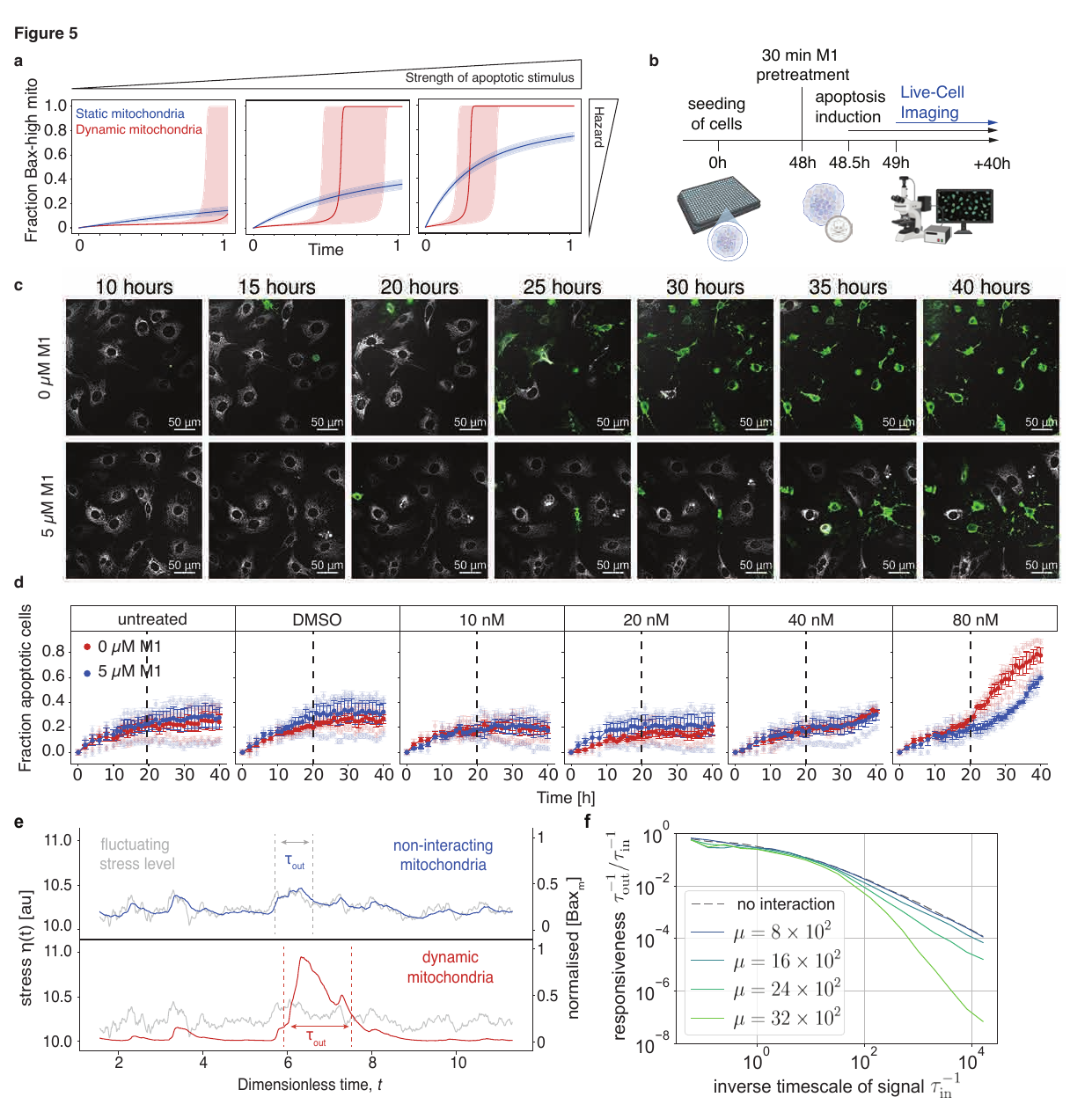}
    \caption{\textbf{a} Predicted time evolution of the fraction of apoptotic cells for low and high apoptotic stimuli. \textbf{b} Experimental setup \textbf{c} Microscopy images of MEF cells treated with ActD to induce apoptosis with or without M1 pretreatment, live stained with TMRE (mitochondria, grey) and Annexin V (green, a marker of apoptosis) and consecutively imaged. \textbf{d} Normalized empirical fraction of  Annexin V-positive cells. Solid dots represent mean $\pm$ SEM of three experimental replicates (shown in light colors). \textbf{e} Exemplary stochastic simulations of the response of the Bax system for static mitochondria (blue) and dynamic mitochondria (red) to time-correlated apoptotic stimuli (grey). \textbf{f} Response of the Bax system to stochastic apoptotic stimuli with correlation time $\tau_{\text{in}}$. Shown is the ratio of the correlation time of mitochondrial Bax concentrations, $\tau_\text{out}$, and $\tau_\text{in}$ the auto-correlation time of stress signal $\eta(t)$. The physiologically plausible simulation parameter for and e,f are the same as Fig.~S11(b). Time is de-dimensionlized by normalizing with the timescale of Bax translocation, Supplemental Theory Fig.~S7.}
    \label{fig:experiment-timecourse}
\end{figure}

In the second set of experiments we next tested the kinetic properties of the quasi-particle. To this end we investigated the response of the Bax system to perturbations in the form of a constant stimulus which biologically describes the concentration of stress  mediators in the cytosol and the ensuing recruitment rate of Bax proteins to the MOM. The effect of such a stimulus is a tilt in the effective bistable potential governing mitochondrial Bax concentrations (Supplemental Theory Figure Fig.~{S6}). Equation~\eqref{eq:motion-limit} then predicts a sigmoidal response of mitochondrial Bax concentrations to the stimulus (Supplemental Fig.~S9,~S10,~S11, and  Supplemental Theory~3.3.). This means that the response to the stimulus is suppressed on short time scales and enhanced on long time scales compared to the case of static mitochondria. The characteristic time that separates both temporal regimes of the response kinetics decreases with increasing stimulus strength (Fig.~\ref{fig:experiment-timecourse}a). 

To test whether such kinetics emerges in cells we subjected cells to apoptotic stimuli of varying strength by administering ActD at different concentrations with or without mitochondrial fission inhibiting M1 pretreatment. We subsequently performed live imaging  to monitor the fraction of apoptotic cells labelled by Annexin V (Figure~\ref{fig:experiment-timecourse}b). We found that for a high concentration of ActD, cells without M1 pretreatment showed a stronger response to the apoptotic stimulus compared to cells pretreated with M1. Further, for low concentrations of ActD, cells not preptreated responded significantly weaker compared to M1 treated cells at early timepoints ($t<20\text{}$); We performed a Wilcoxon signed-rank test comparing the two M1 treatment groups while we aggregated the data over all ActD treatment groups and over all early timepoints within the time window $t\in[0\text{h}, 20\text{h}]$ which yielded $p=0.008$. This suggests that the accumulation dynamics of active apoptotic effects (i.e. the Bax accumulation dynamics) in response to an apoptotic stimuli follows a sigmoidal response dynamics in time as predicted by the kinetic equation of the quasi particle, Eq.~\eqref{eq:motion-limit}.

The suppression of the response to stimuli on short time scale with the simultaneous enhancement of the response on long time scales resembles the behaviour of a low-pass filter used in electronics to suppress noise. To investigate whether the Bax system may establish a kinetic low pass filter of metabolite fluctuations we studied its response to a time-correlated stochastic signal $\eta(t)$.  Figure~\ref{fig:experiment-timecourse}e shows exemplary trajectories of a stochastic signal and ensuing Bax (protein) concentrations obtained from full stochastic simulations for physiological parameters (Methods). While in a hypothetical system with static mitochondria, the Bax concentration follows the stochastic signal, for a system with mitochondrial dynamics fast fluctuations in the stochastic signal are suppressed while slow trends tend to be followed and amplified. More general,  the exponent characterizing the suppression of fast fluctuations increases with the rate of mitochondrial fusion and fission (Fig.~\ref{fig:experiment-timecourse}f).

\section*{Discussion}

Biological systems are inherently dynamic, with processes occurring across multiple spatial and temporal scales. Compartmentalization is a characteristic feature of many biological systems which underlies fundamental subcellular processes involved in cellular signaling and decision-making. We demonstrated that interactions between compartments gives rise to emergent quasi-particle kinetics characterized by unique kinetic properties: a suppressed response to extrinsic cues on short timescales and a facilitated, sigmoidal response on longer, persistent timescales.

Although the chemical reaction dynamics within dynamic compartments and the dynamics of the compartments themselves occur on vastly different spatial scales, their timescales can become intertwined. Specifically, macroscopic changes driven by stochastic fluctuations in the chemical reactions enclosed within these compartments often occur on the same timescale as the dynamics of the compartments. A relevant example for this is the stochastic accumulation of Bax proteins on the mitochondrial outer membrane~\cite{cao_mitochondrial_2022} and mitochondrial fusion and fission cycles \cite{bhola_spatial_2009,cagalinec_principles_2013,eisner_mitochondrial_2014}. This coupling between the two processes creates a kinetic regime, where the emergent behavior cannot be fully understood by considering either spatial scale  in isolation.

The biological relevance of the regulation of apoptosis and mitochondrial dynamics is ubiquitous. The canonical relevance lies in morphogenesis of tissues and organs in development as well as removal of superfluous or damaged cells \cite{newton_cell_2024}. Furthermore, erroneous activation or inhibition of apoptosis can contribute to the development of disease \cite{vitale_apoptotic_2023}, including cancer \cite{vitale_apoptotic_2023}, ischemic heart disease \cite{vitale_apoptotic_2023}, and stroke \cite{walther_reinventing_2023,mergenthaler_stroke_2024}, among others. This makes apoptosis a desirable target for therapeutic intervention in medicine \cite{niu_small-molecule_2017,li_too_2021,pogmore_pharmacological_2021,diepstraten_manipulation_2022}. 

The execution of apoptosis is regulated by complex molecular mechanisms \cite{kale_bcl-2_2018,czabotar_mechanisms_2023,lovell_membrane_2008,billen_bcl-xl_2008,bogner_allosteric_2020}. Next to Bax, the structurally and functionally closely related protein Bak, which resides in the MOM, contributes to execution of apoptosis by pore formation \cite{kale_bcl-2_2018,czabotar_mechanisms_2023}, and homomeric or heteromeric pores seem to follow different kinetics leading to MOMP \cite{cosentino_interplay_2022}.  Furthermore, complex protein interactions among the Bcl2 family members in the cytosol or the MOM regulate the events leading to execution of apoptosis \cite{kale_bcl-2_2018,czabotar_mechanisms_2023,lovell_membrane_2008,billen_bcl-xl_2008,bogner_allosteric_2020}. However, the model proposed herein fully accounts for the regulation leading to the accumulation of Bax (or Bak) particles in the MOM and considers the particle concentration of these pro-apoptotic effectors in the MOM as prerequisite for pore formation leading to MOMP.

Beyond their classical function in providing energy for the cell, mitochondria are the integrators for various signaling pathways in cells \cite{tabara_molecular_2024}. Dynamic mitochondria are known to enable cells to cope with various states of stress \cite{tabara_molecular_2024,quintana-cabrera_determinants_2023}, and it is established that mitochondrial shape dynamically changes when cells die \cite{tabara_molecular_2024}. In the present study, we emphasize the interaction between dynamic mitochondria and Bax proteins, which collaboratively play a pivotal role in helping the cell determine which death-promoting stimuli are significant enough to execute apoptosis. This introduces a novel concept in biology, highlighting how cellular organelles and proteins dynamically interact to fine-tune life-or-death decisions within the cell.

We derived these results under general conditions: that of white Gaussian noise and the mixing of the time scale of the microscopic and compartment dynamics. We therefore expect that the properties of quasi-particle kinetics may also be relevant in the context of other organelle-associated cell decisions, such as the regulation of protein synthesis mediated by the translocation of mTORC1 to lysosomes \cite{menon_spatial_2014,angarola_coordination_2020,carroll_control_2016}, the maturational dynamics of endosomes \cite{stoorvogel_late_1991,foret_general_2012,wandinger-ness_rab_2014}, or the regulation of oxidative phosphorylation in mitochondria \cite{selivanov_bistability_2009,selivanov_multistationary_2012,nath_thermodynamic_2016}.
Sigmoidal responses are ubiquitous in biological systems and in important building block of their function \cite{kholodenko_cell-signalling_2006,hinczewski_cellular_2014,zhang_ultrasensitive_2013}. In molecular systems, they usually emerge from cooperative binding of multiple proteins. Dynamic compartmentalization allows to generate a sigmoidal response with a Hill tuneable coefficient that depends on kinetic parameters of compartment fusion and fission.

\section*{Methods}

\subsection*{Cell Culture}
The human induced pluripotent stem cell (hiPSC) line BIHi250-A (\href{https://hpscreg.eu/cell-line/BIHi250-A}{https://hpscreg.eu/cell-line/BIHi250-A}) was cultivated in Essential 8 medium prepared according to the original recipe \cite{chen_chemically_2011} and grown on Geltrex (Thermo, A1413302) coated tissue culture plates. Media change was performed daily for six days with one double-feed on day 7. Cells were passaged when confluency reached approximately 80\%. Mouse embryonic fibroblasts (MEFs) immortalized by limited dilution were grown in tissue culture flasks and cultivated in DMEM/F-12 (Thermo, 11320033) supplemented with 5\% FBS (PAN Biotech, P30-3030M) and passaged when a confluency of approximately 80\% was reached. All cell cultures were kept at 37°C and 5\% CO\textsubscript{2}.

\subsection*{M1 Mitochondrial Fission Inhibitor and Actinomycin D Treatment}
For mitochondrial fission inhibition with M1, MEFs were seeded into wells of a 384-well Greiner µClear Imaging plate (Greiner, 781091) at a density of 24,000 cells/cm\textsuperscript{2}. 24 hours after seeding, media was changed. For M1 pre-treatment, M1 (Merck, SML0629-25MG) stock solution (50 mM) was diluted in PBS without calcium and magnesium (PBS-/-, Thermo, 14200075) and sonicated (40 kHz, 5 minutes, room temperature) and then added to respective wells at a final concentration of 5 µM. After 30 minutes of M1 pre-treatment, actinomycin D (ActD, Selleckchem, S8964) was added at concentrations from 0 to 80 nM. Immediately after addition of ActD, tetramethylrhodamine-ethylester (TMRE, Thermo, T669) was added to each well at a final concentration of 50 nM and Annexin-V-FITC (ImmunoTools, 31490013X2) according to the manufacturer's instructions.

\subsection*{BAX Immunofluorescence}
For BAX localization, hiPSCs were detached using TrypLE\textsuperscript{TM} Select (Thermo, 12563011) and seeded as single cells into Geltrex-coated 384-well Greiner µClear plates at a density of 24,000 cells/cm² in E8 medium containing 5 µM Y-27632 Rho-associated, coiled-coil containing protein kinase (ROCK) inhibitor (ROCKi, Selleckchem, S1049). After 24 hours, ROCKi was removed by replacing the medium. Following an additional 24-hour culture period, hiPSCs were treated with 12.5 nM ActD for 12 or 24 hours. A subset of cells was pre-treated with the mitochondrial fission inhibitor M1 for 30 minutes prior to ActD incubation as described above. At the experimental endpoint, cells were fixed with 4\% paraformaldehyde (PFA, Carl Roth, 30525-89-4) for 15 minutes at 37°C and washed twice with PBS\textsuperscript{-/-} at room temperature (RT). Blocking and permeabilization were performed using a blocking buffer consisting of 5\% normal donkey serum (NDS, Milipore, S30-100ML) and 0.3\% saponin (Sigma, 47036) in PBS\textsuperscript{-/-} for 45 minutes at RT. Primary antibodies (COX IV (4D11-B3-E8) mouse mAB, Cell Signaling Technology, 11967; Bax (D2E11) rabbit mAb, Cell Signaling Technology, 5023) were diluted in blocking buffer at 1:500 and 1:200, respectively, and incubated with cells for 1 hour at RT. After three 5-minute washes with PBS\textsuperscript{-/-} at RT, secondary antibodies (donkey anti-mouse (HCA)-Alexa 647, Thermo, A31571; donkey anti-rabbit (HCA)-Alexa 488, Thermo, A32790) were diluted 1:500 in blocking buffer and incubated for 2 hours at RT, protected from light. Following three additional 5-minute washes at RT, the final wash included 10 µM Hoechst 33342 (Thermo, H1399). Cells were protected from light until imaging.

\subsection*{Imaging}
\subsubsection*{Live-Cell Imaging}
Live-cell imaging was performed on a spinning-disk confocal microscope (Revvity Opera Phenix) with a 40x water immersion objective (NA=1.1), operated by Harmony 5.2 (Revvity) software. Digital phase contrast (DPC) was acquired at 50\% laser power and 700 ms exposure time and created with an upper plane at 5 µm, a lower plane at -3 µm, a filter of 1.0, and speckle scale of 10 µm. TMRE was excited with a 561nm laser with a power of 100\%, 80 ms exposure time, and at a height of 0 µm. Annexin-V-FITC was imaged using a 488 nm laser, 50\% laser power, 40 ms exposure time, and imaging height of 0 µm. Sequential channel order was set to acquire the DPC first, followed by TMRE and then Annexin-V-FITC. Two separate sequences for consecutive image acquisition were set. Sequence 1 consisted of 6 timepoints with intervals of 2 hours. Sequence 2 comprised 30 timepoints with 1-hour intervals. This resulted in 2-hour intervals throughout the first 10 hours of the experiment followed by 30 hours with images taken hourly. For every well, 12 out of 81 fields were selected in which images were taken.

\subsubsection*{BAX Immunofluorescence Imaging}
Imaging was performed on a spinning-disk confocal microscope (Revvity Opera Phenix) with a 63x water immersion objective (NA=1.15), operated by Harmony 5.2. Hoechst was imaged with a 375 nm laser with 40\% laser power and 40 ms exposure time, COX IV was imaged with a 647 nm laser, 80\% laser power for 80 ms, and BAX was imaged with a 488 nm laser at 80\% laser power and 80 ms exposure time. All channels were acquired at an imaging height of 2 µm. Lasers for Hoechst and COX IV were excited together followed by a second exposure with excitation at 488 nm. 12 out of 225 fields were imaged per well.

\subsection*{Image Processing}
Image processing was performed with SignalsImageArtist 1.3 (Revvity). Figures with microscopy images were prepared with OMERO.figure v4.4.3 on OMERO plus (Glencoe) . 

\subsubsection*{Live-Cell Imaging}
All images were subjected to a basic flatfield correction and brightfield correction. Images were pre-processed with filter-combinations to eliminate background noise. TMRE images were pre-processed with a sliding parabola (curvature 1), followed by Gaussian smoothing (width 2 pixels). The structures of interest (mitochondria) were identified by segmentation in an image calculated by subtracting the image pre-processed with sliding parabola and Gaussian smoothing from an image preprocessed with  sliding parabola alone. Annexin-V-FITC images were also pre-processed with a sliding parabola (curvature 0.1, Gaussian smoothing 3 pixels width), but no further image calculation. Cells were identified using the \textit{Find Cells} Building block (\textit{Method C}) on the unfiltered TMRE channel as an estimate for the cell area. Morphology and intensity properties of identified cells were calculated and used to only select objects with a cell roundness above 0.3. Furthermore, objects touching the image border were excluded from analysis. Identification of mitochondria was achieved by \textit{Find Image Region} based on the pre-processed / calculated TMRE images and restricted to the population of previously identified cells. Objects were identified based on an absolute threshold with the lowest intensity $\geq$ 6. Morphology and intensity properties for mitochondrial objects were calculated. Objects were eliminated when the area was $<$ 16 px\textsuperscript{2}, when the object length was < 4 px, and when the intensity was $<$ 270. To identify Annexin-V-FITC positive vesicles (AV-vesicles), the \textit{Find Image Region} building block was utilized based on the pre-processed Annexin-V-FITC channel. Within the region of previously identified cells, AV-vesicles were identified by an absolute threshold with the lowest intensity $\geq$ 50. Morphological and intensity properties of AV-vesicles were calculated. All calculated morphology and intensity properties were exported for secondary analysis.
 Raw data were annotated with treatment group and timepoint information based on well identifiers. Cells with an Annexin-V vesicle area > 0 $\mu\text{m}^2$ were classified as apoptotic, while those with a vesicle area of 0 $\mu\text{m}^2$ were classified as non-apoptotic. Statistical analysis was performed in \texttt{python 3.7.13}. 
 For each treatment group and timepoint, the mean apoptotic cell count and its corresponding standard error were calculated using the \texttt{agg} function in \texttt{pandas 1.3.5}. 
 For each treatment group and replica, the fraction of apoptotic cells was normalized by subtracting the fraction of apoptotic cells at the timepoint of first observation ($t=0\text{h}$).
 The normalized fraction of apoptotic cells was plotted over time for each treatment group using visualization libraries \texttt{seaborn 0.12.2} and \texttt{matplotlib 3.2.2}. Additionally, the unweighted mean apoptotic cell count and its standard error were computed across all replicates for each treatment group and timepoint. These aggregated metrics were visualized along with the individual timepoint analysis to enable direct comparison. A Wilcoxon signed-rank test was performed to compare the aggregated data from all ActD treatments combined against the two M1 treatment within the time window $t\in[0\text{h}, 20\text{h}]$.
 
\subsubsection*{Immunofluorescence Imaging }
Brightfield correction and advanced flatfield correction were applied to all images. Channels were individually pre-processed with combinations of different filters to eliminate background and enhance signal of interest. For BAX signal, images were filtered with a sliding parabola at a curvature of 1 before performing a Gaussian smoothing with a width of 3 px. The image pre-processed with sliding parabola and Gaussian smoothing from an image pre-processed with sliding parabola alone to further eliminate background. The COX IV channel was filtered in the same way with a sliding parabola curvature of 10. Hoechst was pre-processed similarly, with a sliding parabola curvature of 0.1 and a Gaussian smooth width of 2 px. Nuclei were identified by \textit{Find Nuclei} building block (\textit{Method B}). After calculation of morphology and intensity properties, only those nuclei were accepted for analysis that showed a mean intensity > 36, a nucleus area > 43 µm\textsuperscript{2}, and a nuclear roundness above 0.7. Cells were identified by a selected region around the nucleus. This region was created with an outer border of -5 µm. COX IV signal was used as an estimation of mitochondrial location within the cell and segmented by the building block \textit{Find Image Region} based on the pre-processed COX IV channel. The segmentation of the COX IV image region was limited to the region of interest previously defined as cells. Based on calculated morphology and intensity properties, COX IV objects were included in the analysis when the object area was > 0.4µm\textsuperscript{2} and the intensity > 22. BAX particles were identified with \textit{Find Image Region} based on the pre-processed BAX channel. BAX particle segmentation was limited to regions identified as COX IV positive.
Raw data were annotated with treatment group and time point information based on well identifiers. Statistical analysis was performed using \texttt{python 3.7.13}. For each cell, the median Bax intensity within its COX IV-positive regions was calculated using \texttt{pandas 1.3.5}. Additionally, the standard deviation of Bax intensity within each COX IV-positive region was calculated, weighted by the area of the region. This weighted standard deviation was aggregated at the cell level and stored as a cell attribute.
The heavily skewed distribution of the standard deviation of the median Bax intensity per cell was visualized as normalized density plots of the natural logarithm of the standard deviation, using \texttt{matplotlib 3.2.2}.
For each ActD treatment, a two-sided Mann-Whitney U-test was performed against the M1 treatment on the standard deviation of the median Bax intensity.

\subsubsection*{Simulations}
To study the multi-scale dynamics of compartmentalized stochastic reaction systems, we used standard SSA (stochastic simulation algorithm) to sample random trajectories of the system dynamics in the compartment and multiplexed the compartment dynamics by stochastically informing the results from the SSA in different compartment with each other. All simulations were implemented in \texttt{python 3.7.13}.

The chemical reactions within each compartment were simulated in parallel with the stochastic evolution of the compartments, with updates occurring at discrete intervals. This separation of timescales allowed for efficient parallelization of the simulation, significantly reducing execution time while maintaining accuracy. Memory management was optimized by discretizing compartment sizes and pre-allocating storage for concentration vectors and reaction matrices. 

\section*{Acknowledgements}
We thank F. J\"ulicher for the helpful feedback, the members of the BIH at Charité Research IT for maintaining the SIMA and OMERO platforms, and all members of the involved groups for critical discussions. This work was supported by the Core Unit pluripotent Stem Cells and Organoids (CUSCO) of the Berlin Institute of Health (BIH) at Charité – Universitätsmedizin Berlin. 
This project has received funding from the European Research Council (ERC, grant agreement no. 950349, S.R.) and under grant agreement 825161 (P.M. and H.S.) under the European Union’s Horizon 2020 research and innovation program, and in part by the Einstein Foundation Berlin (EJF-2020-602, EVF-2021-619, EVF-2021-619-2, EVF-BUA-2022-694 to P.M.; EC3R to P.M. and H.S.), the Else Kröner-Fresenius Stiftung (2019\_A34, P.M.), the Volkswagen Stiftung (9A866, P.M.), and the Stiftung Charité (StC-VF-2024-59, P.M.). P.M. is Einstein Junior Fellow funded by the Einstein Foundation Berlin.


\section*{Code availability}
Simulation routines are described in the Supplemental Theory in Section~4. Code snippets are available from the corresponding author upon reasonable request.

\section*{Competing interests}
The authors declare no competing interests.




\section*{Supplementary material}

# Universal quasi-particle kinetics control the cell death decision: Supplemental Theory

FELIX J. MEIGEL AND STEFFEN RULANDS

## 1. INTRODUCTORY REMARKS

In this supplemental theory, we explore whether multi-scale fluctuations within dynamically compartmentalized systems can give rise to emergent collective behaviors. To investigate this, we employ a foundational model of compartmentalized stochastic systems, characterized by a hierarchical structure of interacting non-equilibrium processes occurring across distinct spatial scales. We analyze the dynamics of interacting entities within dynamic compartments. Without compartmental dynamics, the stochastic interactions of these entities can be effectively studied using Master equations. Here, we illustrate how the dynamics of the compartments themselves lead to collective phenomena, offering a novel perspective on the role of multi-scale fluctuations in non-equilibrium systems. By employing the framework of population balance equations, we describe the likelihood of individual compartments occupying specific states within the system phase space. Utilizing this, we provide an analytical approach to understand these collective behaviors. As our tendency is describing biological systems, we specifically consider chemical reaction kinetics as enclosed dynamics in the compartments, noting here that our theory is more general an can be easily adapted to other enclosed dynamics.

Specifically, we demonstrate that the interplay of compartment fusion and fragmentation results in the localization of probability distributions within a concentration phase space. This localization gives rise to a collective degree of freedom, which encapsulates the collective behavior of these localized probability distributions. Remarkably, this collective motion mirrors the behavior of a single compartment within the same phase space. We draw parallels between these ensemble dynamics and the concepts of collective excitations and quasi-particle solutions found in many-body theory. This chapter focuses on establishing an effective theoretical framework for describing emergent dynamics in open compartmentalized systems, where mass conservation is not assumed. As a result, the discussion here is primarily technical. Here, we will apply this theory to organelle-associated signaling pathways, with a particular focus on mechanisms underlying cell death decision-making.

This supplemental theory closely follows the theory developed in [1] chapter 2 and chapter 3. The text in this supplemental theory is a reduced version focusing only on the aspects relevant for the content of this publication. As a consequence, phrases in this suppemental theory are partial similar or indentical in wording to [1].

## 2. DERIVATION AND CHARACTERISATION OF QUASI-PARTICLE DYNAMICS IN DYNAMICALLY COMPARTMENTALIZED SYSTEMS.

### 2.1. Description of Chemical Reaction Kinetics as Stochastic Systems

In this supplemental theory, we focus on stochastic chemical reaction networks kinetics, accounting for low copy numbers and intrinsic and extrinsic noises sources on the subcellular scale [2–5]. We consider a well stirred mixture of of $N \geq 1$ chemical species $\{S_1, S_2, ..., S_N\}$ inside a fixed volume $\Omega$ that chemically interact at a constant temperature, through $M \geq 1$ reaction channels $\{R_1, R_2, ..., R_M\}$. We define with $X_i(t)$ the number of $S_i$ molecules in the system. Each reaction is defined by

$$R_j \equiv \sum_i^N s_{ij} S_i \xrightarrow{k_j} \sum_i^N r_{ij} S_i, \tag{S1}$$

where $k_j$ is the rate of the reaction, and $s_{ij}$ and $r_{ij}$ stoichiometric coefficients. The state change vector $\vec{V}_j = (V_{1j}, ..., V_{Nj})$ is defined by its components $V_{ij} = r_{ij} - s_{ij}$. While $\{S_1, S_2, ..., S_N\}$ defines the molecules interacting in the signalling pathway, $\{R_1, R_2, ..., R_M\}$ define the dynamics of the signalling pathway. We introduce the dependence on an external signal $\eta(t)$ in this general definition of chemical reaction networks by setting a dependence between the reaction rate and the signal $k_j(\eta(t))$ to a subset $j \in R_{\text{sig}}$. This formalism is equivalent to either externally manipulating the kinetic properties of chemical reactions, or externally controlling the concentration of a chemical species. The propensity of each reaction is given by $a_j(\vec{X}(t)) = k_j h(\vec{X}(t))$, where $h(\vec{X}(t))$ is the number of distinct combinations of $R_j$. The time evolution of joint probability function $P(\vec{X}, t)$ is then given by

$$\frac{\partial}{\partial t} P(\vec{X}, t) = \sum_j^M \left[ a_j(\vec{X} - \vec{V}_j) P(\vec{X} - \vec{V}_j, t) - a_j(\vec{X}) P(\vec{X}(t)) \right], \tag{S2}$$

see also the appendix on the Master equation formalism in [1]. We introduce a change of variables, as we express the state of the system in terms of a concentration $\vec{x}(t) = \vec{X}(t)/\Omega$ and formally $\vec{v} = \vec{V}/\Omega$. By rescaling the propensity while omitting combinatorial factors arising from identical species, we find $\tilde{a}_j = \Omega a_j$. The probability function is rescaled by $\tilde{P}(\vec{x}, t) = \Omega^N P(\vec{X}, t)$. We series expand Eq. (S2) around $\vec{x}(t)$ with the small variable $v_{ij}$, which is also referred to the Kramers-Moyal expansion [2, 3]

$$\frac{\partial}{\partial t} \tilde{P}(\vec{x}, t) = \sum_{n=1}^{\infty} (-1)^n \left(\frac{1}{\Omega}\right)^{n-1} \sum_{\vec{m} \in Q(n)} \frac{1}{\prod_i^N m_i!} \frac{\partial^n}{\prod_i^N \partial x_i^{m_i}} \left[ \left[ \sum_{j=1}^M \prod_i^N V_{ij}^{m_i} \tilde{a}_j(\vec{x}) \right] \tilde{P}(\vec{x}, t) \right]. \tag{S3}$$

For large volumes $\Omega^{-n+1}$ higher order terms vanish. The Fokker-Planck approximation corresponds to the truncation of the Kramers-Moyal expansion at the second order. To this end, we define the drift vector $\vec{F}$ and the noise matrix $\mathbf{D}$

via

$$F_i(\vec{x},t,\eta(t)) = \sum_{j=1}^{M} V_{ij}\tilde{a}_j(\vec{x},t,\eta(t)), \quad D_{ik}(\vec{x},t,\eta(t)) = \frac{1}{\Omega}\sum_{j=1}^{M} V_{ij}V_{kj}a_j(\vec{x},t,\eta(t)),$$
(S4)

where the subscript $\eta(t)$ refers to the dependence of the reaction rate on the external signal $\eta(t)$. Note, that here the noise matrix $\mathbf{D}_{\eta(t)}$ describes dispersive dynamics in the concentration phase space and not in a spatial sense, as we have excluded any spatial notion from our analysis by assuming well-mixed conditions. Making use of the drift vector $\vec{F}_{\eta(t)}$ and the noise matrix $\mathbf{D}_{\eta(t)}$, we find

$$\frac{\partial}{\partial t}\tilde{P}(\vec{x},t) \approx -\sum_{i=1}^{N}\frac{\partial}{\partial x_i}\left[F_i(\vec{x},t,\eta(t))\tilde{P}(\vec{x},t)\right] + \frac{1}{2}\sum_{i,k}^{N}\frac{\partial}{\partial x_i}\frac{\partial}{\partial x_k}\left[D_{ik}(\vec{x},t,\eta(t))\tilde{P}(\vec{x},t)\right].$$
$$= -\nabla \cdot \vec{F}_\eta \tilde{P}(\vec{x},t) + \frac{1}{2}\nabla \cdot \left(\nabla \cdot \mathbf{D}_{\eta(t)}\right)^\top \tilde{P}(\vec{x},t),$$
(S5)

where the second line is equal to the first line by making use of a vector notion. The formulation in terms of a multi-dimensional Fokker-Planck Equation allows for an equivalent formulation in terms of Itô stochastic differential equations. For this, we make use of the Cholesky decomposition $\mathbf{D}(\vec{x},t) = \mathbf{C}(\vec{x},t)\mathbf{C}^\top(\vec{x},t)$ to find

$$\frac{d}{dt}\vec{x} = \vec{F} + \mathbf{C}\vec{\xi}(t),$$
(S6)

where $\vec{\xi}(t)$ is an (uncorrelated) Gaussian white noise vector. This set of coupled stochastic differential equations is also referred to as Chemical Langevin Equations. In the thermodynamic limit, $\Omega \to \infty$, the noise vanishes and the system is described by a set of coupled deterministic differential equations, which corresponds to a formulation on the basis of reaction rates. Notably, both the drift vector $\vec{F}$ and the noise matrix $\mathbf{D}$ encode both the chemical reaction rates $k_j$ and the current state of the system. Here, we formally allowed for the creation and annihilation of species from a bath. As such, our system is not mass-conserving.

### 2.2. Compartmentalised stochastic systems within the framework of Master equations

To capture the multi-scale nature of compartmentalized stochastic reaction kinetics, we examine both the chemical reactions occurring within compartments and the dynamics of the compartments themselves. At the molecular level, we track changes in the concentrations of chemical species over time within individual compartments. At the mesoscale, we account for processes such as compartment growth, shrinkage, fragmentation, and fusion. Using the robust framework of Master equations, we define a microscopic model and leverage this framework to explore the system's behavior through numerical simulations.

We formally define the state of a compartmentalised stochastic reaction kinetics system by a finite set of compartments $\mathfrak{S}$. Each compartment is characterised

3

by a concentration vector $\vec{c}_i$ and macroscopic compartment properties $\vec{o}_i$,

$$\mathbb{S} = \begin{bmatrix} \vdots \\ [\vec{c}_i, \vec{o}_i] \\ \vdots \end{bmatrix}. \tag{S7}$$

with a total number of $n$ compartments. Compartment properties account for characteristics like size, mass, shape, and spatial position. The size of a compartment is defined as $v_i = \vec{o}_{1;i}$. The concentration vector $\vec{c}_i$ specifies the composition of molecular species in this compartment, where $\vec{c}_i v_i = \vec{n}_i$ is the number of molecular species. We take into account the density dependencies of chemical reactions by considering concentrations rather than copy numbers. Moreover, we assume that molecular species diffuse quickly within the compartment, disregarding concentration gradients and assuming well-mixed conditions.

We distinguish different classes of transitions between states $\mathcal{T}_{\mathbb{S} \to \mathbb{S}'}$, accounting for the dynamics on the scale of stochastic reactions and compartment dynamics. On the scale of stochastic reactions (molecular scale), changes in the concentration of molecular species are due to chemical reactions, such as chemical modifications of proteins, complex formation and reversible oligomerisation processes. Formally, a chemical reaction $C_j + C_k + C_l \xrightarrow{k_\alpha} C_n + C_m$ reduces the concentrations of reactant molecular species $c_{i;j}, c_{i;k}$, and $c_{i;l}$ and increases the concentrations of product species $c_{i;m}$ and $c_{i;n}$ at a rate $k_\alpha$ on compartment $i$. The full dynamics are encoded in a chemical reaction network. We associate each reaction with a transition rate $\mathcal{T}_{k_\alpha, i}$. For an individual compartment, the dynamics are described by a chemical Master equation, see Eq. (S2), also compare for example with [6]. We formally collect transition rates only affecting the molecular level as $\mathcal{Q}$, and acknowledge external signals $\vec{\eta}(t)$ perturbing the rates of chemical reactions by writing $\mathcal{Q}_{\eta(t)}$.

On top of stochastic chemical reaction dynamics in each compartment, compartments can also undertake compartment dynamics, such as fusion, fragmentation, growth, movement and degradation. These compartment processes are formally defined as those which affect the properties $\vec{o}_i$ of compartments and can be further classified into two distinct classes. Macroscopic compartment properties, such as shape and movement, can only be changed by compartment dynamics, resulting in transitions between different states $\mathcal{T}_{\vec{o}_i \to \vec{o}'_i}$. We summarise these processes formally with operator $\mathcal{S}$.

Compartment processes such as compartment growth and shrinkage, degradation and synthesis, or compartment fusion and fragmentation, however, affect both the macroscopic properties of compartments and the concentrations of molecular species $(\vec{c}_i, \vec{o}_i)$. For example, compartment growth leads to a decrease in the density of molecular species, which is reflected in a change in concentrations. Consequently, the growth and shrinkage of compartments affect the collision rates and thus the reaction rates. Such changes in reaction rates are consistently taken into consideration when dealing with chemical reactions by using concentration instead of occupation numbers.

In contrast, when a compartment is degraded, both the compartment itself and its content are removed from set $\mathbb{S}$. This naturally affects both the compartment properties and the number of molecular species. The fusion of two compartments with different concentrations $\vec{c}_i$ and $\vec{c}_j$ results in a single compartment with an averaged concentration. This is supported by the assumption of well-mixed conditions, accounting for instantaneous mixing. In analogous terms, also the synthesis of compartments and the fragmentation of compartments changes both compartment properties $\vec{o}_i$ and the concentration compositions $\vec{c}_i$. Summarising all the transitions with respect to both macroscopic compartment properties and the concentration of molecular species yields the operator $\mathcal{R}$.

Accounting for the stochasticity of compartment dynamics and chemical reactions at finite numbers, we define the probability of finding the system in a specified state $P(\mathbb{S})$, which evolves according to a Master equation $\partial_t P(\mathbb{S}) = \mathcal{L} P(\mathbb{S})$. $\mathcal{L}$ accounts for transitions either due to chemical reactions or compartment dynamics; it is composed of the transitions defined in the classes $\mathcal{Q}_{\eta(t)}$, $\mathcal{S}$, and $\mathcal{R}$, and is in symbolic notation:

$$\partial_t P(\mathbb{S}) = \left( \mathcal{Q}_{\eta(t)} + \mathcal{S} + \mathcal{R} \right) P(\mathbb{S}). \tag{S8}$$

We distinguish intra-scale fluxes ($\mathcal{Q}_{\eta(t)} P(\mathbb{S})$, $\mathcal{S} P(\mathbb{S})$) and inter-scale fluxes ($\mathcal{R} P(\mathbb{S})$). Intra-scale fluxes only affect the dynamics on the specified scale. While we allow the dependence of inter-scale fluxes on the properties of the compartment on a different scale[1], inter-scale fluxes $\mathcal{R} P(\mathbb{S})$ directly set the dynamics on the different spatial scales into an interplay. While we formally defined transitions in $\mathcal{R}$ as compartment dynamics, these transitions not only depend on both compartment properties and the chemical composition $(\vec{c}_i, \vec{o}_i)$, but directly change the composition of chemical species $\vec{c}_i$. An analogy can be drawn to hybrid systems, wherein microscopic dynamics are perturbed by *jump* perturbations. In the multi-scale setup, the stochastic reaction dynamics constitute the level of microscopic dynamics which are perturbed by time-discrete compartment dynamics.

Note, that here we refer to *open* compartmentalised systems, as we do not impose the conservation of mass, neither on the level of compartment dynamics nor on the level of the stochastic reaction kinetics dynamics. This, for example, allows for the continuous binding and unbinding of molecules to and from a compartment.

For further analysis, we specify how compartment fusion and fragmentation change the concentrations within each compartment. With regard to compartment fusion, the copy numbers of the molecular species $\vec{n}_i$ and $\vec{n}_j$ are added together to give $\vec{n}_l = \vec{n}_i + \vec{n}_j$ and the compartments sizes $v_l = v_i + v_j$, which induces changes in the concentration vectors $\vec{c}_l = (v_i \vec{c}_i + v_j \vec{c}_j)/(v_i + v_j)$. Conversely, for compartment fragmentation, the number of molecular species is randomly distributed across each daughter compartment. If the mother compartment is described by $(\vec{c}_l, \vec{o}_l)$, we consider a random splitting into the two daughter compartments, with $v_i + v_j = v_l$, $(v_i, v_j < v_l)$, where $v_i$ $B(v_j)$ is a

---

[1] For example, we account for the compartment size by considering concentrations. Vice versa, we could set also a dependence of the compartment processes in $\mathcal{S}$ on the concentration compositions, for example by linking the concentration to the motility of the compartment.

random variable based on the initial size $v_j$ specified by a break-up distribution $B(v_j)$. Furthermore, each molecular species is subjected to a binomial splitting procedure, with $n_{i;i} = \text{Binom}(n_{l;i}, v_i/v_j)$, where the success probability is proportional to the size of compartment $i$ ($p = v_i/v_j$), and $n_{j;i} = n_{l;i} - n_{i;i}$. The concentration compositions on each daughter compartment then follow from scalar multiplication with the respective sizes of the compartments.

While we have only presented a symbolic notation of the dynamics above, a full Master equation can be directly translated from it. This Master equation, however, is lengthy. We detailed out the terms in [1].

### 2.3. Compartment dynamics within the framework of population balance equations

Compartment dynamics involve processes such as compartment remodeling, fusion, and fragmentation. We identify two levels of population dynamics: a kinetic approach using Master equations and a statistical approach employing smooth distribution functions for large populations. The latter is derived from the Master equation under appropriate population size limits [7]. This approach results in describing the compartment dynamics effectively in terms of population balance equations using smooth distribution functions, analogous to Smoluchowski aggregation-fragmentation processes[2].

We define a phase space $\mathbb{D}$ composed by the physical space of coordinates $\vec{r}$ and the property space $\vec{\zeta} = (\xi_1, \xi_2, ...)$ which refers to compartment properties like size, shape, or the state of the internal dynamics. Here, we assume that compartment properties are described by continuous variables, as $x_i \in \mathbb{R}$. We describe the ensemble of compartments by the number density distribution function $f(\vec{r}, \vec{\zeta}, t)$ in the phase space $\mathbb{D}$, where $f(\vec{r}, \vec{\zeta}, t) \mathrm{d}\vec{r} \mathrm{d}\vec{\zeta}$ refers to the ensemble of entities in the range $(\mathrm{d}\vec{r}, \mathrm{d}\vec{\zeta})$ around $(\vec{r}, \vec{\zeta})$. The conservation of mass then implies

$$\int_\mathcal{D} \frac{D}{Dt} f(\vec{r}, \vec{\zeta}, t) \mathrm{d}\vec{r} \mathrm{d}\vec{\zeta} = \int_\mathcal{D} (\Sigma_\mathrm{B} - \Sigma_\mathrm{D}) \mathrm{d}\vec{r} \mathrm{d}\vec{\zeta}, \qquad (S9)$$

where $D/Dt$ refers to the substantial derivative, $\Sigma_\mathrm{B}$ and $\Sigma_\mathrm{D}$ to birth and death term respectively, and the integration was taken over an arbitrary volume $\mathcal{D}$. Approximating both the spatial dynamics and the internal dynamics of the compartments in terms of Fokker-Planck equations, the drift and diffusive fluxes in both the real space and property space read

$$\vec{d}_r = \vec{F}_r f - D_r \nabla_r f \quad \text{and} \quad \vec{d}_\zeta = \vec{F}_\zeta f - D_\zeta \nabla_\zeta f, \qquad (S10)$$

where the subscript refers to either the real space or the property space, $\vec{d}$ is the flux displacement vector, $\vec{F}$ is the drift vector and the matrix $D$ comprises the diffusion coefficients. In a differential form, the conservation of mass then implies

$$\frac{\partial f}{\partial t} + \nabla_r \cdot (\vec{F}_r f) - \nabla_r \cdot (D_r \nabla_r f) + \nabla_\zeta \cdot (\vec{F}_\zeta f) - \nabla_\zeta \cdot (D_\zeta \nabla_\zeta f) = \Sigma_\mathrm{B} - \Sigma_\mathrm{D}. \qquad (S11)$$

---

[2]Also known as Smoluchowski's coagulation formalism [8].

In this formulation, we already account for the multi-scale character of dynamic compartmentalisation, as we explicitly account for both the dynamics of compartments, as well as for the internal dynamics. Next, we will extend on birth and death terms, which allows accounting for both the synthesis and degradation, as well as for the fusion and fragmentation of compartments. To this end, we employ Smoluchowski's coagulation formalism and start by focusing on the fusion of compartments.

Formally, the fusion of two compartments results in a removal of the two fusing compartments and the addition of the fused compartments to the ensemble, and thus manifests in coupled birth and death terms. For the birth term, we need to define how compartment properties $\vec{\zeta}$ change upon the fusion of compartments. We formally define $\vec{\zeta} = Y(\vec{\zeta}', \vec{\zeta}'')$ for the compartment properties $\vec{\zeta}$ of a compartment resulting from the fusion of two compartments with $\vec{\zeta}'$ and $\vec{\zeta}''$. In the continuum limit, the birth and the death term yield

$$\Sigma_{D,\text{fus}}(\vec{r}, \vec{\zeta}, t) = \int d\vec{r}' d\vec{\zeta}'\, a(\vec{r}', \vec{\zeta}', \vec{r}, \vec{\zeta}, t) f^{[2]}(\vec{r}', \vec{\zeta}', \vec{r}, \vec{\zeta}, t),$$

$$\Sigma_{B,\text{fus}}(\vec{r}, \vec{\zeta}, t) = \frac{1}{2} \int d\vec{r}' d\vec{\zeta}' \int d\vec{\zeta}''\, a(\vec{r}', \vec{\zeta}', \vec{r}, \vec{\zeta}'', t) f^{[2]}(\vec{r}', \vec{\zeta}', \vec{r}, \vec{\zeta}'', t) \delta(Y(\vec{\zeta}', \vec{\zeta}'') - \vec{\zeta}),$$
(S12)

where $a(\vec{r}', \vec{\zeta}', \vec{r}, \vec{\zeta}, t)$ is the fusion rate of two compartments specified by $(\vec{r}', \vec{\zeta}')$ and $(\vec{r}, \vec{\zeta})$, and $f^{[2]}(\vec{r}', \vec{\zeta}', \vec{r}, \vec{\zeta}, t)$ is the entity-pair distribution function. $\delta(\vec{x})$ refers to the multi-dimensional $\delta$-function. The factor $1/2$ in the birth term corrects for doubling counting. Based on this formulation, a number of approximations are applied: A mean-field approximation is applied as the entity-pair distribution is expressed as the product of the number density $f^{[2]}(\vec{r}', \vec{\zeta}', \vec{r}, \vec{\zeta}, t) \approx f(\vec{r}', \vec{\zeta}', t) f(\vec{r}, \vec{\zeta}, t)$. As compartments need to be in close spatial proximity for a fusion, local homogeneity is assumed with $f(\vec{r}', \vec{\zeta}', t) \approx f(\vec{r}, \vec{\zeta}', t)$. Carrying out the integration of $\vec{r}'$ over $a(\vec{r}', \vec{\zeta}', \vec{r}, \vec{\zeta}, t)$ reduces the fusion rate to $\kappa(\vec{r}, \vec{\zeta}', \vec{\zeta}, t)$, which is referred to as the collision kernel. This yields

$$\Sigma_{D,\text{fus}}(\vec{r}, \vec{\zeta}, t) = \int d\vec{\zeta}'\, \kappa(\vec{r}, \vec{\zeta}', \vec{\zeta}, t) f(\vec{r}, \vec{\zeta}', t) f(\vec{r}, \vec{\zeta}, t),$$

$$\Sigma_{B,\text{fus}}(\vec{r}, \vec{\zeta}, t) = \frac{1}{2} \int d\vec{\zeta}' d\vec{\zeta}''\, \kappa(\vec{r}, \vec{\zeta}', \vec{\zeta}'', t) f(\vec{r}, \vec{\zeta}', t) f(\vec{r}, \vec{\zeta}'', t) \delta(Y(\vec{\zeta}', \vec{\zeta}'') - \vec{\zeta}).$$
(S13)

Analogously, the fragmentation of compartments also results in both birth and death terms. Here, we define the break-up kernel $b(\vec{r}, \vec{\zeta}', \vec{\zeta}'', t)$ as rate at which a compartment with properties $\vec{\zeta} = Y(\vec{\zeta}', \vec{\zeta}'')$ splits in compartments specified by $\vec{\zeta}'$ and $\vec{\zeta}''$,

$$\Sigma_{D,\text{frag}}(\vec{r}, \vec{\zeta}, t) = f(\vec{r}, \vec{\zeta}, t) \int d\vec{\zeta}' d\vec{\zeta}''\, b(\vec{r}, \vec{\zeta}', \vec{\zeta}, t) \delta(Y(\vec{\zeta}', \vec{\zeta}'') - \vec{\zeta}),$$

$$\Sigma_{B,\text{frag}}(\vec{r}, \vec{\zeta}, t) = \int d\vec{\zeta}'\, b(\vec{r}, \vec{\zeta}', \vec{\zeta}, t) f(\vec{r}, Y(\vec{\zeta}', \vec{\zeta}), t).$$
(S14)

Taken together, this formalism encompasses various compartment dynamics, such as movement $\vec{d}_r$, internal changes $\vec{d}_\zeta$, growth/shrinkage $\vec{d}$, degra-

dation/synthesis $\Sigma_B$ and $\Sigma_D$ and fusion/fragmentation, expressed as integral terms in $\Sigma_B$ and $\Sigma_D$.

In the following, we consider the continuum limit of an infinite number of compartments and describe the compartments using a number density distribution function[3]. We identify the property vector $\vec{\zeta} = (\vec{c}, \vec{o})$ as the combination of the concentration vector and the compartment properties. In our analysis, we neglect the spatial evolution of the system and move to a mean-field description. In the context of organelle-associated signalling pathways, this approximation is motivated by the rapid movement of organelles through the cytosol. The spatial translocation of organelles is facilitated by molecular motors on the cytoskeleton, leading to rapid spatial reorganization and changes in the neighbourhood of organelles. We assume independence between the chemical composition and shape of compartments, thus formally marginalising over the compartment shape. The compartment properties are then completely described by their size, $\vec{o}_i \equiv v_i$.

A key assumption of the subsequent analysis is that the chemical reactions network is approximated as set of coupled stochastic differential equations, with the stochastic fluxes described using a Fokker-Planck approximation, as outlined in above. To account for variations in compartment size, the noise matrix $\mathbf{D}_{\text{eff}} = \mathbf{D}/v_i$ is made to depend on the size of the respective compartment. In particular, the drift and diffusion flux on the left side of the population balance equation are associated with the chemical reaction dynamics. This leads to the simplification of the population balance equation to

$$\boxed{\frac{\partial f}{\partial t} = -\nabla_c(\vec{F}_{\eta(t)}f) + \nabla_c\left(\left(\nabla_c \mathbf{D}_{\eta(t),\text{eff}}\right)^\top f\right) + \Sigma_B - \Sigma_D,} \qquad (S15)$$

where the number density distribution function $f(\vec{c}, v, t)$ defines the frequency to find compartments with a specified concentration composition $\vec{c}$ and size $s$ at time $t$. The birth and the death term ($\Sigma_B, \Sigma_D$) summarise the compartment dynamics.

Compartment degradation and synthesis are, by definition, birth and death terms. In contrast, compartment growth involves both birth and death terms, as compartments are formally removed and replaced with compartments containing the same chemical species $\vec{n}_i$, but with an altered size. If compartment growth is considered a continuous process, it can be represented by an additional drift that changes both the concentration and size of the compartments $-\nabla_{c,v} \cdot (\vec{F}_{\text{growth}} f)$. Similarly, fusion and fragmentation also contain coupled birth and death terms. Next, we approximate the fusion and fragmentation terms by first focusing on a simplified system, in which compartment dynamics are restricted to fusion and fragmentation, and the reaction kinetics are reduced to 1-dimensional dynamics with $\vec{c} = c$. Assessing the quality of the approximation of the fusion and fragmentation flux first in this simplified setting, we subsequently relax on the assumptions to account for the full complexity of

---
[3]Note, that in this approximation, we make no assumption about the timescale of processes, but continuous fluxes arise due to the assumption of the continuum limit in the number of compartments.

Eq. (S15). This approach allows us to keep track of and assess the quality of the approximations we make.

## 2.4. Flux Approximation of Compartment Fusion

In order to build an intuitive understanding of how fusion and fragmentation affect one-dimensional ensemble statistics, we denote the rate of compartment fusion by $\mu$ and the rate of compartment fragmentation by $\varphi$. To begin, we consider the limit of fast compartment fragmentation, $\mu/\varphi \ll 1$. We set a delta-distributed break-up probability $B(v) = \delta(v' - 0.5v)$, such that compartments always split in equal-sized daughters. In order to prevent compartments from splitting into infinitesimal dust, we limit the fragmentation kernel with a Heaviside step function $\Theta(v - v_0)$, so that only compartments larger than $v_0$ undergo fragmentation. The compartment dynamics are thus given by compartments that immediately fragment after a fusion event, such that we effectively find a delta-peaked size distribution $p(v, t_0) = \delta(v_0)$. In this limit, we neglect the compartment size and assume $f(\vec{c}, v, t) \equiv f(\vec{c}, t)$. At the same time, we still assume that the fused compartment state is preserved sufficiently long to allow for a complete mixing in the fused compartment state. With these assumptions, we symbolically write fusion and fragmentation in terms of two compartments as

$$(\vec{c}_i, \vec{c}_j) \xrightarrow{\text{fus. \& frag.}} (\vec{c}_l + \vec{\xi}, \vec{c}_l - \vec{\xi}), \quad \text{with} \quad \vec{c}_l = \frac{\vec{c}_i + \vec{c}_j}{2}, \tag{S16}$$

where $\vec{c}_l$ is the averaged concentration and $\vec{\xi}$ account for the imperfect splitting. Assuming a small Gaussian contribution $\vec{\xi}$ to be negligible[4] in comparison with the noise induced by chemical binding dynamics, we here set it to zero.

Setting the fusion kernel independent of compartment size and composition, $K(\vec{c}, \vec{c}', v, v') = \mu$, we initially focus on one-dimensional systems ($\vec{c} \to c$). We will relax these assumptions in the subsequent analysis. Under this approximation, we find that the fusion and subsequent fragmentation of compartments follows the Smoluchowski aggregation formalism, with source and sink terms describing the respective *jump into* and *jump out of* a state $c$ by

$$\begin{aligned}\frac{\partial f(c,t)}{\partial t} =& \frac{\partial}{\partial c}\left(-F(c,t,\eta(t))f(c,t) + \frac{1}{2}\frac{\partial}{\partial c}\left(D(c,t,\eta(t))f(c,t)\right)\right) \\ &+ \frac{\mu}{2}\int_0^\infty dc_1 f(c_1,t) \int_0^\infty dc_2 f(c_2,t) \delta\left(\frac{1}{2}(c_1+c_2) - c\right) \\ &- \mu f(c,t).\end{aligned} \tag{S17}$$

By making use of a mean-field approximation, we assumed that the two-point number density function, $f(c_1, c_2, t) = f(c_1, t)f(c_2, t)$, is the product of the one-point number density functions. Furthermore, we assumed that the processes of creation and removal of compartments are coupled, as for every two compartments removed due to fragmentation, two new compartments are created.

---

[4]By construction, the noise induced by imperfect splitting is following a binomial distribution, which has a variance $\sigma^2 = np(1-p)$ and vanishes like the chemical reaction noise with $1/\sqrt{n}$. Assuming that fusion and fragmentation happens on slower time-scales than chemical reactions renders noise due to chemical reactions larger than noise induced by imperfect splitting.

Therefore, instead of computing the jumps in and out of points in phase space, we made use of Gauss' theorem and considered the derivative of the virtual flux $J_{\text{fus}}$. This yields

$$\frac{\partial f(c,t)}{\partial t} = \frac{\partial}{\partial c}\left(-F(c,t,\eta(t))f(c,t) + \frac{1}{2}\frac{\partial}{\partial c}\left(D(c,t,\eta(t))f(c,t)\right)\right) - \frac{\partial}{\partial c}J_{\text{fus}}(c), \tag{S18}$$

with the fusion flux $J_{\text{fus}}(c)$ defined by

$$\frac{J_{\text{fus}}(c)}{\mu} = \frac{1}{2}\int dc_1 \int dc_2 f(c_1,t)f(c_2,t)\Theta((c_1-c)(c-c_2))\text{sgn}\left(c'-c\right), \tag{S19}$$

where we defined $c' = (c_1 + c_2)/2$. Here, $\Theta(c)$ refers to the Heaviside Theta function and $\text{sgn}(c)$ to the sign function. Next, we cast the definition of the $\Theta(c)$ and $\text{sgn}(c)$ in two index sets, over which the two-point number density is integrated

$$I_+(c) = \{(c_1,c_2)|c_1 < c \wedge \frac{c_1+c_2}{2} > c\},$$
$$I_-(c) = \{(c_1,c_2)|c_1 > c \wedge \frac{c_1+c_2}{2} < c\},$$

which yields

$$\frac{J_{\text{fus}}(c)}{\mu} = \int_{I_+(c)} dc_1 dc_2 f(c_1,t)f(c_2,t) - \int_{I_-(c)} dc_1 dc_2 f(c_1,t)f(c_2,t). \tag{S20}$$

We perform a variable transformation to spherical coordinates

$$c_1 = \cos(\theta)r + c,$$
$$c_2 = \sin(\theta)r + c,$$

as we express $c_1$ and $c_2$ as functions of radius coordinate $r$ and an angle coordinate $\theta$. In these coordinates, the fusion flux yields

$$\frac{J_{\text{fus}}(c)}{\mu} = \int_{\pi/2}^{3\pi/4} d\theta \int_0^\infty drr f(\cos(\theta)r+c,t)f(\sin(\theta)r+c,t)$$
$$- f(-\cos(\theta)r+c,t)f(-\sin(\theta)r+c,t). \tag{S21}$$

In this formulation, we find symmetries of double integral, as the integral over the radius vanishes for $\theta = 3\pi/4$. Conversely, if $f(\tilde{c}+c,t)^2 \geq f(\tilde{c}+2c,t)f(\tilde{c},t)$, $\forall \tilde{c}, c \in \mathbb{D}$, the integral over the radius has the maximal absolute value for $\theta = \pi/2$ and decreases monotonically in the range $\theta \in [\pi/2, 3\pi/4]$ by construction; Here $\mathbb{D}$ refers to the support of the distribution $f$. Note, that this condition is in particular true for $f(c) \propto e^{-V(c)}$, and $V(c)$ a convex function. We next make use of these symmetries, as we approximate the double integral. To this end, we perform the integral over the angle before the integral radial coordinate. For brevity of notation, we introduce

$$g(r,\theta,c,t) \equiv f(\cos(\theta)r+c,t)f(\sin(\theta)r+c,t) - f(-\cos(\theta)r+c,t)f(-\sin(\theta)r+c,t). \tag{S22}$$

For the integral over $\theta$, we linearise the $g(r,\theta,c,t)$ in $\theta$ for $\theta \in [\pi/2, 3\pi/4]$. $|g(r,\theta,c,t)|$ is a monotonically decreasing function in $\theta \in [\pi/2, 3\pi/4]$ by construction, if $f(m+x,t)^2 \geq f(m+2x,t)f(m,t), \forall m,x \in \mathbb{D}$. We approximate the integral over $\theta \in [\pi/2, 3\pi/4]$ by a triangle, as we linearise $|g(r,\theta,c,t)|$ at $\theta = \pi/2$ and compute the intersection with $h(\theta) = 0$. The intersection is approximately given by

$$\Delta\theta \approx \left|\frac{f(c,t)}{f'(c,t)}\right| \frac{r}{r + \left|\frac{f(c,t)}{f'(c,t)}\right|}, \tag{S23}$$

where $f'(c,t) = \partial_c f(c,t)$ refers to the derivative with respect to the concentration. This approximation for $\Delta\theta$ renders an approximation for the double integral as

$$\frac{J_{\text{fus}}(c)}{\mu} \approx \int_0^\infty dr \frac{1}{2} \Delta\theta g(r, \pi/2, c, t) = \frac{1}{2} \left|\frac{f(c,t)}{f'(c,t)}\right| \int_0^\infty dr \frac{r}{r + \left|\frac{f(c,t)}{f'(c,t)}\right|} g(r, \pi/2, c, t). \tag{S24}$$

We note, that for $c$ a constant, the fraction in the integral over the radius renders a constant.

We introduce the *short range approximation*

$$\boxed{\frac{J_{\text{fus, sr}}(c)}{\mu} \approx \frac{1}{2} \int_0^\infty dr r g(r, \pi/2, c, t) = \frac{1}{2} f(c,t) \int_\mathbb{D} dc' f(c',t)(c'-c),} \tag{S25}$$

which is especially suited to estimate the flux close to the mean of the distribution $f(c)$. By comparing the Fokker-Planck equation for a many-body system with two-particle interaction potentials, we note a structural equivalence, see for example [9]. We thus identify the short-range approximation with an effective two-body interaction potential $w_{i,j;sr} = (c_i - c_j)^2$ and the stable tail long-range approximation with $w_{i,j;stlr} = |c_i - c_j|$. Our findings suggest that the combination of compartment fusion and subsequent fragmentation act as a steady attractive force in the concentration phase-space between all compartments.

In order to simplify the approximation further, we define a normalisation factor $N(t) = \int dc f(c,t)$. We normalise $p(c,t) = f(c,t)/N(t)$ to a probability distribution, and define $N(t)\langle c(t)\rangle = \int dc c f(c,t)$ as the mean of the distribution $f(c,t)$. Furthermore, we define the cumulative distribution $\Phi(c,t) = \int_\infty^c dc' f(c')$, and, by performing the integral in the fusion flux approximations, we find that the short-range approximation describes a steady attraction towards the mean

$$J_{\text{fus, sr}}(c) \approx \frac{\mu}{2} N f(c,t)(\langle c \rangle - c). \tag{S26}$$

In the absence of drifts, $F_{\eta(t)}(c,t) = 0$, and constant diffusion, $D_{\eta(t)}(c,t) = D$, the short-range approximation of the fusion flux admits analytical solutions. We make use of analytical solutions to test the goodness of the approximations. To this end, we simulate a finite-sized ensemble of Brownian random walkers, which undergo random averaging events due to fusion and subsequent immediate fragmentation, as introduced above. With these approximations, the

population balance equation then yields

$$\partial_t f(c,t) = \partial_c \left[ -J_{\text{fus}} + D \partial_c f(c,t) \right]. \tag{S27}$$

Without any loss of generality, we normalize $\int_{-\infty}^{\infty} dc f(c) = 1$ and initialize the random walkers at $p(c, t_0) = \delta(c)$ at the origin. Our analysis predicts that the ensemble statistics will approach a steady state with finite variance, as opposed to an ensemble of random walkers in the absence of fusion and fragment. The simulations reveal that the variance increases linearly with time in the absence of compartment fusion and fragmentation, whereas the variance reaches a fixed value if compartment fusion and fragmentation are taken into consideration. The steady-state distribution for the short-range approximation is

$$f_{\text{sr; ss}}(c) = \sqrt{\frac{\mu}{4\pi D}} \exp\left(-\frac{\mu}{4D} c^2\right), \tag{S28}$$

which predicts a variance $\sigma_{\text{sr}}^2 = 2D/\mu$.

We test our theoretical predictions through numerical simulations. We consider a finite set of random walkers which undergo pair-wise averaging events at a rate of $\mu$. Fig. S1 (a) shows that, in agreement with our prediction, the dynamics approach a steady state when averaging is present, i.e. when compartment fusion and fragmentation occurs. In contrast, the variance in the set of random walkers without interactions shows the expected diffusive spreading and no steady state is approached. Fig. S1 (b) demonstrates that we correctly predict the variance of the steady state distribution as a function of the fusion rates. Note that our prediction admits no free fit parameters. Fig. S1 (c,d) show that we correctly predict the shape of the distribution $f(c,t)$. The short-range approximation correctly describes the shape close to the mean of the distribution, while the long-range stable tail approximation is well-suited to assess the tail of the distribution.

### *2.4.1. Accounting for finite fragmentation rates*

Relaxing on the fast fragmentation limit, we account for compartments of varying sizes. Here, we assume that the break-up probability is uniformly distributed over the compartment volume. The fusion flux integral is adjusted by accounting for the number density $f(c, v, t)$, which additionally account for the compartment size. Adjusting the averaging by rescaling the variable $x'$ accordingly, and tracking the masses that move across position $c$, this yields:

$$\frac{\tilde{J}_{\text{fus}}(c, v_1, v_2)}{\mu} = \int dc_1 \int dc_2\, v_1 \left[ f(c_1, v_1, t) f(c_2, v_2, t) \Theta((c - c_1)(c' - c)) \operatorname{sgn}(c' - c) \right], \tag{S29}$$

with $c' = (v_1 c_1 + v_2 c_2)/(v_1 + v_2)$. Here, we kept the compartment masses $v_1$ and $v_2$ fixed. The full fusion flux is then computed by integrating over the masses as

$$J_{\text{fus}}(c) = \int dv_1 \int dv_2\, \tilde{J}_{\text{fus}}(c, v_1, v_2). \tag{S30}$$

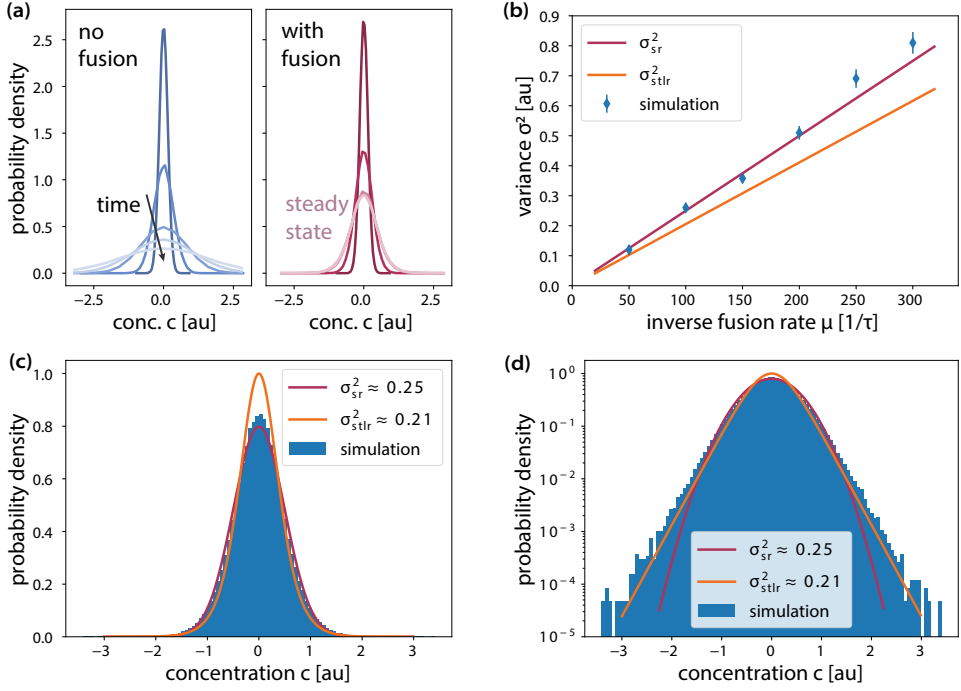

**Fig. S1. Corroboration of the fusion flux approximation by full stochastic numerical simulations** Numerical simulations are conducted by simulating random walkers in a flat potential with constant diffusion coefficient using over-damped Langevin equations. Random walkers are initialised at time $t = 0$ at position $c_0 = 0$. Compartment fusion and subsequent fragmentation is modelled as stochastic, time-discrete averaging events between two randomly chosen random walkers. $\mu$ refers to the rate of the averaging events. All results are present in simulation units, where $\tau$ refers to the time-discrete updating step. (a) While the random walkers without "compartment fusion and fragmentation" disperse, the random averaging events give rise to steady state distribution with finite dispersion. (b) The short-range approximation *sr* is presented in Eq. (S26). The long-range approximation *stlr* is explained in [1]. Note, that the estimates are fully defined by the simulation parameters and admit no free fit parameter. Both approximations correctly predict the functional dependence of the fusion rate $\mu$. Furthermore, the long-range approximations give a good quantitative assessment of the variance. Errorbars show the standard deviation. (c) The short-range approximation furthermore captures the shape of the steady state well close to the mean of the distribution. (d) The short-range approximation does not adequately predict the tail of the distribution, as it is expected by construction, see also [1] and [10].

Here, the subscript 'v' refers to *size*. Accounting for different masses directly translates into the two index sets

$$I_+(c, v_1, v_2) = \{(c_1, c_2) | c_1 < c \wedge \frac{c_1 v_1 + c_2 v_2}{v_1 + v_2} > c\},$$

$$I_-(c, v_1, v_2) = \{(c_1, c_2) | c_1 > c \wedge \frac{c_1 v_1 + c_2 v_2}{v_1 + v_2} < c\},$$

which yields

$$\frac{\tilde{J}_{\text{fus}}(c)}{\mu} = \int_{I_+} dc_1 dc_2 \, v_1 f(c_1, v_1, t) f(c_2, v_2, t) - \int_{I_-} dc_1 dc_2 \, v_1 f(c_1, v_1, t) f(c_2, v_2, t). \tag{S31}$$

We find, that the approximation of the fusion flux accounting for different masses follows the analogous structure as in the fast fragmentation limit discussed above, if the two compartment masses $v_1$ and $v_2$ are considered fixed. When converting to polar coordinates, we find the analogous equation Eq. (S21) with the integration boundary of the integral over the angle varied to $\theta \in [\pi/2, \pi/2 + \alpha]$, where the upper boundary is determined by $\alpha = \arctan(v_2/v_1)$. Analogous to the triangle approximation discussed above, we approximate that both fractions $v_2/v_1 > 1$ and $v_2/v_1 < 1$ effectively reduce the value of the integral. In the picture of the triangle, for $v_2/v_1 < 1$, we find that the upper boundary $\pi/2 + \alpha < 3\pi/4$. In this case, the triangle is not integrated over the full baseline. Conversely, for $v_2/v_1 > 1$ the integration includes negative values. We approximate each integral $\mathcal{I}_\theta(r, \theta, c, v_1, v_2, t)$ by

$$\mathcal{I}_\theta(r, \theta, c, v_1, v_2, t) \approx \mathcal{I}_\theta(r, \pi/4, c, v_1, v_2, t) \left(1 - \left(\arctan\left(\frac{v_2}{v_1}\right) \frac{4}{\pi} n - 1\right)^2\right)$$

$$= \mathcal{I}_\theta(r, \pi/4, c, v_1, v_2, t) \Lambda(v_1, v_2) \tag{S32}$$

which describes an adjustment of the triangle area by the fraction of the reduced or extended baseline. With this, the approximation of the fusion flux shows the same structure as in Eq. (S24),

$$\frac{J_{\text{fus, v}}(c)}{\mu} \approx \int dv_1 \int dv_2 \, v_1 \Lambda(v_1, v_2) \frac{1}{2} \left|\frac{f(c, v_2, t)}{f'(c, v_2, t)}\right| \int_0^\infty dr \frac{r}{r + \left|\frac{f(c,v_2,t)}{f'(c,v_2,t)}\right|} g(r, \pi/2, c, v_1, v_2, t). \tag{S33}$$

While this approximation show the same functional form as in the case of compartments of the same size, the above equation is analytically difficult to capture if the joint distribution $f(c, v_1, t)$ of mass and concentration is considered. In the next step, we further simplify this expression by marginalizing over the different compartment sizes.

### 2.4.2. Marginalizing over compartment sizes

To allow for an analytic treatment of Eq. (S33), we further approximate $\frac{J_{\text{fus},v}(c)}{\mu}$ by marginalizing over the different compartment sizes. To this end, we employ a mean-field approximation, approximating $f(c, v, t) = f(c, t) m(v, t)$. We assume

14

compartment fusion to be independent of the compartment sizes and the fragmentation rate to be proportional to the size of the splitting compartment. The distribution $f(c,t)$ then describes the mass density, that is the number of compartments with concentration $c$ at time $t$ in units of multiples of compartment building block mass. Under these assumptions, the size distribution $m(v,t)$ is approximately given by an exponential distribution $m(v,t) \propto a^2 e^{-a(t)v}$, where $a^2$ adjust for measuring the $m(v,t)$ in terms of multiples of the building blocks. With these assumptions, the fusion flux approximates to

$$\frac{J_{\text{fus, mv}}(c)}{\mu} \approx \frac{a}{3} \left| \frac{f(c,t)}{f'(c,t)} \right| \int_0^\infty dr \frac{r}{r + \left| \frac{f(c,t)}{f'(c,t)} \right|} g(r, \pi/2, c, t). \tag{S34}$$

Here the subscript 'mv' indicates the marginalisation over the compartment size. We find that the fusion flux marginalised over different compartment sizes is proportional to the fusion flux in the approximation of singular compartment size $J_{\text{fus, mv}}(c) \propto J_{\text{fus}}(c)$ and rescaled only by a numerical factor $\gamma(\langle v \rangle)$ which is a function of the average mass.

To consistently account for the marginalisation of the number density $f(c,v,t) \to f(c,t) = \int dv f(c,v,t)$ in the population balance equation in Eq. (S15), it is necessary to also marginalise the other terms. The drift term can be drawn out of the integral, whereas the diffusion $D_{\eta(t)}(c,v)$ has a size dependence of $D_{\eta(t)}(c,v) = D_{\eta(t)}(c)/v$. Taking advantage of the same size distribution as for the marginalisation of the fusion flux, $m(v,t) \propto a^2 \exp[-a(t)v]$, and assuming a mean-field approximation, we obtain

$$D_{\eta(t),\text{eff}}(c) = \int dv\, m(v,t) \frac{D_{\eta(t)}(c)}{v} \approx \frac{D_{\eta(t)}(c)}{\langle v \rangle}. \tag{S35}$$

Furthermore, the birth and death terms in Eq. (S15), which take into account compartment degradation, synthesis and growth, must be marginalised under the same assumption in order to obtain a self-consistent approximation of the population balance equation.

We tested the approximations of marginalising over different sizes using the same model considered in the fast fragmentation limit and Fig. S1. This model featured an absence of a drift term, a constant diffusion coefficient $D(c) = D$, a constant fusion kernel $K(c',c,v',v,t) = \mu$, and finite fragmentation rates $\phi < \infty$. We rescaled the diffusion coefficient to $D(c,v,t) = D/v$ and predicted an effective diffusion constant that scales inversely with the average system size, i.e. $D_{\text{eff}} \propto 1/\langle v \rangle$. Additionally, we defined an effective fusion rate which summarised all the prefactors in front of the approximation of the fusion integral and found a reciprocal scaling $\mu_{\text{eff}} \propto 1/\langle v \rangle$. Therefore, we predicted that the distribution $f(c,v,t)$ should remain fixed as the fragmentation rate $\varphi$ was varied. As expected, our results demonstrated that the short-range approximation in Eq. (S26 accurately approximated the marginalised number density $f(c,t)$, see Fig. S2 (a). Furthermore, in line with our prediction, the variance remained approximately constant, see Fig. S2 (b).

The results of the marginalisation presented in this section rely upon the assumption of a size distribution. Should other compartment fusion and fragmentation dynamics be considered, the approximations made in the context of

15

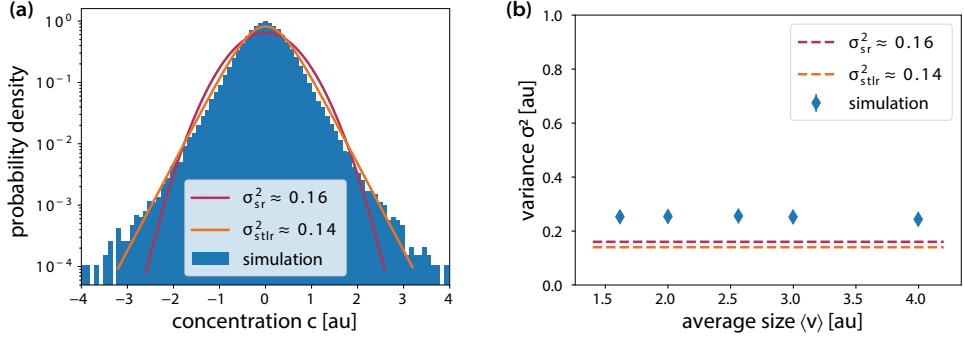

**Fig. S2. Corroboration of the fusion flux approximation accounting for the marginalisation over compartment sizes** Simulations are analogous to Fig. S1, but additional finite fragmentation rates $\varphi < \infty$ are assumed, results in compartments with various sizes. While the enclosed dynamics are modelled by Langevin Equations, the compartment dynamics simulated as full stochastic processes, also compare with appendix 4. All results are present in simulation units, where $\tau$ refers to the time-discrete updating step. (a) Considering the normalised mass density distribution function $f(c,t) / \int f(c,t) dc$, both the short-ange approximation in Eq. (S26) and long-range approximation presented in [1] give qualitative good estimates of the steady-state distribution. (b) Based on our analysis, we predict the variance of the steady state to be independent of the average compartment size $\langle v \rangle$. Errorbars show the standard deviation. We verify this prediction with numerical simulations. Compartment size is evaluated in multiples of the compartment unit size.

the marginalisation need to be reviewed and adjusted accordingly. The approximation holds well in the model studied in Fig. S2, however, additional processes which introduce a correlation between the concentration and the size distribution reduce the efficacy of the mean-field approximation. This is especially true when fusion kernels with an explicit concentration dependence are taken into account, in which case the size marginalisation should be applied with caution. Thus far, we have only focused on effective one-dimensional models. In the next stage, we will generalise the approximation to include $d > 1$ dimensional dynamics.

### 2.4.3. Accounting for multi-variate concentration vectors

The approximation of the fusion flux in Eq. (S26) relied on the assumption of a 1-dimensional dynamics in the concentration $c$. We note, that the fusion of two compartments creates the virtual displacement of compartment masses along a 1-dimensional line in the concentration phase space. We used in the approximation of the double integral, that this agrees with the dimension of the phase space for 1-dimensional dynamics.

We note, that also in higher dimensional dynamics the fusion of two compartments creates a virtual displacement along a one-dimensional line. To approximate the fusion flux at a specified point $\vec{J}_{\text{fus}}(\vec{c}^*)$ in the concentration phase space, we hence need to account for mass displacements due compartment fusion across $\vec{c}^{*\top}$ in all possible directions. To this end, we consider

16

a parametrisation of a $d$-dimensional concentration phase space in spherical coordinates, making use of a radial coordinate $r$ and $d-1$ angle coordinates $\phi_1, ..., \phi_{d-1}$, where $r \in [0, \infty)$, $\phi_1, ..., \phi_{d-2} \in [0, \pi]$, and $\phi_{d-1} \in [0, 2\pi)$, as

$$c_1 = c_1^* + r\cos(\phi_1)$$
$$c_2 = c_2^* + r\sin(\phi_1)\cos(\phi_2)$$
$$...$$
$$c_{d-1} = c_{d-1}^* + r\sin(\phi_1)\cdots\cos(\phi_{d-1})$$
$$c_d = c_d^* + r\sin(\phi_1)\cdots\sin(\phi_{d-1}).$$

We define a line in the multi-dimensional phase space, by fixing all angle variables $\phi_1, ..., \phi_{d-1}$ and allowing the radial coordinate to assume values in $r \in (-\infty, \infty)$, as $\vec{l}(r, \phi_1, ..., \phi_{d-1})$. For the brevity of notation, we collect the angle variables in a vector $\vec{\alpha} = (\phi_1, ..., \phi_{d-1})^\top$, and write $\vec{l}_{\vec{\alpha}}(r)$. For every line, we define a unit vector $\vec{e}_\alpha$, which points in the same direction as the line $\vec{l}_{\vec{\alpha}}(r)$. Along every line $\vec{l}_{\vec{\alpha}}(r)$, the virtual displacement of compartments attribute to the total flux $\vec{J}_{\text{fus}}(\vec{c}^*)$. To account for this, we define the fusion flux along a specified line as $\vec{J}_{\text{fus},\vec{\alpha}}(\vec{c}^*)$. Note, that the fusion flux along the one-dimensional line can be approximated by Eq. (S26). Instead of varying the coordinated one-dimensional concentration $c$, here the radial coordinate $r$ is varied, while the angle variables $\vec{a}$ are considered fixed. By construction, the flux decomposes into $\vec{J}_{\text{fus},\vec{\alpha}}(\vec{c}^*) = J_{\text{fus},\vec{\alpha}}(\vec{c}^*) \cdot \vec{e}_\alpha$. The total flux is recovered by

$$\vec{J}_{\text{fus}}(\vec{c}^*) = \int d\vec{\alpha}\, J_{\text{fus},\vec{\alpha}}(\vec{c}^*) \cdot \vec{e}_\alpha, \tag{S36}$$

where all angles are integrated over the range $\phi_i \in [0, \pi]$ to cover the full phase space.

The definition in Eq. (S36) correctly accounts for correlations in the different molecular species in the concentration vector $\vec{c}$, yet it does not admit further analytical treatment. We next approximate Eq. (S36), by setting a mean-field approximation on the concentrations, such that $f(\vec{c}, v, t) = \prod_j^d f_j(c_j, v, t)$. We set the formal independence of averaging events due to compartment fusion. In this approximation, the fusion flux factorizes, such that each component is

$$\vec{J}_{\text{fus},i}(\vec{c}^*) \approx \mu \frac{1}{3\langle v \rangle} \left| \frac{f(\vec{c}^*, t)}{f'(\vec{c}^*, t)} \right| \int_0^\infty dr \frac{r}{r + \left|\frac{f(\vec{c}^*,t)}{f'(\vec{c}^*,t)}\right|} g_i(r, \pi/2, c_i^*, t) \left( \prod_{j \neq i} f_j(c_j^*, t) \right) \tag{S37}$$

in analogy to Eq. (S24), and $g_i(r, \pi/2, \vec{c}_i^*, t)$ constructed with the single component number density $f_i(c_i, t)$. Here, we made use of the marginalisation of different compartment sizes. Note, that this in particular allows for the simple generalisation of the short-range approximation of the fusion flux in Eq. (S26), which yields

$$\boxed{\vec{J}_{\text{fus, sr}}(\vec{c}) \approx -\frac{\mu}{3\langle v \rangle} f(\vec{c}, t) \nabla_c \int_\mathbb{D} d\vec{x}\, f(\vec{x}, t)\, (\vec{x} - \vec{c})^2\,.} \tag{S38}$$

17

## 2.5. Qualitative effects of compartment dynamics

In order to assess the effects of different compartment dynamics qualitatively, we begin by discussing the dynamics in the absence of compartment dynamics. The chemical reaction kinetics on each compartment are described by the changes in the molecular species $\vec{c}$ of the compartmentalised stochastic reaction kinetics system. As introduced above, on each compartment, we observe a stochastic trajectory that depicts the change in the concentration of different molecular species as specified by a chemical reaction network. In the approximation of chemical Langevin equations, the changes in the trajectory are split into two qualitative contributions: the drift term describes deterministic changes in the concentration vector $\vec{c}$, and the diffusion term accounts for the intrinsic stochasticity of chemical reactions at small concentrations and deviations from the deterministic trajectory.

As we consider an ensemble of several compartments, there are variations in the concentration vector $\vec{c}_i$ due to the stochastic noise. To describe this ensemble, we use a number density distribution function $f(\vec{c}, t)$, which estimates the amount of compartment mass with a specified concentration $\vec{c}$ at time $t$. The temporal changes in the number density are described by a population balance equation. The noise causes the distribution $f(\vec{c}, t)$ to spread over the concentration phase space, the details of which are determined by the interplay of the drift and the diffusion flux. As the variance among the compartments increases, the dynamics are said to *explore* the concentration phase space. The collective configuration of the compartments as an ensemble gives the state of the system S, so a full knowledge of the distribution function $f(\vec{c}, t)$ is formally needed to assess this state.

The fragmentation of compartments creates a binomial splitting in every molecular species, thereby increasing the variance in the distribution $f(\vec{c}, t)$ by construction. We approximate the effects of compartment fragmentation by adding an additional diffusion flux. Specifically, in a description where we have marginalised over compartment sizes, we interpret the effects of compartment fragmentation as a correction to the diffusion flux of the chemical reaction dynamics. We thus assess that the fragmentation flux yields no qualitatively different dynamics.

For compartment growth and shrinkage, the growth flux is fully defined at the level of individual compartments; Similar to the deterministic drift flux of the chemical reaction kinetics. We interpret fluxes due to the growth and shrinkage of compartments as corrections to the drift flux of the chemical reaction kinetics. Also, we assess that these compartment dynamics result in no qualitatively different dynamics.

Degradation and synthesis terms cannot be expressed within the framework of fluxes as they are, by definition, birth and death terms. Note that by controlling source and sink terms, the distribution $f(\vec{c}, t)$ can be remodelled into any arbitrary shape. However, if the distribution $f(\vec{c}, t)$ is effectively determined by the fine-tuning of birth and death terms, the dynamics of the system are dictated by the regulation of synthesis and degradation of compartments, and the chemical reaction kinetics would only have marginal effects. If the degradation and synthesis terms are found to dominate the dynamics, the model assumptions

should be carefully reconsidered. It is also worth noting that some specific synthesis and degradation dynamics can be effectively modelled as boundary conditions of the population balance equation: for example, the degradation of compartments upon the accumulation of a critical amount of molecular species $\vec{c}^*$ can be treated as an absorbing boundary condition. Here, we are in particular interested in homeostatic conditions, where we have no explicit coupling between the characteristics of a compartment and its synthesis or degradation. In this case, the synthesis and degradation yield no qualitative effects on the dynamics of the system.

Compartment fusion affects the dynamics of the compartmentalised stochastic reaction kinetics systems in a qualitatively different way. we derived an approximation of the fusion flux, which we termed the *short range approximation* of the fusion flux, and which has the form of an effective two-body interaction potential. Identifying the fusion flux with a two-body interaction potential allows us to interpret the effects of the fusion flux by drawing analogies to other physical systems. Furthermore, we can directly assess a qualitative difference between the fusion flux and the drift flux due to the chemical reaction kinetics: while the drift flux is determined by the molecular species of the chemical reactions in a specific compartment, the fusion flux depends on the full distribution $f(\vec{c}, t)$.

Compartment fusion introduces unique two-body interactions that distinguish it from other compartment dynamics, which typically affect individual compartments independently. Fusion establishes correlations among compartments, breaking the assumption of statistical independence, and leads to the formation of a single, fully-fused compartment in the absence of fragmentation. When both fusion and fragmentation are present, their combined effects drive individual compartments toward the mean of the distribution $f(\vec{c}, t)$, counteracting dispersion caused by diffusive flux and stochastic reaction kinetics. This balance results in a steady ensemble variance, with the localized ensemble moving collectively through the concentration phase space while maintaining an approximately fixed shape.

We recognise that equations analogous to Eq. (S15) in the absence of birth and death terms, and using the approximation of the fusion flux with Eq. (S26), have been considered in the literature under the name of *McKean-Vlasov Equations* [11–15]. These equations have been particularly studied in the context of collective dynamics of stochastic many-body physics, for example, to describe collective excitations in nuclei plasmas [11, 16, 16–18], notice here also the relation to the Vlasov-Fokker-Planck equation. The linkage to these physical systems is mainly due to the interpretation of compartment fusion and fragmentation giving rise to an effective two-body interaction potential; Compare also with [9].

Although our collective degree of freedom emerges from a classical rather than a quantum-mechanical framework, it exhibits similarities to a quasi-particle in quantum mechanics. While we recognize the semantic limitations of calling it a quasi-particle, this terminology highlights the particle-like motion of the collective degree of freedom within the concentration phase space. Importantly, this collective degree of freedom demonstrates qualitatively distinct kinetic behavior compared to individual compartments in the absence of fusion and fragmentation, underscoring its nature as a fundamentally many-body phenomenon. For clarity, we will refer to this collective degree of freedom as a quasi-particle

moving forward.

## 2.6. Equations of motion for the quasi-particle

To analyze the quasi-particle's kinetics, we use the short-range approximation of the fusion flux, which describes steady attraction to the distribution mean. We characterize the quasi-particle using the mean, multivariate median, and variance of $f(\vec{c}, t)$, focusing on ensemble statistics near the center of the distribution. Additionally, we assume that the further compartment dynamics are independent of $\vec{c}$, homeostatic conditions hold such that the total compartment mass is conserved and that the diffusion coefficients $\mathbf{D}_{\text{eff},\eta(t)} = D_{\text{eff},\eta(t)}$ remain constant.

For brevity of notation, we summarise the population balance equation to

$$\partial_t f(\vec{c}, t | \vec{c}_0, t_0) = \nabla[-\vec{F}[\vec{c}, f(\vec{c}, t | \vec{c}_0, t_0)] f(\vec{c}, t | \vec{c}_0, t_0)] + D\nabla^2 f(\vec{c}, t | \vec{c}_0, t_0), \quad \text{(S39)}$$

where $(\vec{c}_0, t_0)$ refers to the initial condition of the distribution. Here, $\vec{F}$ refers to the composite of the approximation of the fusion flux and the deterministic drift component of the stochastic chemical reaction kinetics, whilst the diffusion term accounts for both the diffusion noise in the stochastic reactions and the noise induced by compartment fragmentation. For brevity of notation, we set $f(\vec{c}, t | \vec{c}_0, t_0) \equiv f(\vec{c}, t)$. We define the multivariate generalisation of the median $\vec{m}$, as the point in the concentration phase space where the stable tail long-range approximation of the fusion flux vanishes. Furthermore, we define a multivariate generalisation of the cumulative distribution $\vec{\Phi}$, where each component is

$$\Phi_i(\vec{x}) = \int_{-\infty}^{\infty} \prod_{j \neq i}^{d} dc_j \int_{\infty}^{x} dc_i f(\vec{c}, t) = \int_{\infty}^{x} dc_i q_i(c_i, t). \quad \text{(S40)}$$

In the second equation, we introduced the marginalised density per component $q_i(c_i, t)$ and $d$ refers to the dimensionality of $\vec{c}$. As $\Phi_i$ is by construction a monotonically increasing function, we define the inverse function of the generalised cumulative distribution as $\vec{\Phi}^{-1}$. We define the multi-variate generalisation of the median $\vec{m} = \vec{\Phi}^{-1}(N/2\vec{e}_\mathbb{1})$, where $N = \int d\vec{c} f(\vec{c}, t)$ and $\vec{e}_\mathbb{1}$ is the vector of ones. Applying a mean-field approximation on the flux $\vec{F}$, the integration of the population balance equation yields

$$\partial_t \Phi_i(\vec{c}, t) \approx -F_i[\vec{c}, f(\vec{c}, t)] q_i(c_i, t) + D \partial_{c_i} q_i(c_i, t). \quad \text{(S41)}$$

We note that $\vec{m}$ is a function of the distribution $f(\vec{c}, t)$. We next derive how $\vec{m}$ changes as $f(\vec{c}, t)$ evolves,

$$\partial_t \vec{m} = \partial_t \vec{\Phi}^{-1} \left( \frac{N}{2d} \vec{e}_\mathbb{1} \right). \quad \text{(S42)}$$

Let $\vec{m}_0$ refer to the multi-variate generalised median at time $t_0$. To obtain the temporal derivative of $\vec{m}$, we linearise $\vec{\Phi}(\vec{c}, t)$ around a point $\vec{m}_0$ and obtain for each component

$$\begin{aligned}\Phi_i(\vec{m}, t) =& \Phi_i(\vec{m}_0, t_0) + \nabla_c \Phi_i(\vec{m}_0, t_0) \cdot (\vec{m} - \vec{m}_0) \\ &+ \partial_t \Phi_i(\vec{m}_0, t_0) \cdot (t - t_0) + \mathcal{O}(\vec{c}^2, t^2).\end{aligned} \quad \text{(S43)}$$

Not that by construction $\partial_{c_i}\Phi_j = 0$ if $i \neq j$. Solving the equation componentwise, $m_i$ is

$$m_i \approx -\frac{\partial_t \Phi_i(\vec{m}_0, t_0) \cdot (t - t_0)}{q_i(m_{0,i})} + m_{0,i}, \qquad (S44)$$

which directly yields

$$\partial_t m_i \approx -\frac{\partial_t \Phi_i(\vec{m}_0, t_{m,0})}{q_i(m_{0,i})}. \qquad (S45)$$

This temporal derivative gives the instantaneous response for $\vec{m}_0$ at time $t_{m,0}$, and is an approximation for later time points. Using the approximation for $\partial_t \Phi_i(\vec{c}, t)$, the temporal derivative of the generalised median further simplifies to

$$\partial_t m_i \approx F_i[\vec{m}, f(\vec{m}, t)] + \frac{D \partial_{m_i} q_i(m_i, t)}{q_i(m_i, t)}. \qquad (S46)$$

For $||F_i|| \gg ||D||$, the first term on the right-hand side of the equation dominates. Note, that $D$ vanishes for large systems sizes due to the rescaling by the system size. Inserting the definition of the drift flux due to the chemical reaction kinetics and the fusion flux, the effective equation of motion for the generalised median is

$$\partial_t \vec{m} = \vec{F}_{\eta(t)}(\vec{m}) + \Lambda(\langle \vec{c} \rangle - \vec{m}). \qquad (S47)$$

The drift term $\vec{F}_{\eta(t)}$ refers to the chemical reaction kinetics, where $\eta(t)$ denotes the external signal. The short-range approximation is used for the fusion flux and all prefactors are collected in $\Lambda$. The difference between the mean $\langle \vec{c} \rangle$ and the multivariate generalisation of the median $\vec{m}$ quantifies the skewness of the distribution $f(\vec{c}, t)$ as an analogue to a rescaled version of the non-parametric skew in one-dimension. This skewness parameter is referred to as $\vec{s} = \langle \vec{c} \rangle - \vec{m}$ and the fusion flux gives rise to an additional force that pulls the median in the direction of $\vec{s}$. The rate of compartment fusion $\Lambda \propto \mu$ is directly proportional to the strength of this force. We now proceed to derive the temporal derivative $\partial \langle m \rangle$ of $\vec{s}$ over time.

Integrating the population balance equation, we assume that $f(\vec{c}, t)$ decay exponentially or faster in $\vec{c} \to \infty$ and find the temporal evolution of the mean is

$$\partial_t \langle c \rangle = \int_{-\infty}^{\infty} d\vec{c}\, \vec{F}[\vec{c}, f(\vec{c}, t)] f(\vec{c}, t) = \sum_{\alpha} \frac{D^{\alpha} f(\langle \vec{c} \rangle, t)}{\alpha!} \int_{-\infty}^{\infty} d\vec{c}\, (\vec{c} - \langle \vec{c} \rangle)^{\alpha}. \qquad (S48)$$

Making use of the series expansion of $\vec{F}$ around $\langle \vec{c} \rangle$ with the multi-index notation $\vec{x}^{\alpha} = x_1^{\alpha_1} \cdots x_d^{\alpha_d}$ and the mixed partial derivatives $D^{\alpha} = \partial_{x_1^{\alpha_1}} \cdots \partial_{x_d^{\alpha_d}}$, we assume the existence of a localised ensemble state with finite dispersion. We approximate that the vector field spanned by the drift term $\vec{F}$ is well-approximated around $\langle \vec{c} \rangle$ on the scale of the finite dispersion of the localised ensemble state by a second-order expansion of $\vec{F}$. With this, $\partial_t \langle \vec{c} \rangle$ approximates to

$$\partial_t \langle \vec{c} \rangle_i \approx F_i(\langle \vec{c} \rangle, t) + \frac{1}{2}\text{tr}(H_{F_i} K_{\vec{c},\vec{c}}^{\top}). \qquad (S49)$$

21

Note that the first-order derivative vanishes due to the definition of $\langle \vec{c} \rangle$. Here, we made use of the Hessian matrix $H_{F_i}$ of the $i$-component of $\vec{F}$ and the variance matrix $K_{\vec{c},\vec{c}}$. Assuming a mean-field approximation analogous the approximations made for $\partial_t \vec{m}$ in Eq. (S47), the expression simplifies to

$$\partial_t \langle \vec{c} \rangle \approx \vec{F}(\langle \vec{c} \rangle, t) + \frac{1}{2}(\nabla_c^2)\vec{F}(\langle \vec{c} \rangle, t) \sum_i \sigma_{ii}^2, \tag{S50}$$

where $\sigma_{ii}^2$ refers to the variance of the marginalised distribution $q_i(c_i, t)$. Note, that the sum $\sum_i \sigma_{ii}^2$ refers to the variance of $f(\vec{c}, t)$, which we approximated as fixed for the localised ensemble state. For brevity of notation, we define $\gamma = \sum_i \sigma_{ii}^2$. The motion of the mean is thus driven by the interplay of the force field and its second derivative.

Assuming the existence of a localised ensemble state, we find that the multivariate generalisation of the median $\vec{m}$ is steadily attracted to the mean $\langle \vec{c} \rangle$. We hence consider $\vec{s}$ a small parameter and further approximate

$$\vec{F}(\langle \vec{c} \rangle, t) \approx \vec{F}(\vec{m}, t) + (\vec{s}\nabla_c)\vec{F}(\vec{m}, t), \quad \text{and} \tag{S51}$$

$$\frac{\gamma}{2}(\nabla_c^2)\vec{F}(\langle \vec{c} \rangle, t) \approx \frac{\gamma}{2}(\nabla_c^2)\vec{F}(\langle \vec{m} \rangle, t) + \frac{\gamma}{2}(\vec{s}\nabla_c)(\nabla_c^2)\vec{F}(\langle \vec{m} \rangle, t). \tag{S52}$$

With this, we find for the temporal evolution of the skewness of the distribution

$$\partial_t \vec{s} = -\Lambda \vec{s} + (\vec{s}\nabla_c)\vec{F}(\vec{m}, t) + \frac{\gamma}{2}(\vec{s}\nabla_c)(\nabla_c^2)\vec{F}(\langle \vec{m} \rangle, t) + \frac{\gamma}{2}(\nabla_c^2)\vec{F}(\langle \vec{m} \rangle, t). \tag{S53}$$

In the limit of fast compartment fusion and fragmentation, $\Lambda > ||\vec{F}||$, implying that macroscopic concentration changes happen on slower or similar timescales as compartment fusion and fragmentation, we derive a set of two simple coupled differential equation describing the kinetics of the quasi-particle

$$\partial_t \vec{m} = \vec{F}_{\eta(t)}(\vec{m}) + \Lambda \vec{s} \tag{S54a}$$

$$\partial_t \vec{s} = -\Lambda \vec{s} + \frac{\gamma}{2}(\nabla_c^2)\vec{F}(\vec{m}, t). \tag{S54b}$$

For one-dimensional dynamics in the concentration phase space, the equations further simplify to

$$\partial m = F_{\eta(t)} + \Lambda s \tag{S55a}$$

$$\partial s = -\Lambda s + \gamma(\partial_{cc} F_{\eta(t)}). \tag{S55b}$$

We interpret $\vec{m}$ as the position of the quasi-particle and $\vec{s}$ as the internal deformation of the localised ensemble state, which we will elaborate on in the next section. While, the motion of a point particle is dictated solely by the value of $\vec{F}_{\eta(t)}(\vec{m}, t)$, the motion of the quasi-particle is additionally affected by a force proportional to the skewness $\vec{s}$. This skewness $\vec{s}$ is driven by the second derivative of the vector field $\vec{F}_{\eta(t)}$, which can be linked to the curvature of $\vec{F}_{\eta(t)}$. As the quasi-particle degree of freedom has a finite dispersion, it not only responds to the local value of $\vec{F}_{\eta(t)}(\vec{m}, t)$, but also to the local neighbourhood of $\vec{m}$, the size of which is determined by the dispersion of the localised ensemble state. This further sets the dynamics of the quasi-particle apart from ensemble dynamics

without compartment fusion and fragmentation, where the generalised median or the mean formally depends on the full distribution $f(\vec{c}, t)$. Finally, we assume the limit of fast relaxation in $\partial_t \vec{s} = 0$, which yields:

$$\partial_t \vec{m} = \vec{F}_{\eta(t)}(\vec{m}) + \frac{\gamma}{2}(\nabla_c^2)\vec{F}(\vec{m}, t). \tag{S56}$$

This simple differential equation allows for the qualitative inference of the dynamics of the quasi-particle through graphical analysis, provided that the dynamics of the force field are qualitatively known.

### 2.7. Mechanic analogue of the quasi-particle

In order to facilitate an intuitive approach to the kinetics described by the effective equations of motion for the quasi-particle, we present a one-dimensional, mechanic analogue, which is described by equivalent equations of motion to Eq. (S55). For this, we consider three point-masses which move with an over-damped motion in a one-dimensional force field and are coupled by ideal springs. For simplicity, we take the one-dimensional force field to be given by a gradient dynamics $F(x) = -V'(x)$. The position of the three point-masses, from left to right, are labelled $x_l, x_c$, and $x_r$, as shown in Fig. S4. We set the point masses to have equal mass $m$ and the two springs to have rest length $l_0$. The over-damped motion of the three point-masses is then described by the equations of motion:

$$\nu \dot{x}_l + m \ddot{x}_l = \tilde{F}(x_l) + \tilde{k}(x_l - x_c - l_0)$$
$$\nu \dot{x}_c + m \ddot{x}_c = \tilde{F}(x_c) + \tilde{k}(x_c - x_r - l_0) - \tilde{k}(x_l - x_c - l_0)$$
$$\nu \dot{x}_r + m \ddot{x}_r = \tilde{F}(x_r) - \tilde{k}(x_c - x_r - l_0),$$

where we used the friction coefficient $\nu$, the rescaled force field $\tilde{F}$ and the spring constant $\tilde{k}$ We define $F = \tilde{F}/\nu$ and $k = 3\tilde{k}/\nu$. In the over-damped limit $m/\nu \ll 1$, the motion of the centre mass $x_c$ is

$$\dot{x}_c = F(x_c) + k(\langle x \rangle - x_c), \tag{S57}$$

where $\langle x \rangle = (x_l + x_c + x_+ x_r)$ is the centre of mass of the system. We define the deviation of the centre mass to the centre of mass $s = \langle x \rangle - x_c$, which we call the internal deformation of the system, as this deviation is associated with the asymmetric extension of the two ideal springs. In turn, $s$ changes according to

$$\dot{s} = -k(\langle x \rangle - x_c) - F(x_c) + \frac{1}{3}(F(x_l) + F(x_c) + F(x_r))$$
$$= -ks + \frac{1}{3}(F(x_r) + F(x_l) - 2F(x_c))$$
$$\approx -ks + l_0^2 \partial_{xx} F(x_c). \tag{S58}$$

We note the formal equivalence between the system of three point masses and the effective equations of motion for the quasi-particle degree of freedom in Eq. (S54), as the centre mass $x_c$ corresponds to the median and the deviation $s$ is linked to the skewness. Analogous to the quasi-particle, the system of three spring masses accounts for the local curvature of the force field, as the deviation

23

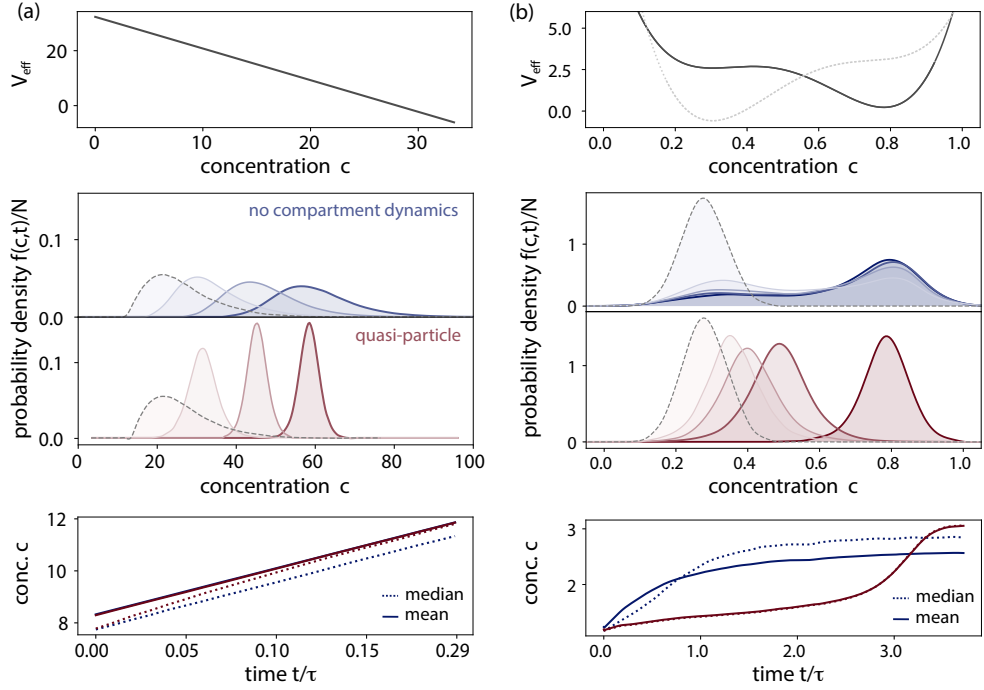

**Fig. S3. The quasi-particle moves with a fixed functional shape through the concentration phase space** Here, full stochastic simulations over a finite set of compartments are conducted. On the microscopic level, the dynamics are given by over-damped Langevin Equations with constant diffusion and a potential specified in the plot. Compartment dynamics are taken as full stochastic process, also compare with appendix 4. The simulations are evaluated in simulation units, where the time is rescaled to fit the relaxation time of the ensemble in absence of compartment fusion and fragmentation. 'Quasi-particle' refers to the ensemble with compartment fusion and fragmentation (red), while 'no compartment dynamics' refers to the ensemble with no compartment fusion and fragmentation (blue). The rate of compartment fusion is set to $\mu^{-1}/\tau = 0.05$ and the average compartment size is $\langle v \rangle = 2$ in multiples of the compartment unit size. (a) Sliding in a skewed potential. The initial configuration is given by a positively skewed distribution. In blue, a system with no compartment fusion and fragmentation shows dispersive dynamics. In contrast, the quasi-particle state emerges due to compartment fusion and fragmentation and slides down the potential in a fixed functional shape. We find that the median of the distribution is effectively attracted to the mean, as the quasi-particle state emerges. (b) Response in a double-well potential. The old potential is represented by a grey dotted line. Also in more complex potentials, the quasi-particle admits an approximately fixed shape as moves through the concentration phase space. Notably, the quasi-particle admits altered response kinetics.

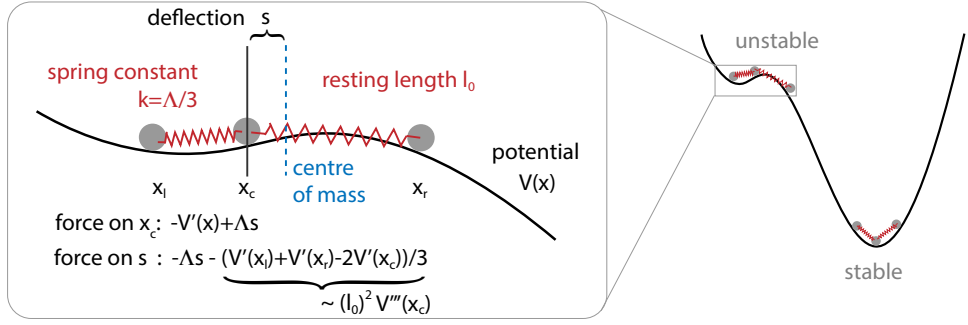

**Fig. S4. Illustration of the mechanic analogue to the quasi-particle** A system of three coupled, over-damped point-masses recreates the dynamics of the quasi-particle in 1-dimension. The springs are analogous to the steady attraction between compartments due to compartment fusion and fragmentation. We here illustrate how the second derivative on the forces arises in the simple analogue of three coupled point-masses. This in particular renders some minima of a potential unstable.

$s$ depends on the second derivative, which is weighted by $l_0^2$. It is worth noting that $\sqrt{l_0^2}$ gives an estimate of the width of the three spring masses, which is analogous to taking the root of the summed variances $\sqrt{\gamma}$ in Eq. (S54), which is an estimate for the width of the localised ensemble state. Focusing on $k$, we find a correspondence between the strength of the springs and the rate of compartment fusion, which sets the strength of attraction between the median and mean for the quasi-particle. We next study the response kinetics to external perturbations that alter the force field $F$.

### 2.8. Response kinetics of the quasi-particle

In section 2.7, the mechanic analogue was introduced in order to illustrate how the kinetics of the quasi-particle differs qualitatively from that of a single point mass. In this section, we make use of this analogue to discuss the response of the quasi-particle to external perturbations. For this, we consider a force field generated by a gradient dynamics $F(x) = -V'(x)$ and restrict our focus to one-dimensional dynamics. Nonetheless, the insights acquired in this section may be readily translated to multi-variate dynamics and general force-fields.

First, we note that the kinetic properties of the three-point mass system are altered in comparison to those of a point particle: The system has different steady states and fixed points, and different response dynamics on long and short timescales. To understand why the three-point mass system has different steady states, consider the asymmetric double-well potential illustrated in Fig. S5. For point masses, both wells of the system are fixed points. However, for the three coupled spring masses, the higher well is not a fixed point of the system. Although the centre mass is located at the bottom of the higher potential well, and experiences no force $F(x_c) = 0$, it still experiences a force due to $s \neq 0$. The left mass slides down the potential and pushes $x_c$ towards the potential barrier at $x^*$, while the right mass pulls the centre mass down into the next potential. From Eq. (S55), we find that, in steady state, $s$ typically has finite values of $|s| > 0$,

25

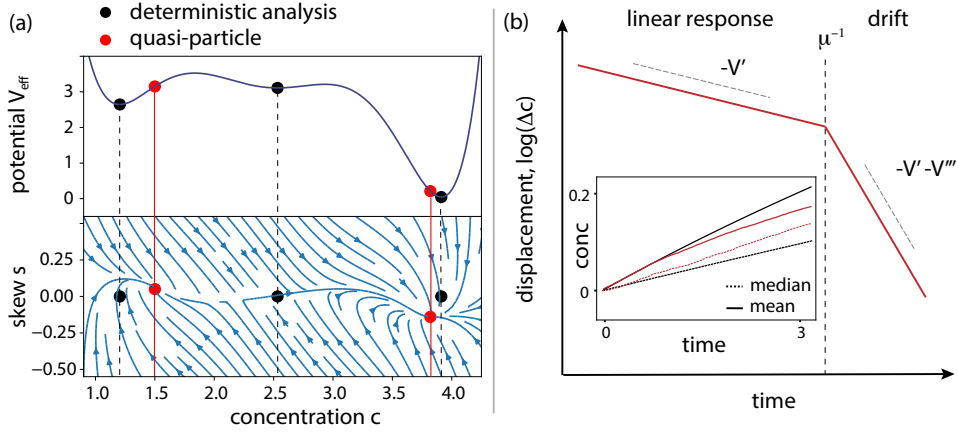

**Fig. S5. The quasi-particle exhibits altered kinetic properties** (a) The stable fixed points of the quasi-particle are altered compared to the analysis of the deterministic system $\vec{F}(\vec{c}, t)$. This is illustrated as the phase-portrait of Eq. (S55) is exemplified for a multi-stable potential. Stable fixed points of the quasi-particle are plotted as red dots, and fixed points of $\vec{F}(\vec{c}, t)$ are plotted as black dots. Note, that we formally set $s = 0$ for the fixed points of $\vec{F}(\vec{c}, t)$. (b) The quasi-particle exhibits altered response kinetics on different timescales, as represented by a schematic. The dotted line shows the transition between a linear response regime, to a regime where effective dynamics arise due to the drift in the direction of the skew. The inset shows simulations analogous to Fig. S3, which shows the response to a perturbation at $t = 0$. Red refers to the quasi-particle while black refers to a compartment ensemble with no fusion and fragmentation dynamics. Note, that both mean and median initially evolve as in the independent ensemble and we see effective dynamics on large time-scales.

thus perturbing the fixed points of the force field $F(x_{c,ss}) = 0$. This allows for the vanishing of fixed points, resulting in the three point-masses typically having a smaller number of stable fixed points than a single point-mass.

The altered fixed points imply changes in the long-term dynamics, as the three-point-mass system approaches different states compared to those of single point-masses. Furthermore, the response kinetics on short timescales are also affected. To investigate the response kinetics, we focus on the relaxation dynamics following a general perturbation from $V(x) \to \tilde{V}(x)$. For the linearised response, we approximate the new potential as follows:

$$-\tilde{F}(x) = \tilde{V}'(x) = \tilde{V}'(m) + \tilde{V}''(m)(x - m) + \mathcal{O}((x - m)^2) \approx \beta_{1,0} + \beta_{1,1}(x - m) \tag{S59}$$

$$-\tilde{F}''(x) = \tilde{V}'''(x) = \tilde{V}'''(m) + \tilde{V}''''(m)(x - m) + \mathcal{O}((x - m)^2) \approx \beta_{3,0} + \beta_{3,1}(x - m). \tag{S60}$$

26

With this, Eq. (S55) simplifies to

$$\frac{\partial}{\partial t}\begin{pmatrix} m \\ s \end{pmatrix} \approx \begin{pmatrix} -\beta_{1,1} & +k \\ -\gamma\beta_{3,1} & -k \end{pmatrix} \begin{pmatrix} m \\ s \end{pmatrix} - \begin{pmatrix} \beta_{1,0} \\ \gamma\beta_{3,0} \end{pmatrix} \quad \text{(S61)}$$

This matrix equations admits results of the general form

$$\boxed{\begin{pmatrix} m(t) \\ s(t) \end{pmatrix} = c_1 e^{\lambda_1 t}\vec{u}_1 + c_2 e^{\lambda_2 t}\vec{u}_2 - \vec{b}t,} \quad \text{(S62)}$$

where $\vec{u}_1, \vec{u}_2$ are eigenvectors of the matrix in Eq. (S61) and $\vec{b}t$ refers to the particular solution of Eq. (S62). In the limit of strong spring constants, $k \gg |\beta_{1,1}|, |\gamma\beta_{3,1}|$, we observe two distinct eigenvalues: $\lambda_1 \approx k + \gamma\beta_{3,1}$ and $\lambda_2 \approx \beta_{1,1} - \gamma\beta_{3,1}$, which indicate two different relaxation scales. On short timescales, $\tau < k^{-1}$, the point masses behave as if they are uncoupled, resulting in a new value of $s$. On longer timescales, $\tau > k^{-1}$, a finite value of $s$ gives rise to an additional drift of $m$.

By comparing to the quasi-particle in the context of compartmentalised stochastic systems, we conclude that the kinetics are altered analogously. Concentration compositions $\vec{c}^*$ that render stable steady states according to the deterministic analysis of the chemical reaction network, $\vec{F}(\vec{c}^*) = 0$, are not necessarily steady states of the quasi-particle. Furthermore, the dynamics on short timescales follow different response kinetics: Compartments respond independently to an external perturbation on timescales shorter than the timescale of compartment fusion and fragmentation. With this, we conclude on the characterisation of the emergent, quasi-particle.

## 3. THE REGULATION OF CELL DEATH IN THE FRAMEWORK OF COMPARTMENTALISED SYSTEMS

In order to apply our finding of the quasi-particle kinetic characterization to cell death decision-making, we next specify on the stochastic reaction dynamics and compartment dynamics. to the end, we discuss the Bcl-2 reaction network in the context of chemical reaction kinetics. Based on this formalism, we then specify on the microscopic model as we consider the dynamics of the intrinsic apoptotic signalling pathway by both accounting for the Bcl-2 reaction kinetics and mitochondrial dynamics at the same time.

### 3.0.1. Bcl-2 signalling pathway

Apoptosis is a form of programmed cell death, by which a cell is fragmented into small vesicles where biomolecules are degraded. It is a highly regulated process that, once initiated, cannot be stopped [19]. The intrinsic apoptotic pathway is activated in response to cell stress. Regulation of the intrinsic pathway is achieved by proteins from the Bcl-2 family, which mediate the release of cytochrome c (Cyt c) from mitochondria [19, 20].

The intrinsic apoptotic signalling pathway is initiated by a variety of stress signals, which converge in the regulation of the abundance of proteins of the

Bcl-2 family. The Bcl-2 protein family comprises a number of proteins that share Bcl-2 homology (BH) domains. Variations in the number of homology regions are observed amongst family members, while four distinct BH domains (BH1, BH2, BH3, and BH4) are distinguished. Additionally, some family members contain trans-membrane domains, which facilitate binding to the mitochondrial outer membrane. The Bcl-2 family is divided into three groups, based on primary function: anti-apoptotic proteins (e.g. Bcl-2, Bcl-XL, Bcl-W, Mcl-1), pro-apoptotic pore-formers (Bax, Bak), and pro-apoptotic BH3-only proteins (e.g. Bid, Bad, Noxa, Puma). Notably, Bax and Bak both exist in active and inactive forms; homo-dimerisation is inhibited in the inactive form. While Bak is anchored to the mitochondrial outer membrane in both active and inactive forms, Bax is located in the cytosol when inactive and binds to the mitochondrial outer membrane upon activation. Pro-apoptotic BH3-only proteins activate Bax and Bak, while anti-apoptotic proteins deactivate them, form hetero-dimers, and/or bind pro-apoptotic BH3-only proteins. The various anti-apoptotic and pro-apoptotic BH3-only proteins differ in terms of binding affinities, preferred localisation (cytosol or mitochondrial outer membrane), and abundance. Abundance is subject to a variety of cellular signals, such as metabolic conditions, nutrient deprivation, DNA damage and radiation, as well as several toxins. Mitochondrial outer membrane permeabilisation (MOMP) is the process of forming pores in the mitochondrial outer membrane, which leads to the leakage of mitochondrial content into the cytosol; this is linked to mitochondrial lysis and degradation.

We consider the accumulation of Bax or Bak in their active form on the outer mitochondrial membrane to be a hallmark of apoptosis induction, leading to the release of Cyt c into the cytosol [19–21].

### 3.1. Modelling the Bcl2 reaction network

Here, we propose a simple model of how Bcl-2 protein interactions mediate the release of Cyt c and set this in relation to the current literature. We then place this model into the context of the population balance equation and specify the qualitative form of deterministic drift vector $F_{\eta(t)}(\vec{c})$ and reaction binding noise diffusion matrix $\mathbf{D}_{\eta(t)}(\vec{c})$ in Eq. (S15).

The temporal progression of mitochondrial outer membrane permeabilisation (MOMP), the dynamics of Bax pore formation mechanism, and the details of the Bcl-2 interaction network are today still subjects of active research [20]. Several models have been proposed to capture the qualitative dynamics of the apoptotic signalling pathway, which mainly focus on the interaction between proteins in the Bcl-2 family [20, 22, 23]. Central to these models is the link between increased concentrations of Bax-Bax and Bak-Bak homo-dimers and increased pore-formation [24, 25]. Here, we briefly study the simplest of such models, which accounts for the auto-catalytic self-activation of Bax by Bax-Bax homo-dimers, and the role of anti-apoptotic proteins and pro-apoptotic BH3-only proteins. We do not distinguish between various pro-apoptotic BH3-only proteins and anti-apoptotic proteins, and collectively refer to the concentration of pro-apoptotic BH3-only proteins as [Act], and to the concentration of anti-apoptotic proteins as [Inhib]. For the sake of simplification, we refer in the following explicitly only to Bax but implicitly always to both Bax and Bak, as indeed the combined concentration of inactive Bax and Bak as well as active Bax

28

and Bak, and Bax-Bax and Bak-Bak homodimers should be considered.

Inactive (cytosolic) $Bax_c$ and Bak is activated, and is transformed to active, membrane-bound $Bax_m$, which then forms membrane-bound homo-dimers $Bax_D$:

$$Bax_c [k_3[\text{Inhib}]] k_1[\text{Act}] + k_2[Bax_D] Bax_m \quad \text{and} \quad 2Bax_m [k_5] k_4 Bax_D, \quad \text{(S63)}$$

where $[X]$ refers to concentrations and we assume well-mixed conditions. Accounting for the conservation of mass for Bax with $[Bax_{\text{tot}}] = [Bax_c] + [Bax_m] + 2[Bax_D]$, the steady state solution in the thermodynamic limit yields

$$\begin{aligned} 0 = \quad & k_1[\text{Act}] \left( [Bax_{\text{tot}}] - [Bax_m] - \frac{2k_4}{k_5}[Bax_m]^2 \right) \\ & + \frac{k_2 k_5}{k_4}[Bax_m]^2 \left( [Bax_{\text{tot}}] - [Bax_m] - \frac{2k_4}{k_5}[Bax_m]^2 \right) - k_3[\text{Inhib}][Bax_m], \end{aligned}$$
(S64)

which is for given $k_1, k_2, k_3, k_4, k_5$, and $[Bax_{\text{tot}}]$ a quartic equation in $[Bax_m]$. Notably, this quartic equation admits analytic solutions. Yet, as the analytic solution is convoluted, we here restrict to investiagting a graphical analysis to emphasize the general features of the solution.

We normalise the concentrations to $[Bax_{\text{tot}}] \equiv 1$ [conc.], where [conc.] refers to a rescaled concentration unit. $\tau$ refers to a time unit. Fixing the reaction rates $k_4/k_5 = 0.1$ [conc.]$^{-1}\tau^{-1}$, $k_2 = 80\,\tau^{-1}$ and $k_1 = k_3 = 1\,\tau^{-1}$, we find regions of bistability, as we vary the concentration of pro-apoptotic and anti-apoptotic proteins, see Fig. S6. For this effective one-dimensional dynamics, we interpret the temporal evolutions of $Bax_m$ as an effective gradient dynamics $\partial_t[Bax_m] = -\nabla_{Bax_m} V([Bax_m])$. This interpretation allows for an intuitive approach to the kinetics of the membrane-bound Bax concentration $[Bax_m]$. We find that $[Bax_m]$ effectively evolves in a bistable potential, in which the concentration of pro-apoptotic and anti-apoptotic proteins sets the skew of the potential. We link the two wells of the bistable potential with two different steady-states of the kinetics, which describe two different functional operating states of the signalling pathway. We refer to the state with low $[Bax_m]$ concentration as *physiological* Bax concentration state, while we term the high concentration state as apoptotic state. We associate the apoptotic state with the increased formation of pores in the mitochondrial membrane and the release of Cyt c. In the context of apoptotic decision-making, we are interested in the transition from the physiological to the apoptotic state. This translates in the picture of the effective bistable potential to a transition from the low-concentration well to the high concentrations state.

Extensive apoptotic models that take into account varying affinities of different pro-apoptotic and anti-apoptotic proteins have qualitatively agreed with the finding of a bistability for membrane-bound Bax [22, 23, 26]. This qualitative result is also robust when subjected to a refined analysis considering the preferential spatial location of proteins of the Bcl-2 family in either the cytosol or the mitochondrial outer membrane. A precise measurement of reaction rates and total concentrations is experimentally challenging and prone to error. Order of magnitude estimates demonstrated that bistability can occur for a

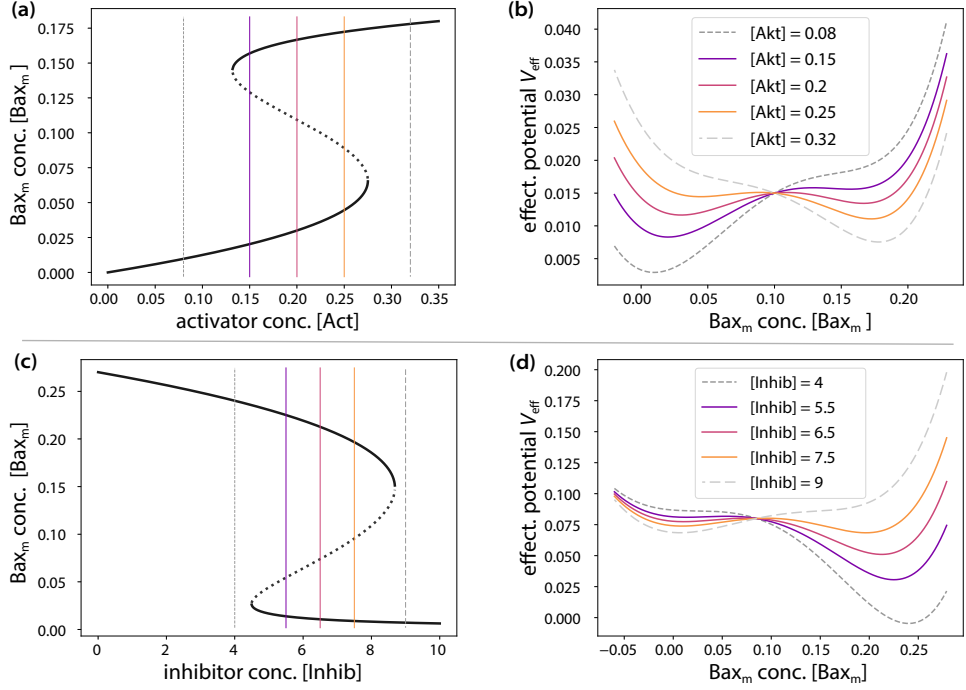

**Fig. S6. The Bax accumulation dynamics are effectively captured by a bistable potential**. For the parameters specified in section 3.1, we systematically vary the concentration of activator (a,b) and inhibitor concentration (c,d). In (a) and (c), we find that the bifurcation diagram shows a cusp-bifurcation. Stable fixed points are indicated by a bold line, while the unstable fixed points are represented by a dashed line. The one-dimensional effective dynamics in $[Bax_m]$ allows for tracking the dynamics as gradient dynamics. With $d_t[Bax_m] = -d_c V_{\text{eff}}([Bax_m])$, the effective potential is plotted in (b) and (d). We find that the effective potential is a bistable potential. Changing the activator and the inhibitor concentration changes the skew of the potential. Different colours correspond to different activator and inhibitor concentrations, accordingly indicated also in (a) and (c). All concentrations are in the effective units [conc.], and $V_{\text{eff}}$ in $[\text{conc.}]^2/\tau$, as introduced in section 3.1.

physiologically plausible parameter choice [22, 26], and footage of bistability for membrane-bound Bax was observed in experiments [23, 26]. Furthermore, the concentrations of Bcl-2 proteins estimated on individual mitochondria is approximately $\sim 20\,\text{nM}$ [27] for the physiological state, which is in agreement with measured estimates of a few hundred Bax proteins per mitochondrion [28]. These estimates suggest that the apoptosis reaction pathway should be treated in a stochastic framework, compare also to [4]. This formally sets the basis for the estimation of the drift term $F_{\eta(t)}(\vec{c})$ and the diffusion term $\mathbf{D}_{\eta(t)}(\vec{c})$ within the context of the population balance equation in Eq. (S15).

Estimating the membrane diffusion of proteins of size $\sim 25\,\text{kDa}$ with $\sim 10\,\mu\text{m}^2\,\text{s}^{-1}$ [29], results in a membrane mixing on the order of seconds. This leads us to consider mitochondria as a well-mixed reaction compartment. We assume fast diffusion of the cytosolic Bcl-2 family proteins with higher molecular

copy numbers than on individual mitochondria and model them as a fluctuating signal $\eta(t)$. We neglect spatial correlation between organelles through diffusion of the cytosolic Bcl-2 proteins and consider independent realisations of stochastic apoptotic dynamics on separated, not-fused mitochondria. Molecular binding noise induces stochastic transitions of individual mitochondria from the low to the high membrane-bound Bax concentration state. Conversely, the transition from the high to the low concentration state is inhibited by the occurrence of MOMP and the subsequent lysis and degradation of mitochondria. Although we can formally extract estimates for the drift term $F_{\eta(t)}(\vec{c})$ and the diffusion term $\mathbf{D}_{\eta(t)}(\vec{c})$ from reaction network models, this requires the precise estimation of reaction rates and physiological concentrations of Bcl-2 family proteins by experimental measurements. Alternatively, a qualitative discussion of the signalling pathway kinetics can be used to estimate how different model refinements qualitatively affect the kinetics. We will investigate this in the next section.

### 3.2. Effective model of the apoptotic signalling pathway

We are interested in investigating how mitochondrial dynamics qualitatively affect the kinetics of the Bcl-2 signalling pathway. Quantitative analysis of apoptosis quickly becomes analytically intractable, is highly specific for given cell types and can be prone to errors due to the estimation of a large number of reaction rates and physiological concentrations, compare for example with [28]. In contrast, a qualitative analysis allows us to evaluate the influence of various parameters and the general effects of mitochondrial dynamics on apoptotic signalling. Therefore, we introduce an effective model for the temporal evolution of the concentration of membrane-bound Bax on each mitochondrion. Our motivation for using an effective model is to enable analytic tractability, which allows us to estimate how model refinements qualitatively affect our findings, as we will show in section 3.3. To ensure plausible results, we estimate the parameters of the effective model by comparison with experiments

In the effective model, each mitochondrion is characterised by the membrane-bound Bax concentration $[\text{Bax}_\text{m}] \equiv \tilde{c}$ and its size $s$, where we introduced $\tilde{c}$ for notational simplicity. Recall, that the membrane-bound Bax concentration $\tilde{c}$ evolves stochastically in a bistable potential, where the skew of the potential is set by the level of stress metabolites in the cell cytosol, as discussed in section 3.1. Consistent with the picture of a bistable potential, two attractive fixed points of the kinetics for $\tilde{c}$ can be distinguished: a low and a high potential well. These concentrations fixed points represent two different equilibrium states of the signalling pathway and are denoted by $\tilde{c}_\text{l}^*$ and $\tilde{c}_\text{h}^*$, respectively. The potential well with the higher Bax concentration, $\tilde{c}_\text{h}^* > \tilde{c}_\text{l}^*$, corresponds to the release of Cyt c from the mitochondrion. We define the effective potential

$$V_\text{eff}(\tilde{c}) = a(\tilde{c} - \tilde{c}_0)^4 - b(\tilde{c} - \tilde{c})^2 + \alpha(\tilde{c} - \tilde{c}_0), \tag{S65}$$

where we used a constant $\tilde{c}_0$ to centre the potential around the origin for notational simplicity. $a, b \in \mathbb{R}^+$ are positive real-valued parameters of the bistable potential. $\alpha \in \mathbb{R}$ refers to the initial skew of the potential, which is the initial apoptotic priming.

The exact functional form of $D(\tilde{c})$ requires precise estimation of molecular concentration and chemical reaction rates in the Bcl-2 signalling pathway, making estimates of $D(\tilde{c})$ highly prone to errors and sensitive to model details. Here, we make use of a strong approximation of $D(\tilde{c})$, which grants analytical tractability and allows us to study how variations in $D(\tilde{c})$ can affect the findings qualitatively. As $\tilde{c}_h^*$ is associated with the release of Cyt c, we are mainly interested in fluctuations around $\tilde{c}_l^*$ and the *escape* into the high concentration state. Consequently, we approximate $D(\tilde{c})$ around $\tilde{c}_l^*$. As a rule of thumb, the diffusion coefficient is proportional to the total amount of proteins, such that $D(\tilde{c}) \sim \tilde{c}$. Assuming that the low concentration well and the barrier of the potential, $\tilde{c}_l^* \sim \tilde{c}_b^*$, are on the same order of magnitude, we approximate to zeroth order $D(\tilde{c}) \approx D(\tilde{c}_l^*) \equiv D$. Additionally, we account for external signals $\eta(t)$, in which we incorporate changes in pro-apoptotic and anti-apoptotic proteins with. Together, the effective model shows as

$$\partial_t \tilde{c} = -\partial_{\tilde{c}} V_{\text{eff}}(\tilde{c}) + \frac{D}{\sqrt{v}} \xi(t) + \eta(t), \tag{S66}$$

where $\xi(t)$ refers to white Gaussian noise and we correct for the size of mitochondria by $\sqrt{v}$. We discuss how modifications of the potential or the noise affect our qualitative findings in section 3.3. For fixed $\eta(t) = \eta$, we rescale the potential, such that the low concentration state is $c_l^* \equiv 0$ and $c_h^* \equiv 1$. In the notation of Eq. (S66), we define weak as stimuli $\eta(t)$ which let the system predominantly remain in a bistable configuration. Conversely, we define strong apoptotic stimuli as stimuli in which the high concentration state $c_h^*$ is the only stable fixed point of the system. In particular, we only speak of apoptotic decision-making in the context of weak apoptotic stimuli, as the system necessarily evolves quickly into the high Bax-concentration state for strong apoptotic stimuli.

To include also on the mitochondrial dynamics, we adapt the notion of hybrid systems as we numerically implement a trajectory version of the multi-scale dynamics. To this end, we define the state of the system analogous to Eq. (S7), as

$$\mathbb{S} = \begin{bmatrix} \vdots \\ [c_i, v_i] \\ \vdots \end{bmatrix}. \tag{S67}$$

We define that mitochondria are fully described by their size $v_i$ and the Bax concentration $c_i$ on their membranes. Symbolically, we define stochastic trajectories of the system as

$$\frac{d}{dt}\mathbb{S}(t) = \mathcal{L}_{\text{Bax}}[\mathbb{S}(t)] + \mathcal{F}_{\text{fus+frag}}[\mathbb{S}(t)]. \tag{S68}$$

Here, $\mathcal{L}_{\text{Bax}}[\mathbb{S}(t)]$ refers to the stochastic dynamics in Eq. (S66) that happen in parallel and independently on the different mitochondria. $\mathcal{F}_{\text{fus+frag}}[\mathbb{S}(t)]$ refers to stochastic, time-discrete events of mitochondrial fusion and fragmentation that alter the system. For the stochastic compartment fusion and fragmentation dynamics, we consider the stochastic rates defined in the context of Smoluchowski

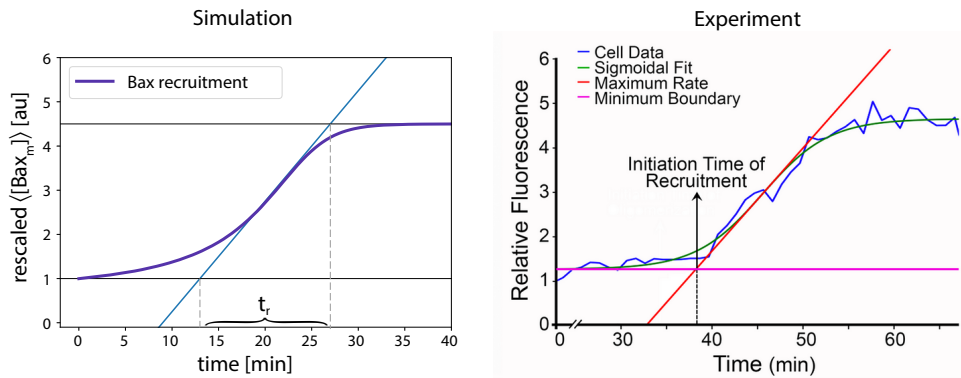

**Fig. S7. Comparison of Blc-2 reaction kinetics between simulated and experimental timescales**. By comparing with experimental data, we adjust the timescale for the translocation from the low to high Bax concentration state on the mitochondrial membrane. (a) Shows the collective translocation from the low to the high Bax concentration state in simulations of the effective model Eq. (S66) in response to a strong apoptotic stimulus. Plotted is the mean Bax concentration on the mitochondrial membrane $\langle [\text{Bax}_m] \rangle$, where the average is drawn over the ensemble of mitochondria. (b) is a Figure presented in [36] in Fig. 2E. Here, they track the Bax accumulation dynamics with fluorescent microscopy of individual mitochondria. The presented line represents an ensemble average. We rescaled (a) to visually fit to (b). By comparison, we estimate the characteristic timescale $t_r = 15$ min as the translocation timescale. We fix all other physiological timescales of the effective model accordingly.

aggregation-fragmentation. We consider this model for the numerical simulation of the system, in line with the numerical routine described in section 4. [30, 31] have demonstrated that the Smoluchowski aggregation-fragmentation dynamics adequately capture the mitochondrial size distribution, as size-independent fusion kernels and fission kernels proportional to the mitochondrial size are considered. We assume that mitochondrial dynamics happen on the timescales of minutes, as measured in [32–35]. With this model, we want to investigate if we can also identify in the context of apoptosis the predicted quasi-particle kinetics. Moreover, we want to assess how mitochondrial dynamics qualitatively affect the response kinetics of the apoptotic signalling pathway and possibly assess the biological functions of mitochondrial dynamics.

For a qualitative analysis, we are interested in order-of-magnitude estimates of the parameters in Eq. (S66), which we find by comparison with experiments. We adjust $a$ and $b$ according to the time-scale of Bax accumulation on the mitochondrial membrane upon a strong apoptotic stimulus. [36] measured the duration of the transition from the low to the high Bax-membrane concentration state in response to the poisoning of the cell with apoptosis-inducing drug STS and found an estimate of $\delta t \sim 15$ min, as illustrated in Fig. S7. By this, we also define the characteristic timescale of our system as $t_r = 15$ min. This timescale gives an order of magnitude estimate for the movement $c_l^*$ to $c_h^*$ if the bistability is demolished by a negative skew $\alpha \ll 0$. This timescale coincides

33

roughly with the measurement that Bax translocates to mitochondria at a rate of $4.7 \pm 0.2 \times 10^{-3}\,\text{s}^{-1}$, as obtained by fluorescent recovery after photo-bleaching (FRAP) experiments [37]. This is consistent with an equilibrium between the on and off rates of Bax.

While an estimate of $D$ based on molecular concentrations and reaction rates is error-prone, we estimate the order of magnitude of this parameter by the escape rate of individual mitochondria from $c_l^*$ to $c_h^*$ in the presence of weak apoptotic stimuli, as done in experiments by [38, 39]. In particular, [38] showed that over a time span of $4\,h$ only a subset of $< 10\%$ mitochondrial mass underwent mitochondrial outer membrane permeabilisation in response to weak apoptotic stimuli, which gives an order of magnitude for $D$ for a given potential $V_{\text{eff}}(c)$.

### 3.3. Quasi-particle kinetics in the regulation of cell death

In equation Eq. (S55) we elucidated on how the dynamics of compartmentalised stochastic reaction systems can be captured by simple equations of motion for a quasi-particle. Here, we display the benefit of the effective equation of motion Eq. (S55) for an intuitive approach to the kinetics of a compartmentalised stochastic system, as we perform a graphical analysis of Eq. (S55).

We discussed above how the membrane-bound Bax concentration $c$ is qualitatively described by stochastic movement in a bistable potential. We use $c$ following a gradient dynamics to increase our understanding of the effective kinetics of $c$. We consider fixed apoptotic stimuli $\eta(t) = \eta$ and a time-invariant effective potential $V_{\text{eff}}(t) = V_{\text{eff}}$. In the absence of apoptotic signals, the potential is skewed to a stable fixed point at low Bax membrane concentrations, $c_l^*$. An apoptotic stimulus will drive it into the bistable region, where both the high and low concentration states, $c_h^*$ and $c_l^*$, are stable fixed points. Recall, that we refer to weak and strong apoptotic stimuli depending on whether the potential is skewed towards the bistable or monostable region, respectively. Of particular interest in the context of apoptotic decision-making are weak apoptotic stimuli, where deregulation of the apoptotic signalling pathway can lead to abnormal cell death or carcinogenic effects.

Next, we perform a graphical analysis of the effective equations of motion, Eq. (S55). For weak apoptotic stimuli, we are interested in the transition from the low Bax concentration state $c_l^*$ to the high Bax concentration state $c_h^*$. In order to qualitatively examine this, we illustrate the kinetics of the quasi-particle in a bistable potential. We do not specify the details of the potential, as this analysis is purely qualitative. Our findings will be refined by considering physiologically plausible parameter estimates after studying the dynamics qualitatively. In Fig. S9 (a), the bistable potential subjected to a weak apoptotic stimulus is shown, along with its gradient and third derivative. The grey line indicates zero-crossings, which correspond to the fixed points of the kinetics for point particles in the case where $\gamma = 0$. The third derivative of the potential in the example shown is a linear function. The third derivative $\gamma/2 \cdot V'''(c)$ reduces the barrier between the two stable fixed points for the quasi-particle kinetics, and thus acts like a perturbation. Note, that the strength of the perturbation is proportional to the variance over all mitochondrial compartments. This effectively reduces the region of attraction of the fixed point at low Bax concentrations for the quasi-particle.

34

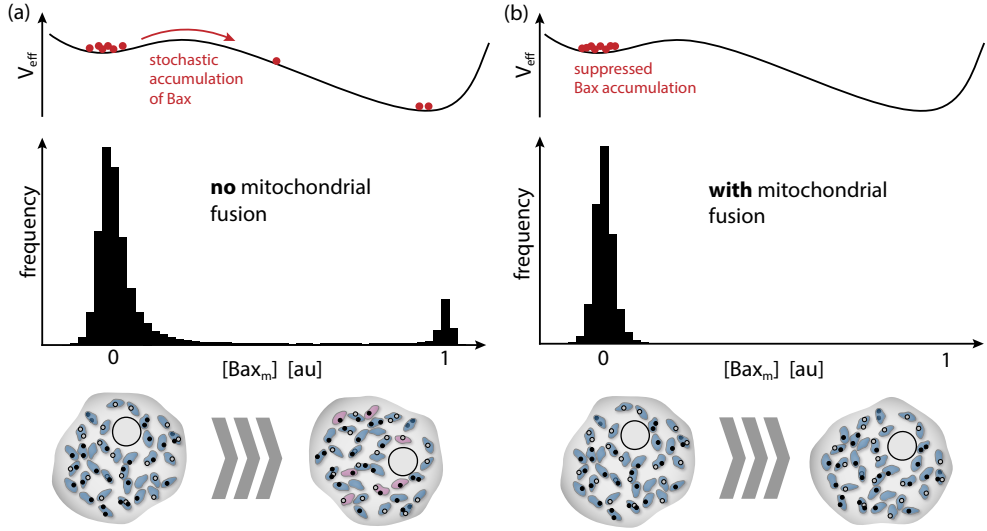

**Fig. S8. Mitochondrial dynamics give rise to a localisation of the apoptotic signalling pathway in the concentration phase space**. We conduct numerical simulations to assess qualitatively the effect of mitochondrial fusion and fragmentation of the simulation. For this we choose physiological parameter conditions, as we match the translocation time-scale to $t_r = 15$ min and set the rate of mitochondrial fusion and fragment to $\mu t_r = 3.3$. We adjusted the effective (molecular) diffusion coefficient to $Dt_r = 6 \times 10^3$ in rescaled concentration units, which results in the absence of mitochondrial fusion and fragmentation to the stochastic switching of a subset $< 10\%$ from the low to the high concentration state, in line with the observation in [38]. In the top plot, we have an illustration of the effective potential, where individual mitochondria are represented as red dots. In the middle, a histogram illustrates the frequency statistics of the mitochondrial ensemble in one-dimensional $[Bax_m]$ concentration space. The bottom illustrates that ensemble statistics in the concentration phase space and the real space are distinctly different. The identical mitochondrial size distributions are considered in (a) and (b). While we see the random accumulation of Bax on mitochondria in ensembles with no mitochondrial fusion and fragmentation dynamics in (a), we find in (b) that mitochondrial fusion and fragmentation dynamics give rise to a localisation of the mitochondrial ensemble in the concentration phase space. The stochastic accumulation of Bax and the subsequent release of the toxin Cyt c is suppressed.

Fig. S9 (b) displays the phase portrait of the equations in Eq. (S55). The red lines represent the nullclines, where both differential equations independently vanish. The crossings of the nullclines indicate fixed points. For finite dispersion above a critical value ($\gamma > \gamma^*$), only the high concentration state ($c_h^*$) remains as a stable fixed point, even though the potential $V_{\text{eff}}(c)$ is still in the bistable region. The stable fixed point is indicated by a black star, and the position of the mitochondrial ensemble in the low-concentration state is indicated by a grey star. Blue arrows show the flow in the concentration phase space. We observe the emergence of a slow centre manifold, which the flows rapidly decay

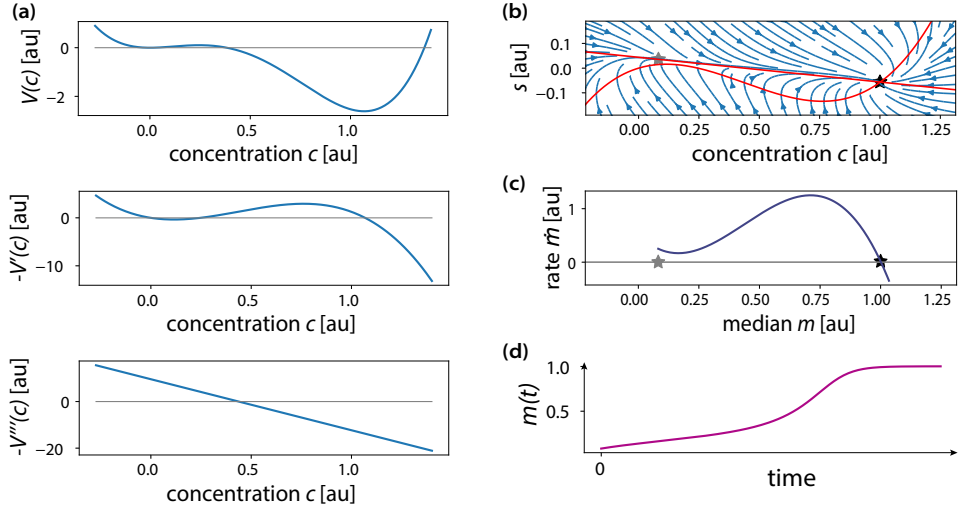

**Fig. S9. Graphical analysis of the equations of motion for the quasi-particle in the context of apoptotic decision-making.** The simple equations of motion in Eq. (S55) allow for graphical analysis, which we applied to the simple qualitative model in Eq. (S66). From this, we can directly read off the qualitative response kinetics of the apoptotic signalling pathway to fixed, weak apoptotic stimuli. Here, we present a qualitative analysis with parameter choices in arbitrary units. (a) Plotted are the effective potential, its gradient and its third derivative. (b) phase-portrait of n Eq. (S55) in a bistable potential. Red lines are the nullclines of the dynamics. The stable fixed point is indicated by a black star. the grey start refers to the ensemble mean at time $t = 0$. (c) Eq. (S47) gives an estimate for the dynamics along the central manifold. (d) Graphical integration of the rates in (c) predicts sigmoidal response kinetics of the localised mitochondrial ensemble in the concentration phase space. There here presented analysis only has a qualitative character and emphasises that our analysis allows us to qualitatively assess the kinetics of the multi-scale system by a simple graphical analysis.

towards. The centre manifold is well approximated by the nullcline where $\partial_t s = 0$. Consequently, the dynamics of $m$ along the centre manifold are well approximated by Eq. (S56). In Fig. S9 (c), we sketch Eq. (S56) from the grey star to the black star. We observe that the rate increases as the position of the mitochondrial ensemble $m$ approaches the stable steady state. By performing a graphical integration, we find that the quasi-particle relaxes to the stable steady state following a sigmoidal relaxation, as shown in Fig. S9 (d). The transition into the high Bax concentration state is initially suppressed on short timescales and facilitated on long timescales after the ensemble has crossed the potential barrier between the two states. Note, however, that this qualitative analysis does not enable us to further assess the timescales which would correspond to *short* and *long* timescales. We elaborate on these timescales by a semi-analytic approach after we study the qualitative differences we expect in the response kinetics due to mitochondrial fusion and fragmentation.

We expect altered fixed points for the quasi-particle in contrast to a hyperfused mitochondrial network or mitochondria with no fusion and fragmentation. To this end, we compare four different hypothetical mitochondrial ensembles, which differ in their dynamics. First, we consider an ensemble with mitochondrial fusion and fragmentation, which exhibits the emergent collective, quasi-particle degree of freedom. Second, we consider an ensemble with no mitochondrial fusion and fragmentation, which has the same mitochondrial size distribution as the ensemble with fusion and fragmentation. We consider as a third ensemble, an organelle ensemble with no fusion and fragmentation, again the same size distribution, but with a rescaled diffusivity, $D_{\text{rs}} < D$. Here, we rescale the diffusivity such that both the organelle ensemble with fusion and fragmentation and this organelle ensemble show the same variance in dispersion in the absence of apoptotic stimuli. As a fourth model, we consider an ensemble where all mitochondria are hyper-fused to build a big, percolated mitochondrial network. We formally set $s \to \infty$, and hence $D \to 0$. We refer to this ensemble as a *point particle* in analogy to the *quasi-particle*. For the case of vanishing diffusivity ($D \to 0$), both states $c_h^*$ and $c_l^*$ are stable fixed points, and transition between the two states is only achieved when the bistable region is left in response to strong apoptotic stimuli. In contrast, when diffusivity is finite, $D > 0$, an ensemble of mitochondria with absent fusion and fragmentation, mitochondria are expected to randomly switch between the high and low concentration states.

We next compare our qualitative predictions of the response kinetics with numerical simulations, as shown in Fig. S10. For this, we again consider the four hypothetical mitochondrial ensembles. Here, we compare the changes in the mean amount of membrane-bound Bax over the mitochondrial ensemble ($\langle c \rangle$) in response to apoptotic stimuli and turn to physiologically plausible parameter choices. Note, that we obtained the physiologically plausible parameter choices by comparing with experimentally measured timescales of the systems, as outlined above. Specifically, we find that the response kinetics for the four different ensembles are qualitatively similar for strong apoptotic stimuli, Fig. S10 (a), but differ for weak apoptotic stimuli, Fig. S10 (b). While the ensemble with no mitochondrial fusion and fragmentation and vanishing diffusivity is stable in low-concentration states, the two ensembles with finite diffusivity show an approximately exponential relaxation. Notably, even on short timescales, a small proportion of mitochondria switch into the high-concentration state in response to weak apoptotic stimuli. This is in contrast to the response of the quasi-particle, where the response to the apoptotic stimulus is suppressed on short timescales and facilitated on long timescales. This is in qualitative agreement with our graphical analysis presented in Fig. S9. In particular, the response is initially inhibited compared to the ensemble with matched ensemble variance and facilitated on long timescales compared to the ensemble in the absence of mitochondrial fusion and fragmentation. On short timescales, the steady fusion and fragmentation dynamics impede the stochastic escape of individual mitochondria to the high-concentration state. On long timescales, after the mean and the median of the ensemble have crossed the potential barrier, mitochondrial fusion and fragmentation facilitate the crossing of the barrier for individual mitochondria.

Further corroborating our qualitative findings, we next examine how well we

37

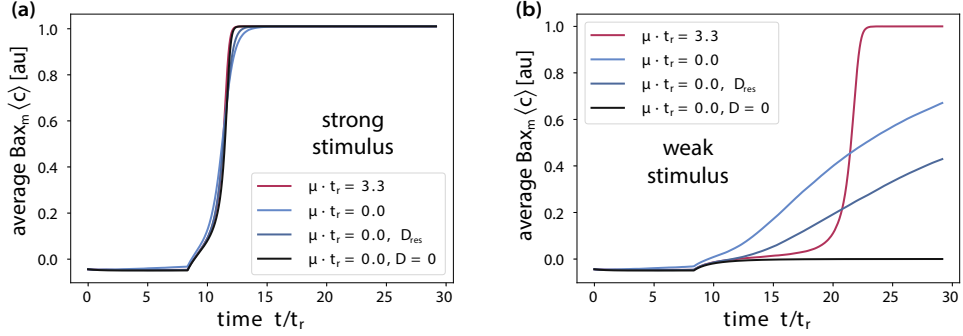

**Fig. S10. Accounting for mitochondrial dynamics results in qualitative different response kinetics in full stochastic simulations**. In full stochastic simulations following the model in Eq. S68 are evaluated to show the response kinetics to a strong (a) and weak (b) apoptotic stimulus. The physiological plausible simulation parameters for (b) are identical to Fig. S8. For (a) the skew of the potential is chosen, such that the bistable region is left. the stimulus is applied at time $t = 9/t_r$. The mean Bax concentration $\langle c \rangle$ is tracked, where the mean is computed over the finite-sized mitochondrial ensemble. Four different ensembles are distinguished, where all ensembles show identical mitochondrial size distribution. Besides an ensemble with and an ensemble without fusion and fragmentation dynamics, we also consider mitochondrial ensemble without fusion and fragmentation dynamics and rescaled (molecular) diffusion constant. See section 3.3 for details. While we find that all ensembles show the same response kinetics for the strong apoptotic stimulus. For weak apoptotic stimuli, we find that the response kinetics are distinctly different. Moreover, we find the full stochastic simulations of the system and the sigmoidal response kinetics, which we predicted by the graphical analysis in Fig. S9.

can capture the qualitative dynamics by solving the effective equations of motion in Eq. (S47) semi-analytically by numerical integration. For this, we compare the result of Eq. (S47) with the full stochastic simulations following Eq. (S68). We test our results with different parametrisations of the potential and different estimates of system parameters. Additionally, we consider a finite ensemble of mitochondria with different sizes, which undergo stochastic, time-discrete events of fusion and fragmentation.

The effective equation Eq. (S56) has one free parameter $\gamma$, which is linked to the variance over the mitochondrial compartment ensemble. We fix this parameter by a single, independent simulation, from which we assess the ensemble variance in the absence of an apoptotic stimulus. Our semi-analytic prediction is thus parameter-free. As Eq. (S56) is a non-linear differential equation, we solve it by direct numerical integration using `odeint` in `python`. The theoretical prediction for the ensemble dynamics, in the absence of mitochondrial fusion and fragmentation, follows Kramer's escape problem for mitochondria of different sizes, see appendix 3.5. Assuming that mean and median are close in the ensemble of interacting mitochondria, we map out the average Bax concentration $\langle c \rangle$ and consider Eq. (S56) as an approximation of the mean Bax concentration.

Figures S11 (a) and (b) compare the ensemble response for two different potential parametrisations with and without mitochondrial fusion and fragmentation. Physiologically plausible parameter choices were made, with the timescale of macroscopic concentration changes linked to experimentally observed timescale, Fig. S7, and the time-scale of mitochondrial fusion and fragmentation set on a timescale of minutes. Time was rescaled consistently with the timescale of macroscopic concentration changes $t_r = 15$ min. A qualitative agreement is observed between the theoretical prediction and the numerical simulations, with the sigmoidal response kinetics correctly predicted by the full stochastic simulations and the numerical integration of Equation (S56).

By systematically altering the simulation parameters, we investigate how the results are qualitatively affected. In Fig. S11 (c), we systematically increase the rate of mitochondrial fusion and fragmentation, while by keeping the size distribution fixed. We predict that the variance over the mitochondrial compartment ensemble will be inversely proportional to the rate of mitochondrial fusion, $\gamma \propto \mu^{-1}$. Our simulation and the semi-analytic solution confirm that a decrease in the fusion rate leads to a decrease in the sigmoidal shape. However, we find that the simulation and the semi-analytic solution begin to deviate at small fusion rates, thus necessitating refinement of Eq. (S56). We improve the quality of the semi-analytic solution again by decreasing the diffusivity $D$, which strengthens the sigmoidal shape and delays the shift from the low concentration state to the high concentration state, see Fig. S11 (d). Lastly, we consider the cases of changing the skew of the potential in Fig. S11 (e) and reducing the strength of the potential globally in Fig. S11 (f). Our findings indicate that lower potential barriers correspond to weaker sigmoidal shapes with a faster transition between the low and the high concentration states, whilst increased potential barriers have the opposite effect. Note, that for all different setups, Eq. (S56) yields reasonable good estimates. Further recall, that the semi-analytic prediction admits no free fit-parameters!

This systematic scan of parameters enables us to generalise our qualitative insight into the sigmoidal translocation dynamics in steady state to general bistable potentials. We find that the qualitative insight for suppressing a response to weak apoptotic stimuli on short timescales and facilitating the response on long timescales is not a result of specifics of the $V_{\text{eff}}(c)$, but is a general feature of the switching between two stable steady states. Additionally, it should be noted that the sigmoidal response is only invoked above a critical value of stimulus strength, as the sigmoidal response requires the vanishing of the fixed point at low Bax membrane concentrations, see Fig. S9 (b). Furthermore, we conclude that the exact time-point of the transition between low and high-concentration states depends on the specifics of the potential, and a quantitative prediction of the transition time requires a precise estimate of the physiological reaction rates and concentrations, which cannot be obtained from the effective model. Nonetheless, we can make a rough estimate of the order of magnitude of the timescales on which the response is suppressed. Linking the timescale of suppression with the timescale of escaping from the lower potential well, we expect this timescale to be of the same order of magnitude, yet several times larger than the translocation time $t_r$, see Fig. S11.

For the analysis in Fig. S11, we neglected the lysis of mitochondria as they

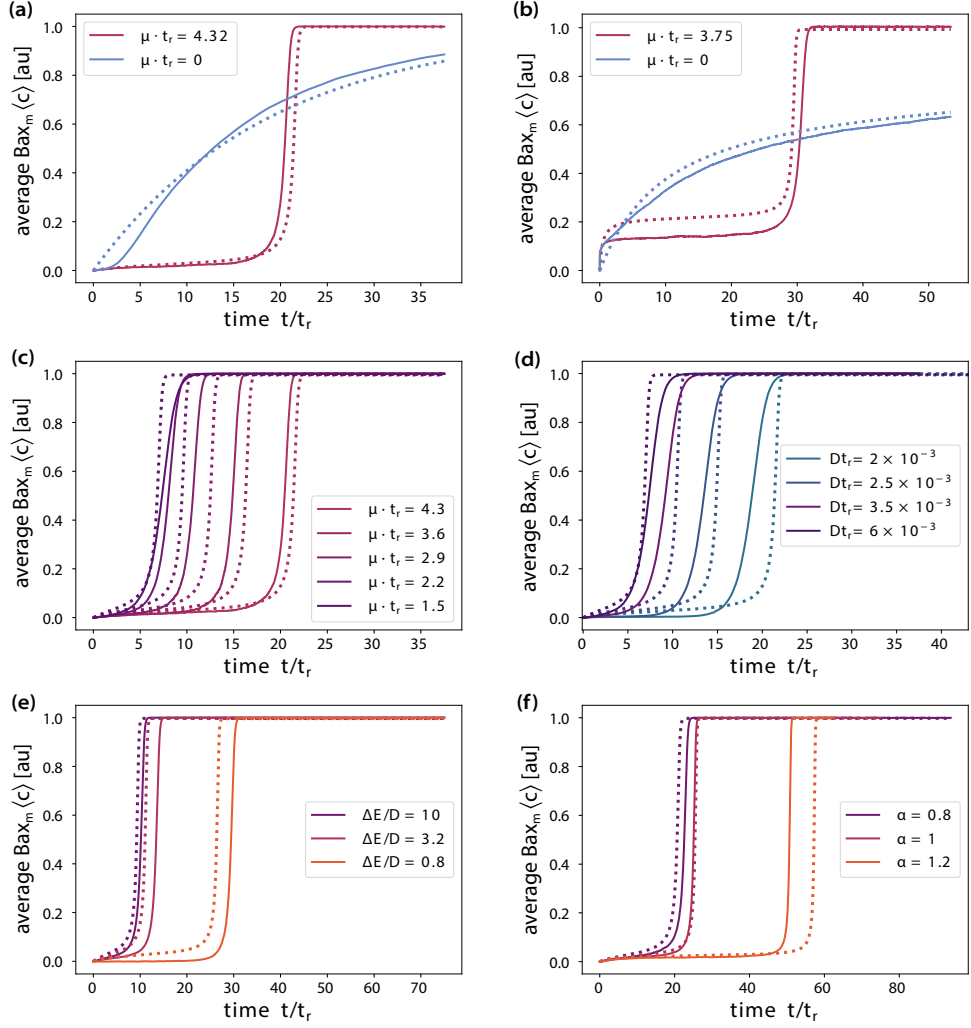

**Fig. S11. The equations of motion for the quasi-particle agrees with full stochastic simulations**. We compare the equations of motion in Eq. (S47) with full stochastic simulations (solid lines) and find quantitative good agreement. We obtain the solution of Eq. (S47) by numerical integration. A single fit parameter is fixed by independent measurement of the ensemble variance in the absence of apoptotic stimuli. The other parameters are read off from the simulation. Both the simulation (solid line) and the semi-analytic result (dotted line) are rescaled by $t_r$. Two different potential parametrisations are considered in (a) with $V'(c)t_r = 0.01(c(2.5 - 3.6c + 0.25c^2))$ and in (b) with $V'(c)t_r = 0.1(c(5 - 1.5c + 0.1c^2))$. For (a), the molecular diffusivity is $Dt_r = 6 \times 10^{-3}$ and (b) is $Dt_r = 2 \times 10^{-2}$. All simulation parameters are in units of the rescaled concentration. Find the model for $\mu = 0$ in appendix 3.5. The simulation parameters in (a) are the same as in Fig. S8 and S10. The average mitochondrial size is $\langle v \rangle = 2$. For(c-f) identical parameters to (a). (c) With reducing mitochondrial fusion and fragmentation rate $\mu$ the sigmoidal shape gets less distinct. (d) For $\mu t_r = 1.5$ the sigmoidalness is more emphasised for lower molecular diffusion constant. (e,f) Increasing the potential barrier gives more distinct sigmoidal responses. For (e) a skew is applied. For (f) the potential is multiplied with the strength parameter $\alpha$.

reach the high concentration state. We now relax this assumption by investigating two different scenarios to account for our current ignorance of the exact temporal dynamics of cytochrome c release and its effects on mitochondrial dynamics [19]. In the first scenario, we assume that mitochondria stop undergoing fusion and fission dynamics and release cytochrome c as they reach the high concentration state $c_h^*$. In the second scenario, we assume that mitochondria release cytochrome c after crossing the potential barrier and cease their fusion and fission activities. For both scenarios, we model the lysis of mitochondria as an absorbing boundary condition in our system. Instead of tracking the mean amount of membrane-bound Bax, we directly quantify the mass fraction of Cyt c releasing mitochondria $m_c$. Recall that the effective attraction to the mean in the approximation Eq. (S26) formally requires us to exclude absorbed mitochondria from the computation of the mean.

In accordance with the framework of Smoluchowski aggregation-fragmentation dynamics, we set the rate of mitochondrial fusion proportional to the mass of mitochondria not releasing Cyt c, $\mu \propto m_0 - m_c$. This can be made intuitive by assuming that, with a reduced number of mitochondria, the mitochondrial density and the rate at which two mitochondria meet decrease.

Intuitively, we can understand the sigmoidal response dynamics by tracking the dynamics from the perspective of individual mitochondria. Initially, the stochastic escape is suppressed as mitochondria are effectively pulled back to the low-concentration state on short timescales. As the mean and median of the distribution have crossed the potential, the dynamics shift and mitochondria remaining in the low concentration state are now actively pulled across the potential barrier by mitochondria compartment fusion and subsequent fragmentation. Also in the case of an absorbing boundary condition at the potential barrier, the stochastic escape is initially suppressed on short timescales, yet the active pulling over the barrier is absent on long timescales. Instead, the sigmoidal shape results from the steady decrease of the mitochondrial fusion rate $\mu$ as the mitochondrial mass are reduced. Consequently, the suppression of the stochastic escape vanishes as the mitochondrial compartments get absorbed.

As previously demonstrated, the sigmoidal response dynamics is a general feature of compartmentalised stochastic reaction kinetics systems when subjected to dynamics in bistable potentials. Our qualitative findings are not affected by the details of the parametrisation of the model in Eq. (S66). We now conclude this chapter by discussing the effects of multiplicative noise. In the case of absorption at the potential barrier, corrections due to multiplicative noise [40, 41] can be straightforwardly accounted for as corrections of the Kramer's escape rate. We do not anticipate that multiplicative noise will qualitatively impact the sigmoidal relaxation. In the absence of absorption, a more refined analysis can formally account for multiplicative noise in Eq. (S54). Nevertheless, we expect that multiplicative noise will have qualitative effects similar to those of an additional drift on the system. For the quasi-particle, the assumption of fixed dispersion should be carefully re-evaluated, as the variance should reflect the multiplicative character of the noise. While multiplicative noise complicates the analysis, we anticipate that the sigmoidal response dynamics will remain in the presence of multiplicative noise.

Concluding, we find that a sigmoidal response is a qualitative characteris-

tic of compartmentalised stochastic reaction kinetics systems if the transition between two stable states is evoked by an external stimulus. In the case of apoptotic decision-making, this results in the initially suppressed response to weak apoptotic stimuli on a short timescale. Yet, conversely, on longer timescales, compartment fusion and fragmentation help to facilitate the response to the kinetics on longer timescales. This section has focused on the response to constant, non-fluctuating external stimuli. The next section will generalise the qualitative insights gained to general, fluctuating signals $\eta(t)$

### 3.4. Fluctuating stimuli and the emergence of a kinetic low-pass filter

In section 3.3, we studied the response dynamics of compartmentalised stochastic reaction kinetics systems applied to the context of apoptotic decision-making for constant apoptotic stimuli. While the response dynamics for strong apoptotic stimuli are only slightly affected by mitochondrial (compartment) dynamics, the kinetics differ strongly when considering mitochondrial dynamics in the context of weak apoptotic stimuli. We depicted the collective ensemble dynamics in the presence of mitochondrial fusion and fragmentation with a quasi-particle-like ensemble configuration and observed sigmoidal relaxation dynamics to the high Bax concentration state in response to weak apoptotic stimuli. We further established that the shape and timing of the transition between low and high Bax concentration states depend on the reaction network kinetics and require detailed quantification of reaction rates and concentrations under physiological conditions.

Considering that the interior of cells constitutes a highly stochastic environment, where proteins are continually synthesised and degraded, we generalise in this section to fluctuating stimuli and show that the interplay of protein complex dynamics and mitochondrial fusion and fragmentation act as a kinetic low-pass filter. This allows cells to suppress responses to fast noise fluctuations while allowing for responses to slow, biologically relevant changes in their environment.

At the molecular level, the reaction network among proteins in the Bcl-2 family mediates the release of the toxin Cyt c for individual mitochondria. As discussed in section 3.0.1, this family includes a variety of pro-apoptotic and anti-apoptotic species, which collectively constitute the cell's stress level and set the strength of an apoptotic stimulus. This, in particular, enables the cell to respond to a wide range of different stress triggers, such as oxidative stress, radiation stress, metabolic stress, toxin-induced stress and DNA damage. Recognising the interior of the cell as a highly stochastic environment, we anticipate that the level of pro- and anti-apoptotic species will vary over time.

To investigate the effects of fluctuating stimuli, we consider that the collective stress level $\eta(t)$ follows a Gaussian stochastic process. We set $\eta(t)$ to follow an Ornstein-Uhlenbeck process, $d\eta = \theta_{\text{OU}}(\eta_0 - \eta)dt + \sqrt{D_{\text{OU}}}dW$, where $W$ is a standard Wiener process. We scale $\theta_{\text{OU}} \propto \tau^{-1}$ and $D_{\text{OU}} \propto \tau^{-1}$, with $\tau$ the timescale of the Ornstein-Uhlenbeck process. This scaling allows us to systematically vary *how fast $\eta(t)$* fluctuates while keeping the amplitude of the fluctuations constant, as $\lim_{t\to\infty} \text{Var}(\eta(t)_\tau) = \text{const.}$. We assume that the average of $\eta(t)$ is at the level of a weak-apoptotic stimulus and $\sqrt{\text{Var}_t(\eta(t))} + \langle\eta(t)\rangle_t < \eta_s$ so that the amplitude of $\eta(t)$ remains in the region of weak apoptotic stimuli in a

$\sigma$-environment. Nevertheless, this allows for the stochastic occurrence of stress levels $\eta$ which correspond to strong apoptotic stimuli.

In Fig. 5 (f) in the main text of the manuscript, we qualitatively compare the responses of two ensembles with and without fusion and fragmentation dynamics to a fluctuating stimulus $\eta(t)$. To this end, we perform full stochastic simulations. We use the same simulation parameter as in Fig. S11 b for $\eta(t) = 0$ and consider the noise as linear skew on the potential. We consider as noise amplitude a variation of $\sigma_\eta = 0.25\Delta E$ of the barrier height. For technical reasons, we neglect the lysis of mitochondria subsequent to the accumulation of membrane-bound Bax. Experimentally, this can be achieved by inhibiting Bax homo-oligomerisation [21]. We study the relaxation of this assumption later in this section. The realisation of the fluctuating stress level $\eta(t)$ is presented as a grey profile line. For the ensemble with no fusion and fragmentation, we find that the mean Bax-membrane concentration in the ensemble $\langle c \rangle$ closely follows the fluctuations of the stress level. This corresponds to an intermediate level of mitochondria in the high Bax concentration state. The situation is different for an ensemble with mitochondrial fusion and fragmentation, where we observe that the Bax concentration predominantly resides in the low Bax membrane concentration state, with only a small subset of mitochondria in the Bax concentration state. When the stress level $\eta(t)$ is persistent on an increased level, we find an abrupt change from the low to the high Bax concentration state and a strong response of the system. This qualitatively generalises the idea of sigmoidal response kinetics for fixed apoptotic stimuli, as demonstrated in section 3.3.

Next, we formalise the observations by systematically varying the timescale of the fluctuating apoptotic stimulus and setting it in relation to the timescale of the ensemble response. We determine the timescale by measuring the auto-correlation time of both the stimulus $\eta(t)$ and the fluctuating mean Bax concentration $\langle c(t) \rangle$. Specifically, the autocorrelation with a lag $\Delta t$ is

$$\rho_{\Delta t} = \frac{\text{Cov}[X_t, X_{t+\Delta t}]}{\text{Var}[X_t]} \tag{S69}$$

and we define the auto-correlation time $\tau$ by approximating $\rho_{\Delta t} \approx \exp(-\Delta t/\tau^{-1})$. Here, we use the same definition to determine both the auto-correlation time of the stimulus $\tau_\eta$ and the ensemble response $\tau_c$. Note that macroscopic changes in the time signal dominate the auto-correlation time, which renders it particularly suitable for the detection of an ensemble transition from the low to the high concentration state. We interpret the ratio $R = \tau_\eta/\tau_c$ as a measure of the responsiveness of the system. If fluctuations in the stimulus $\eta(t)$ are readily translated into a response of the ensemble $\langle c(t) \rangle$, both the stimulus and the ensemble show macroscopic changes on the same timescale, with the ratio close to unity, $R \approx 1$. Conversely, for $R \ll 1$, the timescale of macroscopic concentration changes in the ensemble is much larger than macroscopic changes in the stimulus, implying that fluctuations in the stimulus $\eta(t)$ are translated into a system's response only with a low frequency. In this case, the fluctuations are filtered out, as they are suppressed by the ensemble dynamics. For the case $R \gg 1$, macroscopic concentration changes in the ensemble happen more frequently compared to the changes in the stimulus. In this case, the stimulus has only a marginal effect

on the progression of apoptosis. Note that we expect to occlude the last case by construction[5]. By construction, the timescale of the stimulus $\eta(t)$ is given by $\tau_\eta = \theta_{\text{OU}}^{-1}$. By rescaling the diffusivity in the Ornstein-Uhlenbeck process accordingly to receive a constant variance scaling, we guarantee that changes in the auto-correlation $\tau_c$ are purely evoked by *how fast* the stimulus is fluctuating.

In Fig. 5 (f) in the main text of the manuscript, we further systematically map the responsiveness $R$ over the inverse timescale of the fluctuating stimulus $\eta(t)$ for different rates of mitochondrial fusion and fragmentation $\mu$. Here, we keep the size distribution of mitochondria fixed for all different mitochondrial fusion rates $\mu$. Fast noise fluctuations in $\eta(t)$ correspond to low values of $\tau$ and are shown on right side of the x-axis, while slow changes in $\eta(t)$ are shown on the left side. We find that for all mitochondrial fusion and fragmentation rates considered here, the responsiveness approaches $R = 1$ for slowly varying stimuli $\eta(t)$. In contrast, a mitochondrial ensemble with higher fusion and fragmentation rates approaches lower values in the responsiveness $R_\mu \ll R_{\mu'}$ for higher fusion rates $\mu > \mu'$. We find that mitochondrial dynamics facilitate the suppression of fast noise fluctuations of the stimulus with an order-of-magnitude effect.

Qualitatively, this is in line with the visual observation we made on the trajectory pictures, as the response to stimulus $\eta(t)$ is suppressed on short timescales and facilitated on long timescales. In agreement with the observation that faster rates of mitochondrial fusion correspond to stronger suppression of the response, we find that $R_\mu < R_{\mu'}$ for $\mu < \mu'$ on all timescales $\tau_\eta$. As demonstrated for the response to fixed apoptotic stimuli, the filtering capacity of the system should be viewed as a qualitative result, as the quantitative values are strongly dependent on the specifics of the Bcl-2 reaction network. Additionally, the transition from the responsive dynamics $R(\tau_\eta) \approx 1$ to $R(\tau_\eta') \ll 1$ is sensitive to the strength of fluctuations $\text{Var}_t(\eta(t))$. An order of magnitude estimate suggests that the transition $\tau_\eta^*$ lies on the order of magnitude of the Bax translocation time $t_r$. We conclude that mitochondrial dynamics give rise to a kinetic low-pass filter, allowing cells to repress the response to short, transient stress fluctuations.

For the analysis presented here, we considered the absence of mitochondrial lysis following the release of Cyt c, and allowed for the recovery of individual mitochondria from the high Bax concentration state back to the low Bax concentration state. This formally prevents apoptosis in cells, enabling us to create a large dataset on how Bax accumulation dynamics are affected by mitochondrial fusion and fragmentation. Note that the auto-correlation time is sensitive to macroscopic concentration changes. As such, the kinetic low pass filter quantifies both the accumulation and retro-translocation dynamics of Bax. If mitochondrial lysis is included, only the accumulation dynamics of Bax are quantifiable; however, as discussed [1], we anticipate that this will still yield a sigmoidal response kinetics. Consequently, we expect the qualitative finding of mitochondrial fusion and fragmentation giving rise to a kinetic low-pass filter to be robust in the presence of the progression of the mitochondrial signalling pathway. Next, we interpret this finding in the context of apoptotic decision-making,

---

[5]In experimental setups, $R \gg 1$ hints towards the fact that the wrong stimulus was measured.

focusing on the sensitivity and specificity of the response dynamics.

### 3.5. Modification of Kramer's escape rate due to heterogeneous size distributions

Heterogeneous size distributions of organelle ensembles without organelle dynamics changing the sizes of individual organelles suggest treating organelle subpopulations of identical sizes independently, rather than marginalising over organelle sizes. In the context of the random switching dynamics due to apoptotic stimuli, as presented in Section 3.3, separate Kramer's escape problems can be considered for each subpopulation of a specified organelle size. The backward reaction from the high concentration state to the low concentration state is assumed to be negligible. The Kramer's rate can be computed using standard methods, such as those outlined in [42]. The Kramer's escape rate for the subpopulation of size $s$ is symbolically denoted by $\omega(s)$, and the total mass in a subpopulation is defined as $m_s$. Then, the time evolution of mitochondrial mass in the low concentration state yields

$$m(t) = \sum_s m_s \exp(-\omega(s)t). \tag{S70}$$

Defining the high concentration state at $c_h$ and the low concentration state at $c_l$, and assuming that the Kramer's limit is met for all $s$, we approximate the mean of the distributions in response to an apoptotic stimulus

$$\langle c(t) \rangle \approx c_l \cdot \left( \sum_s \frac{m_s}{m(0)} \exp(-\omega(s)t) \right) + c_h \cdot \left( 1 - \sum_s \frac{m_s}{m(0)} \exp(-\omega(s)t) \right). \tag{S71}$$

For given potentials, this equation yields analytic solutions with no free fit parameter.

### 4. NUMERICAL ROUTINE FOR THE SIMULATION OF COMPARTMENTALISED SYSTEMS

In order to investigate the multi-scale dynamics of compartmentalised stochastic reaction kinetics systems, we sample random trajectories of the dynamics specified by the Master equation (S8). For this purpose, we employ the direct method of the *stochastic simulation algorithm* [43, 44]. This formulation enables us to implement the dynamics straightforwardly. However, due to the multi-scale nature of the dynamics, the computational costs tend to become prohibitively high when large systems with large concentration vectors $\vec{c}_i$ are simulated, as this drastically slows down the run time. Therefore, we introduce additional approximations, which allow for a drastic reduction in computational time by assuming that compartmental dynamics typically proceed at slower timescales than chemical reactions.

In order to implement the dynamics specified by the Master equation Eq. (S8), we associate each possible transition with a rate $\mathcal{T}_{S \to S'} \equiv \mathcal{T}_j$, which is commonly referred to as the propensity of transition $j$ [44]. The ratio $\mathcal{T}_j / \sum_l \mathcal{T}_l$ represents the probability of taking transition $j$ in order to move to the subsequent step,

with the sum running over all possible transitions between state $\mathbb{S}$ and another state $\mathbb{S}'$. By drawing two random uniform numbers $r_1$ and $r_2$ the system changes its state after the time step

$$\tau = \frac{1}{\sum_l \mathcal{T}_l} \log\left(\frac{1}{r_1}\right) \tag{S72}$$

according to the transition specified by the $j$, which fulfils being the smallest integer satisfying

$$\sum_l^j \mathcal{T}_l > r_2 \sum_l^j \mathcal{T}_l. \tag{S73}$$

This routine updates the system in sequential order without making any assumption on the timescale of the specific reactions. Due to the large number of possible transitions, the stochastic evolution of this dynamics typically shows long run times prohibiting extensive simulations of large systems.

In order to enhance the efficiency of our code, we account explicitly for the multi-scale organisation of compartmentalised stochastic reaction kinetics systems. We make a distinction between compartment dynamics and inter-scale fluxes resulting from chemical reactions. The compartment dynamics are simulated using the Master equation, along with a stochastic simulation algorithm, while the concentration $\vec{c}_i$ in each compartment is evolved parallel to this, at intervals of duration $\tau_o$. By splitting the code in this way, it can be parallelised, thus significantly reducing the execution time. This is especially beneficial when the dynamics of the compartments are often slower than those on the molecular level.

We might further optimise the run time of the algorithm by additionally parallelising the compartment dynamics. To do this, we discretise the evolution steps of the compartment dynamics to a fixed $\tau_o$, and assign for each compartment probabilities for undergoing compartment dynamics after the next step, for example fusion with another compartment. This routine accurately recreates the size distribution predicted by the compartment aggregation-fragmentation dynamics. Additionally, for large systems with large concentration vectors $\vec{c}_i$, memory management is optimised by assigning compartments of discrete sizes. For each *elementary* compartment block, memory capacities are pre-allocated, in which the composition vector and the transition matrix of the chemical reaction network are stored and updated. We have implemented the multi-scale dynamics both in `Python` and `C++`; while `Python` allows for simple and intuitive exploration of the phenomena encoded in the multi-scale dynamics, `C++` provides opportunities for efficiency optimisation of the code through controlling and pre-allocating the memory demands.



\end{document}